\documentclass[12pt,a4wide,epsf]{article}
\usepackage{epsf}
\usepackage[hang,nooneline]{subfigure}

\usepackage{graphicx}

\newcommand{\insertplot}[5]{\begin{figure}
 \hfill\hbox to 0.05in{\vbox to #5in{\vfill
 \inputplot{#1}{#4}{#5}}\hfill}
 \hfill\vspace{-.1in}
 \caption{#2}\label{#3}
 \end{figure}}
 \newcommand{\inputplot}[3]{
 \special{ps: plotfile #1}
}
\newcounter{fig}   

\newcommand{\vphi}{\varphi}
\newcommand{\vepsilon}{\varepsilon}
\newcommand{\DS}{\displaystyle}

\usepackage{graphicx}

\begin{document}

\title{
Properties of rotating Einstein-Maxwell-Dilaton black holes
in odd dimensions}

\vspace{1.5truecm}
\author{
{\bf Jose Luis Bl\'azquez-Salcedo$^1$},
{\bf Jutta Kunz$^2$},
{\bf Francisco Navarro-L\'erida$^3$}\\
%
$^1$
Dept.~de F\'{\i}sica Te\'orica II, Ciencias F\'{\i}sicas\\
Universidad Complutense de Madrid, E-28040 Madrid, Spain\\
$^2$ 
Institut f\"ur  Physik, Universit\"at Oldenburg\\ Postfach 2503,
D-26111 Oldenburg, Germany\\
$^3$ 
Dept.~de F\'{\i}sica At\'omica, Molecular y Nuclear, Ciencias F\'{\i}sicas\\
Universidad Complutense de Madrid, E-28040 Madrid, Spain
}

\vspace{1.5truecm}

\date{\today}

\maketitle
\vspace{1.0truecm}

\begin{abstract}
We investigate rotating Einstein-Maxwell-Dilaton (EMd) black holes in odd 
dimensions.  Focusing on black holes with equal-magnitude angular momenta, 
we determine the domain of existence of these black holes. Non-extremal 
black holes reside with the boundaries determined by the static and the 
extremal rotating black holes. 
The extremal EMd black holes show proportionality of their horizon area
and their angular momenta. Thus the charge does not enter.
We also address the Einstein-Maxwell case, where the extremal rotating black 
holes exhibit two branches. On the branch emerging from the Myers-Perry 
solutions their angular momenta are proportional to their horizon area, whereas
on the branch emerging from the static solutions their angular momenta are
proportional to their horizon angular momenta. 
Only subsets of the near-horizon solutions are realized globally.
Investigating the physical properties of these EMd black holes, 
we note that one can learn much
about the extremal rotating solutions from the much simpler static solutions.
The angular momenta of the extremal black holes are proportional to the
area of the static ones for the Kaluza-Klein value of the dilaton
coupling constant, and remain analogous for other values.
The same is found for the horizon angular velocities of the extremal black 
holes, which possess an analogous behavior to the surface gravity of the
static black holes.
The gyromagnetic ratio is rather well approximated by the `static' value,
obtained perturbatively for small angular momenta.
\end{abstract}


\vfill\eject

\section{Introduction}

Discovered in 1986 by Myers and Perry (MP) \cite{Myers:1986un,Myers:2011yc},
the rotating higher-dimensional vacuum black holes 
have been generalized to include various types of matter fields
(see e.g.~\cite{Youm:1997hw,Maeda:2011sh,Marolf:2011zs}),
as motivated by supergravity and string theory.
New analytical black hole solutions can be obtained by a number of
solution generation techniques.
A straightforward method is based on the Kaluza-Klein (KK) reduction.
In the simplest case this leads to electrically
charged Einstein-Maxwell-dilaton (EMd) black holes
for a particular value of the dilaton coupling constant,
$h=h_{\rm KK}$
\cite{Chodos:1980df,Frolov:1987rj,Horne:1992zy,Kunz:2006jd}.
For general dilaton coupling constant $h$, however,
rotating EMd black hole solutions
or their Einstein-Maxwell (EM) counterparts
need either perturbative techniques
\cite{Aliev:2004ec,Aliev:2005npa,Aliev:2006yk,NavarroLerida:2007ez,Sheykhi:2008bs,Allahverdizadeh:2010xx,Allahverdizadeh:2010fn}
or numerical analysis 
\cite{Kunz:2005nm,Kunz:2006eh,Blazquez-Salcedo:2013yba}.

General stationary black holes in $D$ dimensions 
possess $N=[(D-1)/2]$ independent angular momenta $J_i$,
associated with $N$ orthogonal planes of rotation \cite{Myers:1986un},
where $N$ is the integer part of $(D-1)/2$, corresponding to the
rank of the rotation group $SO(D-1)$. 
In odd-$D$ dimensions,
when all $N$ angular momenta have equal-magnitude,
the symmetry of the solutions is enhanced.
The EMd equations then simplify,
leading to cohomogeneity-1 equations.

Here we consider such cohomogeneity-1 EMd black holes.
On the one hand, we solve the coupled system of EMd equations
numerically and study the properties of the black holes
both for extremal and non-extremal solutions.
On the other hand, we derive the near-horizon solutions
for the extremal black holes. 
The physical properties of non-extremal black holes,
like the horizon area, 
the gyromagnetic ratio,
or the surface gravity
then assume values that are bounded by those of the extremal
black holes and those of the static black holes.

For extremal MP black holes in odd-$D$ dimensions
the horizon area $A_{\rm H}$ is proportional to
the magnitude of the equal angular momenta, $J= |J_i|$,
$J = \sqrt{2(D-3)} A_{\rm H}$.
This relation for extremal black holes represents the
limiting case of a more general relation for
MP black holes in terms of the inner and outer horizon areas
of non-extremal black holes
\cite{Cvetic:2010mn}, and was pointed out in four dimensions before
\cite{Ansorg:2007fh,Hennig:2008yw,Ansorg:2008bv,Hennig:2008zy,Ansorg:2009yi,Hennig:2009aa,Ansorg:2010ru}.
In the presence of charge, the relation generalizes, and
the product of the horizon areas 
can typically be written as a sum between the squares of the
angular momenta and some powers of the charges
\cite{Cvetic:2010mn,Ansorg:2007fh,Hennig:2008yw,Ansorg:2008bv,Hennig:2008zy,Ansorg:2009yi,Hennig:2009aa,Ansorg:2010ru,Castro:2012av}.
Area-angular momentum-charge inequalities for stable marginally 
outer trapped surfaces were studied for EMd theory in
\cite{Yazadjiev:2012bx,Yazadjiev:2013hk}.

Considering such area-angular momentum relations for extremal EMd 
and EM black holes,
we find that the EM case is special, since there are
two branches of extremal solutions. The first branch
emerges from the MP black holes and retains the proportionality
between angular momentum and area.
Thus the charge does not enter here.
The second branch emerges from the 
higher-dimensional Reissner-Nordstr\"om (RN) black holes.
Here the proportionality between angular momentum and area is lost.
Instead there is a new proportionality between angular momentum
and horizon angular momentum
along this second branch.
As soon as the dilaton is coupled, however, the second branch
disappears and the proportionality
between angular momentum and area persists for
all extremal EMd solutions, independent of the dilaton
coupling constant $h$.

The paper is organized as follows.
In section 2 we present the action, the parametrization for
the metric and the fields, as well as the general formulae
for the physical properties.
In section 3 we briefly recall 
the analytically known Kaluza-Klein black holes and their properties.
We derive the near-horizon solutions in section 4, discussing
in particular, the two-branch structure in the EM case.
Section 5 contains our numerical results. Here we exhibit
the properties of five-dimensional black holes in detail,
and demonstrate subsequently, that the pattern observed in five dimensions
is generic for higher dimensions.
We present our conclusions in section 6.

\section{Action, Ans\"atze and Charges}

\subsection{Einstein-Maxwell-dilaton action}

We consider the $D$-dimensional EMd action
\begin{equation} \label{EMDac}
I= \int d^Dx\sqrt{-g} \biggl[ R
 -\frac{1}{2}\partial_\mu\phi \, \partial^\mu\phi
 -\frac{1}{4}e^{-2h\phi}F_{\mu\nu}F^{\mu\nu} \biggr ] \ , 
\end{equation}
with curvature scalar $R$,
dilaton field $\phi$,
dilaton coupling constant $h$,
and field strength tensor
$ F_{\mu \nu} = \partial_\mu A_\nu -\partial_\nu A_\mu $,
where $A_\mu $ denotes the gauge potential. 
We choose units such that $16 \pi G_D =1$, where $G_D$ 
is the $D$-dimensional Newton constant.

Variation of the action with respect to the metric
leads to the Einstein equations 
\begin{equation}
R_{\mu\nu}-\frac{1}{2}g_{\mu\nu}R = \frac{1}{2} \, T_{\mu\nu}
\  , \label{ee}
\end{equation}
with stress-energy tensor
\begin{equation}
T_{\mu \nu} = 
  \partial_\mu \phi \, \partial_\nu\phi
 - \frac{1}{2}g_{\mu\nu}\,
 \partial_\rho\phi\, \partial^\rho\phi \,
+
  e^{-2h\phi} \left( F_{\mu\rho} {F_\nu}^\rho 
  - \frac{1}{4} g_{\mu \nu} F_{\rho \sigma} F^{\rho \sigma} \right)  
\ , \label{tmunu}
\end{equation}
whereas variation with respect to the fields yields
for the Maxwell field 
\begin{equation}
\nabla_\mu \left( e^{-2h\phi}\, F^{\mu\nu} \right)  = 0 
\ , \label{feqA}
\end{equation}
and for the dilaton field
\begin{equation}
\nabla_\mu \nabla^\mu\phi
 = -\frac{h}{2} \, e^{-2h\phi} \, F_{\mu\nu}F^{\mu\nu}
\ . \label{edil}
\end{equation}

Special cases correspond to Einstein-Maxwell theory,
where $h=0$,
and to Kaluza-Klein theory, where $h=h_{\rm KK}$,
\begin{equation}
h_{\rm KK} =\sqrt{ \frac{D-1}{2(D-2)} } 
\ . \label{h_KK} 
\end{equation}

\subsection{Ans\"atze}

To obtain stationary black hole solutions,
which represent generalizations of the 
Myers-Perry solutions \cite{Myers:1986un} to EMd theory,
we consider black hole spacetimes with $N$-azimuthal symmetries,
implying the existence of $N+1$ commuting Killing vectors,
\begin{equation}
\xi \equiv \partial_t \ ,  \ \ \
\eta_{(k)} \equiv \partial_{\vphi_k}
\ , \label{killing} \end{equation}
for $k=1, \dots , N$. Hence we are considering black holes with spherical
horizon topology.

While the general EMd black holes possess $N$ independent
angular momenta, we now restrict to black holes whose 
angular momenta have all equal-magnitude.
In odd dimensions,
the metric and the gauge field parametrization then
simplify considerably, and the problem reduces to cohomogeneity-1.
The EMd equations then form
a set of ordinary differential equations,
just as in Einstein-Maxwell theory
\cite{Kunz:2006eh}.

We parametrize the metric as follows
\cite{Kunz:2005nm,Kunz:2006eh}
\begin{eqnarray}
ds^2 &=&-f dt^2 + \frac{m}{f} \left[ dr^2 + r^2 \sum_{i=1}^{N-1}
  \left(\prod_{j=0}^{i-1} \cos^2\theta_j \right) d\theta_i^2\right] \nonumber \\
&&+\frac{n}{f} r^2 \sum_{i=1}^N \left( \prod_{j=0}^{i-1} \cos^2 \theta_j
  \right) \sin^2\theta_i \left(\vepsilon_i d\vphi_i - \frac{\omega}{r}
  dt\right)^2 \nonumber \\
&&+\frac{m-n}{f} r^2 \left\{ \sum_{i=1}^N \left( \prod_{j=0}^{i-1} \cos^2
  \theta_j \right) \sin^2\theta_i  d\vphi_i^2 \right. \nonumber\\
&& -\left. \left[\sum_{i=1}^N \left( \prod_{j=0}^{i-1} \cos^2
  \theta_j \right) \sin^2\theta_i \vepsilon_i d\vphi_i \right]^2 \right\} 
\ , \label{metric} \end{eqnarray}
where $\theta_0 \equiv 0$, $\theta_i \in [0,\pi/2]$
for $i=1,\dots , N-1$,
$\theta_N \equiv \pi/2$, $\vphi_k \in [0,2\pi]$ for $k=1,\dots , N$,
and $\vepsilon_k = \pm 1$ denotes the sense of rotation
in the $k$-th orthogonal plane of rotation.

An adequate parametrization for the gauge potential is given by
\begin{equation}
A_\mu dx^\mu =  a_0 dt + a_\vphi \sum_{i=1}^N \left(\prod_{j=0}^{i-1}
  \cos^2\theta_j\right) \sin^2\theta_i \vepsilon_i d\vphi_i 
\ . \label{gaugepotential}
\end{equation}
Independent of the number of dimensions $D$,
this parametrization involves
four functions for the metric, $f$, $m$, $n$, $\omega$,
two functions for the gauge field, $a_0$, $a_\vphi$,
and one function for the dilaton field, $\phi$,
all of which depend only on the radial coordinate $r$.

\subsection{Global charges}

We consider asymptotically flat solutions.
Thus, asymptotically the metric should approach the Minkowski metric.

The mass $M$ and the angular momenta $J_{(k)}$ of the black holes 
are obtained from the Komar expressions 
associated with the respective Killing vector fields
\begin{equation}
M = -  \frac{D-2}{D-3} \int_{S_{\infty}^{D-2}} \alpha  
\ , \label{Kmass} \end{equation}
\begin{equation}
J_{(k)} =   \int_{S_{\infty}^{D-2}} \beta_{(k)} 
\ , \label{Kang} \end{equation}
where $\alpha_{\mu_1 \dots \mu_{D-2}} \equiv \epsilon_{\mu_1 \dots \mu_{D-2}
  \rho \sigma} \nabla^\rho \xi^\sigma$,
and
$\beta_{ (k) \mu_1 \dots \mu_{D-2}} \equiv \epsilon_{\mu_1 \dots \mu_{D-2}
  \rho \sigma} \nabla^\rho \eta_{(k)}^\sigma$.
Equal-magnitude angular momenta then satisfy $|J_{(k)}|=J$, 
$k=1, \dots , N$.

The electric charge $Q$ is obtained from
\begin{equation}
Q= - \frac{1}{2} \int_{S_{\infty}^{D-2}} e^{-2 h \phi} \ {}^* F 
\ , \label{charge} \end{equation}
with 
${}^* F_{\mu_1 \dots \mu_{D-2}} \equiv  
  \epsilon_{\mu_1 \dots \mu_{D-2} \rho \sigma} F^{\rho \sigma}$.
The magnetic moment $\mu_{\rm mag}$ is determined from the
asymptotic expansion of the gauge potential $a_{\varphi}$.
The gyromagnetic ratio $g$ is then obtained from the
magnetic moment $\mu_{\rm mag}$ via
\begin{equation}
{\mu_{\rm mag}}=g \frac{Q J}{2M}
\ . \label{gyro} 
\end{equation}

The dilaton charge $\Sigma$ is defined via
\begin{equation}
\Sigma = -A(S^{D-2})\lim_{r\rightarrow\infty}{r^{D-2}\frac{d\phi}{dr}} \ , \label{Sigma}
\end{equation}
where $A(S^{D-2})$ is the area of the $S^{D-2}$ sphere.

\subsection{Horizon properties}

The event horizon is located at $r=r_{\rm H}$.
Here the Killing vector $\chi$,
\begin{equation}
\chi = \xi + \Omega \sum_{k=1}^N \vepsilon_k \eta_{(k)} \ ,
\label{chi} \end{equation}
is null,
and $\Omega$ represents the horizon angular velocity.
Without loss of generality, $\Omega$ is assumed
to be non-negative, any negative sign being included in $\vepsilon_k$.

The area of the horizon $A_{\rm H}$ is given by
\begin{equation}
A_{\rm H}=\int_{{\cal H}} \sqrt{|g^{(D-2)}|}=r_{\rm H}^{D-2} A(S^{D-2}) \lim_{r \to r_{\rm H}}
 \sqrt{\frac{m^{D-3} n}{f^{D-2}}} \label{hor_area} \ , \end{equation}
and the surface gravity $\kappa$ reads
\begin{equation}
\kappa^2 = \left. -\frac{1}{2}  
      (\nabla_\mu \chi_\nu) (\nabla^\mu \chi^\nu)
                    \right|_{\cal H}  
 =  \lim_{r \to r_{\rm H}} \frac{f}{(r-r_{\rm H}) \sqrt{m}}
\ . \label{kappa} \end{equation}

The horizon mass $M_{\rm H}$ and horizon angular momenta
$J_{{\rm H} (k)}$ are given by
\begin{equation}
M_{\rm H} = - \frac{D-2}{D-3} \int_{{\cal H}} \alpha 
\ , \label{Hmass} \end{equation}
\begin{equation}
J_{{\rm H} (k)} =   \int_{{\cal H}} \beta_{(k)} \ 
\ , \label{Hang} \end{equation}
where ${\cal H}$ represents the surface of the horizon.
For equal-magnitude angular momenta
$|J_{{\rm H} (k)}| =J_{\rm H}$, $k=1, \dots , N$.

The electric charge $Q$ can also be determined at the horizon via
\begin{equation}
Q= - \frac{1}{2} \int_{{\cal H}}  e^{-2 h \phi} \ {}^* F
\ . \label{Hcharge} 
\end{equation}
The horizon electrostatic potential $\Phi_{\rm H}$ is defined by
\begin{equation}
\Phi_{\rm H} = \left. \chi^\mu A_\mu \right|_{r=r_{\rm H}} 
\ . \label{Phi} 
\end{equation}
$\Phi_{\rm H}$ is constant at the horizon.

\subsection{Mass formula}

The Smarr mass formula for Einstein-Maxwell black holes
with $N$ equal-magnitude angular momenta 
reads \cite{Kunz:2005nm,Gauntlett:1998fz} 
\begin{equation}
\frac{D-3}{D-2} M = 2 \kappa A_{\rm H} + N \Omega
J  + \frac{D-3}{D-2} \Phi_{\rm H} Q  \ . \label{smarr}
\end{equation}
This mass formula holds also in the presence of a dilaton field
\cite{Kunz:2006jd}.

Introducing the dilaton charge into the mass formula via
\cite{Kunz:2006jd}
\begin{equation}
\frac{\Sigma}{h} = - \Phi_{\rm H}Q 
\ , \label{dilaton_rel} 
\end{equation}
Eq.~(\ref{smarr}) yields the Smarr mass formula
\begin{equation}
M= 2\frac{D-2}{D-3}\kappa A_{\rm H} + \frac{D-2}{D-3}
N \Omega J + 2 \Phi_{\rm H}Q + \frac{\Sigma}{h} \ , \label{mass2}
\end{equation}
known to hold also for non-Abelian black holes (in $D=4$) 
\cite{Kleihaus:2002tc}.

\subsection{Scaling symmetry}

We note that the solutions have a scaling
symmetry, e.g.,
\begin{equation}
 \tilde M= \tau^{D-3} M \, \ \ \ 
 \tilde J_i= \tau^{D-2} J_i \, \ \ \ 
 \tilde Q = \tau^{D-3} Q \ ,
\label{scale1} \end{equation}
\begin{equation}
 \tilde r_{\rm H}=\tau r_{\rm H} \ ,\ \ \ 
 \tilde \Omega = \Omega/\tau \ ,\ \ \ 
 \tilde \kappa = \kappa/\tau \ ,
\label{scaling} \end{equation}
etc.

Let us therefore introduce scaled quantities,
where we scale with respect to appropriate powers
of the mass. These scaled quantities include
the scaled angular momentum
$j=|J|/M^{(D-2)/(D-3)}$,
the scaled charge $q=|Q|/M$,
the scaled area $a_{\rm H} = A_{\rm H}/M^{(D-2)/(D-3)}$,
the scaled surface gravity $\bar \kappa = \kappa M^{1/(D-3)}$,
and the scaled horizon angular velocity $\bar \Omega = \Omega M^{1/(D-3)}$.

In terms of the scaled quantities, the Smarr relation Eq.~(\ref{smarr}) reads
\begin{equation}
1= 2\frac{D-2}{D-3}\bar \kappa a_{\rm H} + \frac{D-2}{D-3}
N \bar \Omega j +  \Phi_{\rm H}q \ . \label{mass3}
\end{equation}

\section{Kaluza-Klein black holes} \label{sec_KK}

A straightforward method to obtain charged rotating black hole
solutions in $D$ dimensions is based on the Kaluza-Klein reduction.
Here one embeds
a $D$-dimensional vacuum solution in $D+1$ dimensions,
performs a boost transformation,
and then reduces the solution to $D$ dimensions
\cite{Chodos:1980df,Frolov:1987rj,Horne:1992zy,Llatas:1996gh,Kunz:2006jd}.

The KK black holes, obtained in this way
for the particular value of the dilaton coupling constant $h=h_{\rm KK}$,
\begin{equation}
h_{\rm KK}=\frac{D-1}{\sqrt{2(D-1)(D-2)}} = (D-1) \iota 
\ , \label{h_def} 
\end{equation}
also satisfy the quadratic relation
\cite{Gibbons:1985ac,Rasheed:1995zv,Kunz:2006jd}
\begin{equation}
\DS \frac{Q^2}{M -\DS \frac{2(D-2)\iota}{D-3}\Sigma} = - 2 (D-3) \iota \Sigma 
\ . \label{quadratic_relation} 
\end{equation}
This relation determines the dilaton charge
in terms of the mass and the electric charge,
while the angular momenta do not enter.

In contrast, the static black hole solutions are known analytically
for arbitrary value of the dilaton coupling constant 
\cite{Gibbons:1987ps,Garfinkle:1990qj,Gregory:1992kr}.
It should be interesting to find a generalization of 
relation Eq.~(\ref{quadratic_relation})
for general values of the coupling constant $h$. 

\subsection{Physical properties}\label{propKK}

Let us briefly recall some properties of these KK black holes.
In terms of parameters $m$, $a_i$, and $\alpha$, their mass $M$, angular
momenta $J_i$, charge $Q$, magnetic momenta $\mu_{{\rm mag}, i}$, and dilaton charge
$\Sigma$ are given by
\begin{equation}
M=m\left(1+(D-3)\cosh^2\alpha\right) A(S^{D-2}) \ , \label{mass}
\end{equation}
\begin{equation}
J_i = 2 m\, a_i \cosh\alpha A(S^{D-2}) \ , 
\label{angular_momenta}
\end{equation}
\begin{equation}
Q=(D-3)m\,\sinh\alpha\cosh\alpha A(S^{D-2}) \ , \label{electric_charge}
\end{equation}
\begin{equation}
\mu_{{\rm mag}, i} = (D-3) m\, a_i \sinh\alpha A(S^{D-2}) \ , 
\label{magntic_moments}
\end{equation}
\begin{equation}
\Sigma = - \frac{(D-3) m\, \sinh^2\alpha}{2 (D-2) \iota}  A(S^{D-2}) \ , \label{dilaton_charge}
\end{equation}
where $m$ and $a_i$ determine the MP mass and angular momenta in the limit,
where the boost parameter vanishes, $\alpha=0$.

The gyromagnetic ratios $g_i$ are given by
\begin{equation}
g_i = \frac{2M}{Q J_i}\mu_{{\rm mag}, i}  = (D-3) + \frac{1}{\cosh^2\alpha}
\equiv g \  . 
\label{gyromagnetic_ratios}
\end{equation}
Thus the gyromagnetic ratio $g$ depends only
on the charge-to-mass ratio, $q=Q/M$. 
It does not depend on the angular momenta.
Therefore, for a given value of $q$,
the gyromagnetic ratio is the same
for all rotating black holes.
It ranges between $g=D-2$ in the limit of
vanishing charge, $q \to 0$, and $g=D-3$
in the limit of maximal charge, $|q| \to 1$.

The event horizon of the KK black holes is characterized as the
largest non-negative root $\rho=\rho_{\rm H}$ of 
\begin{equation}
\Delta \equiv \prod_{i=1}^N (\rho^2 +a_i^2) -m \rho^{2-\vepsilon} \ ,
\label{Delta}
\end{equation}
where $\vepsilon$ takes the values $\vepsilon=0, 1$ for odd and even
dimensions, respectively. Notice that $\rho$ is a Boyer-Lindquist type of radial coordinate.  
The (constant) horizon angular velocities $\Omega_i$,
the horizon area $A_{\rm H}$, 
the surface gravity $\kappa$,
and the horizon electrostatic potential $\Phi_{\rm H}$
are given by
\begin{equation}
\Omega_i = \frac{a_i}{(\rho_{\rm H}^2+a_i^2) \cosh\alpha} \ ,  
\label{Omegas}
\end{equation}
\begin{equation}
A_{\rm H} = \frac{\cosh\alpha}{\rho_{\rm H}^{1-\vepsilon}} A(S^{D-2}) \prod_{i=1}^N
(\rho_{\rm H}^2 +a_i^2) \ , 
\label{area_hor}
\end{equation}
\begin{equation}
\left. \kappa = \frac{\Delta_{,\rho}}{2 m \rho_{\rm H}^{2-\vepsilon} \cosh\alpha}
\right|_{\rho=\rho_{\rm H}} \ , 
\label{surf_gravity}
\end{equation}
\begin{equation}
\Phi_{\rm H} = \frac{\sinh\alpha}{\cosh\alpha} \ .
\label{electros_pot}
\end{equation}
Thus, like $g$, the horizon electrostatic potential $\Phi_{\rm H}$
depends only on the charge-to-mass ratio $q$,
with $0 \le \Phi_{\rm H} \le 1$.

These KK black holes satisfy the general Smarr formula
\begin{equation}
M= 2\frac{D-2}{D-3}\kappa A_{\rm H} + \frac{D-2}{D-3}\sum_{i=1}^N
\Omega_i J_i + \Phi_{\rm H} Q \ , \label{smarrKK}
\end{equation}
which also holds for arbitrary dilaton coupling constant $h$.

\subsection{Extremal solutions}\label{sec_KK_rel}

We now turn to the extremal rotating KK black holes.
We focus on
black holes with equal-magnitude angular momenta, $J_i=J$, $i=1,...,N$.
First, we present the metric 
\begin{eqnarray} 
ds^2 &=& \left( 1 + \frac{m\sinh^2{\alpha}}{\rho^{\vepsilon}(\rho^2+a^2)^{N-1}}
\right)^{\frac{1}{D-2}} \left\{
-dt^2 + \frac{\rho^2(\rho^2+a^2)^{N-1}}{(\rho^2+a^2)^{N}-m\rho^{2-\vepsilon}}d\rho^2     \right. \nonumber
\\ &+&   \left.  (\rho^2+a^2)\sum_{i=1}^N(d\mu_i^2 + \mu_i^2d\phi_i^2) +
\vepsilon \rho^2 d\nu^2 + \frac{m\left(\cosh^2{\alpha}dt - a \displaystyle \sum_{i=1}^N
  \mu_i^2d\phi_i\right)^2}{\rho^{\vepsilon}(\rho^2+a^2)^{N-1}+m\sinh^2{\alpha}} 
\right\} \ ,
\label{KK_metric_extremal_equal_ai}
\end{eqnarray}
where $\mu_i$, the direction cosines, can be chosen in odd $D$-dimensions
\begin{equation}
\mu_i= \sin \theta_i \prod_{j=0}^{i-1} \cos \theta_j \ ,
\end{equation}
where $\theta_{0}=0$, $\theta_{N}=\pi/2$, and in even $D$-dimensions
\begin{equation}
\mu_i= \sin \theta_{(i+1)} \prod_{j=0}^{i} \cos \theta_j \ ,
\end{equation}
where $\theta_{0}=0$, $\theta_{N+1}=\pi/2$, and $\mu_0 = \nu$. Note that
$\displaystyle \sum_{i=1}^N
\mu_i^2 + \varepsilon \nu^2 = 1$.

The 
gauge potential is given by
\begin{equation}
A =
\frac{m\sinh{\alpha}}{\rho^{\vepsilon}(\rho^2+a^2)^{N-1}+m\sinh^2{\alpha}}\left(
  \cosh^2{\alpha}dt - a\sum_{i=1}^N \mu_id\phi_i \right) \ .
\label{KK_A_extremal_equal_ai}
\end{equation}
Finally, the dilaton is given by the following expression
\begin{equation}
\phi = -\frac{D-1}{2(D-2)}\log\left( 1 + \frac{m\sinh^2{\alpha}}{\rho^{\vepsilon}(\rho^2+a^2)^{N-1}}
\right) \ .
\label{KK_dil_extremal_equal_ai}
\end{equation}
These extremal black holes are characterized only by two free parameters, 
since the angular momentum parameter $a$ and the mass parameter $m$
can be given in terms of the horizon radius $\rho_{\rm H}$:
\begin{eqnarray}
{\rm even\ } D:
&& a^2=(2N-1)\, \rho_{\rm H}^2 \nonumber \ , \\
&& m=(2N)^N \rho_{\rm H}^{2N-1} \nonumber \ , \\ 
{\rm odd\ }  D:
&& a^2=(N-1)\, \rho_{\rm H}^2 \nonumber \ , \\
&& m=N^N \rho_{\rm H}^{2N-2} \ .
\end{eqnarray}
The relation between area and angular momentum
for extremal black holes then follows
\begin{eqnarray}
{\rm even\ } D:
&& J = 2\sqrt{D-3} A_{\rm H} \nonumber \ , \\
{\rm odd\ }  D:
&& J = \sqrt{2(D-3)} A_{\rm H} \ .
\label{ah_rel1} 
\end{eqnarray}

Considering the Smarr relation Eq.~(\ref{smarr}), 
we note that on the one hand for extremal black holes we have
vanishing $\bar \kappa$, thus
\begin{equation}
1 =  \left. \frac{D-2}{D-3} N \bar \Omega j \right|_{\rm ex}
+ \Phi_{\rm H} q \ .
\end{equation}
On the other hand, for static black holes
we have vanishing $\bar \Omega$,
thus
\begin{equation}
1 =  \left. 2\frac{D-2}{D-3}\bar \kappa a_{\rm H} \right|_{\rm st}
+ \Phi_{\rm H} q \ .
\end{equation}
Since the horizon electrostatic potential $\Phi_{\rm H}$
is independent of the angular momentum,
we obtain for a given $q$ the interesting relation
\begin{equation}
\left. N \bar \Omega j \right|_{\rm ex} \ = 
\left. \phantom{\bar \Omega}
2 \bar \kappa a_{\rm H} \right|_{\rm st} \ . 
\label{rel_kaOm}
\end{equation}

The scaled area $a_{\rm H}$ of the static solutions is proportional to the
scaled angular momenta $j$ of the extremal solutions,  coinciding in five
dimensions. For odd dimensions we find
\begin{equation}
 j_{\rm ex} =
 2N^{\frac{-N}{2(N-1)}}(N-1)^{1/2}a_{\rm H, st} \ ,
\label{jah_odd}
\end{equation}
while for even dimensions
\begin{equation}
 j_{\rm ex} =
 2^{\frac{N-1}{2N-1}}N^{\frac{-N}{2N-1}}(2N-1)^{1/2}a_{\rm
    H, st} \ .
\label{jah_even}
\end{equation}

Since  
$j_{\rm ex}$ and $a_{\rm H,st}$
are proportional,
we obtain a relation between 
the scaled horizon velocity $ \bar \Omega_{\rm ex}$
of the extremal rotating black holes
and the scaled surface gravity $\bar \kappa_{\rm st} $
of the static black holes. For odd dimensions it reads
\begin{equation}
\bar \kappa_{\rm st}  =  N^{\frac{N-2}{2(N-1)}}(N-1)^{1/2}\bar \Omega_{\rm ex} \ ,
\label{kappaOm_odd}
\end{equation}
and for even dimensions
\begin{equation}
\bar \kappa_{\rm st}  =
2^{\frac{-N}{2N-1}}N^{\frac{N-1}{2N-1}}(2N-1)^{1/2}\bar \Omega_{\rm ex} \ .
\label{kappaOm_even}
\end{equation}

Again, in five dimensions, $\bar \kappa_{\rm st}$ and
$\bar \Omega_{\rm ex}$ coincide, while they
are proportional in higher dimensions.

These KK relations hold also for vacuum black holes, obtained in the limit
$q \to 0$.

\section{Near-horizon solutions}

The existence of extremal EMd and EM black hole solutions
is supported by the existence of exact solutions,
describing a rotating squashed $AdS_2\times S^{D-2}$ spacetime.
These solutions correspond to the neighborhood of the event horizon of 
extremal black holes.

To obtain these exact near-horizon solutions for odd $D$-dimensions, we
parametrize 
the metric as follows
\begin{eqnarray}
ds^2 &=&v_1\left( \frac{dr^2}{r^2} - r^2 dt^2 \right)
  + v_2 \sum_{i=1}^{N-1}
  \left(\prod_{j=0}^{i-1} \cos^2\theta_j \right) d\theta_i^2 \nonumber \\
&&+v_2 v_3 \sum_{i=1}^N \left( \prod_{j=0}^{i-1} \cos^2 \theta_j
  \right) \sin^2\theta_i \left(\vepsilon_i d\vphi_i - kr
  dt\right)^2 \nonumber \\
&&+v_2(1-v_3) \left\{ \sum_{i=1}^N \left( \prod_{j=0}^{i-1} \cos^2
  \theta_j \right) \sin^2\theta_i  d\vphi_i^2 \right. \nonumber\\
&& -\left. \left[\sum_{i=1}^N \left( \prod_{j=0}^{i-1} \cos^2
  \theta_j \right) \sin^2\theta_i \vepsilon_i d\vphi_i \right]^2 \right\} 
\ . \label{metric_near} \end{eqnarray}
Here we have employed a new radial coordinate
such that the horizon is located at $r=0$.
This position of the horizon can always be obtained by 
a transformation $r\to r-r_{\rm H}$.

Moreover, we have shifted to a corrotating frame
such that the angular velocity vanishes on the horizon.
The corresponding parametrization for the gauge potential
in the corrotating frame reads
\begin{equation}
A_\mu dx^\mu =  (q_1-q_2 k) r dt + q_2 \sum_{i=1}^N \left(\prod_{j=0}^{i-1}
  \cos^2\theta_j\right) \sin^2\theta_i \vepsilon_i d\vphi_i 
\ . \label{gaugepotential_near}
\end{equation}
The dilaton field is simply given by $\phi=u$.

The parameters $k$, $v_i$,  $q_i$, and $u$ are constants, which
satisfy a set of algebraic relations.
In what follows we choose to determine them by using the near-horizon
formalism proposed in \cite{Astefanesei:2006dd,Goldstein:2007km}.

To that end let us denote by $f(k,v_1,v_2,v_3,q_1,q_2,u)$
the Lagrangian density $\sqrt{-g} {\cal L}$
evaluated for the near-horizon geometry Eq.~(\ref{metric_near}) and potential
Eq.~(\ref{gaugepotential_near})
and integrated over the angular coordinates,
\begin{eqnarray}
\label{at2}
f(k,v_1,v_2,v_3,q_1,q_2,u)
&=&\int d \theta_1\dots d\theta_{N-1} d \varphi_1 \dots d\varphi_N \sqrt{-g} {\cal L} \ .
\end{eqnarray}

The field equations then follow from the variation of this functional.
In particular, the derivative with respect to $k$ and $q_1$
yields the angular momenta and the charge,
\begin{equation}
\label{at3}
 \frac{\partial f}{\partial k}=2J \ , \quad
 \frac{\partial f}{\partial q_1}=Q \ ,
\end{equation}
respectively, while the derivative with respect to the remaining parameters vanishes,
\begin{equation}
 \frac{\partial f}{\partial v_i}=0  , \  i=1,2,3   \ , \quad
 \frac{\partial f}{\partial q_2}=0 \ , \quad
\frac{\partial f}{\partial u}=0 \ .
\label{at0}
\end{equation}

The entropy function is obtained by taking the
Legendre transform of the above integral with respect to the
the parameter $k$, associated with all equal-magnitude
angular momenta, $J_1=\dots = J_N=J$,
and with respect to the parameter $q_1$, associated with the charge $Q$,
\begin{equation}
\label{at4}
{\cal E}(J,k,Q,q_1,q_2,v_1,v_2,v_3,u)=
2 \pi     \left[  N J k + Q q_1 - f(k,v_1,v_2,v_3,q_1,q_2,u) \right]\ .
\end{equation}
Then the entropy associated with the black holes can be calculated by
evaluating this function at the extremum, $S={\cal E}_{extremum}$.

\subsection{Generic dilaton coupling}\label{subsec_EMD_near}

To obtain the near-horizon solutions, the set of equations Eqs.~(\ref{at3})-(\ref{at0})
must be solved.
For a given spacetime dimension $D=2N+1$,
the solutions can be expressed in terms of three 
independent parameters, $v_1$, $q_2$, and $u$,
\begin{eqnarray}
\label{sol1_genD}
v_2 &=& 2(N-1)N v_1 \ , \nonumber\\
v_3 &=& N -\frac{2q_2^2e^{-2hu}}{N(N-1)v_1} \ , \nonumber\\
q_1 &=& 0 \ , \nonumber\\
k &=& \frac{1}{N\sqrt{N-1}} \ , \nonumber\\
J &=& \frac{2^{N+1}\sqrt{2N^{2}(N-1)v_1-4q_2^2e^{-2hu}}}{
   N(N-1)^{3/2}(N-2)!v1}(\pi N (N-1)v_1)^{N} , \nonumber\\
Q &=&  -\frac{2^{N+2}\sqrt{2N^{2}(N-1)v_1-4q_2^2e^{-2hu}}}{
  N(N-1)^{3/2}N!}(\pi N (N-1)v_1)^{N}\frac{q_2 e^{-2hu}}{v_1^2} \ .
\label{sol_nh_dilaton}
\end{eqnarray}
Interestingly, $q_1$ vanishes and $k$ is constant.

Let us now inspect the horizon charges 
obtained from this solution, Eq.~(\ref{sol1_genD}).
The horizon mass is given by
\begin{equation}
M_{\rm H} = \frac{(D-1)(D-2)}{2(D-3)} \Omega J_{\rm H} \ ,
\end{equation}
and the horizon angular momenta by
\begin{equation}
J_{\rm H} = 2^{N}\sqrt{N-1}(N-2)!\left(N(N-1)\right)^{N-3}\pi^N
\left(2N^2(N-1)v_1-4q_2^2 e^{-2hu} \right)^{3/2} \ .
\end{equation}
The horizon area is obtained from the entropy function, Eq.~(\ref{at4})
\begin{equation}
A_{\rm H} = \frac{\cal E}{4\pi} = \frac{J}{\sqrt{2(D-3)}} \ .
\end{equation}
Thus relation  Eq.~(\ref{ah_rel1}) holds for generic values of
the dilaton coupling constant $h$ ($h\ne 0$).
Extremal dilatonic black holes thus show proportionality of their horizon area
and their angular momenta.

\begin{figure}[t!]
\begin{center}
\mbox{\hspace*{-2.0cm}
\subfigure[][]{
\includegraphics[height=.45\textheight, angle =270]{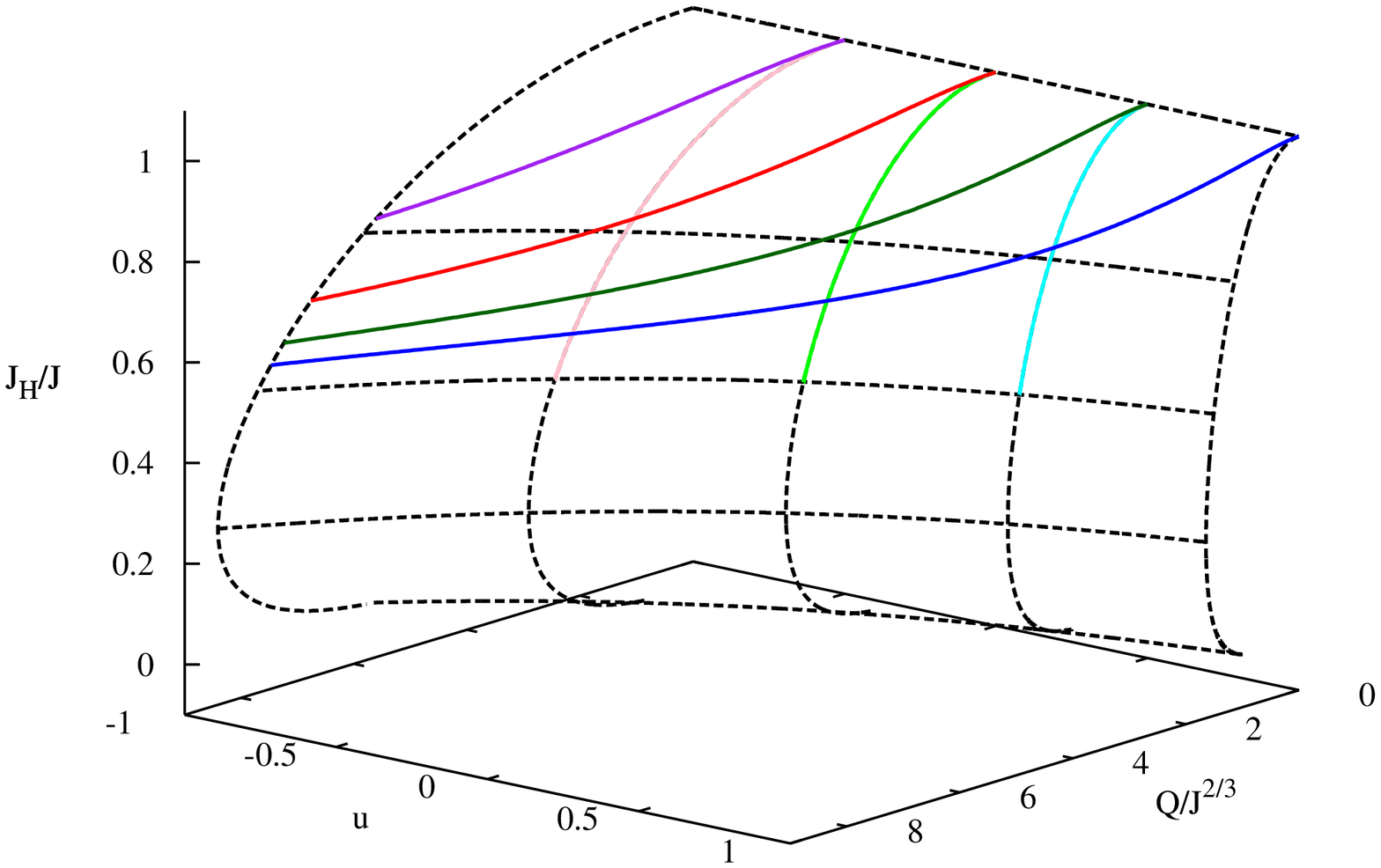}
\label{fig1a}
}
\hspace*{-1.0cm}
\subfigure[][]{
\includegraphics[height=.38\textheight, angle =270]{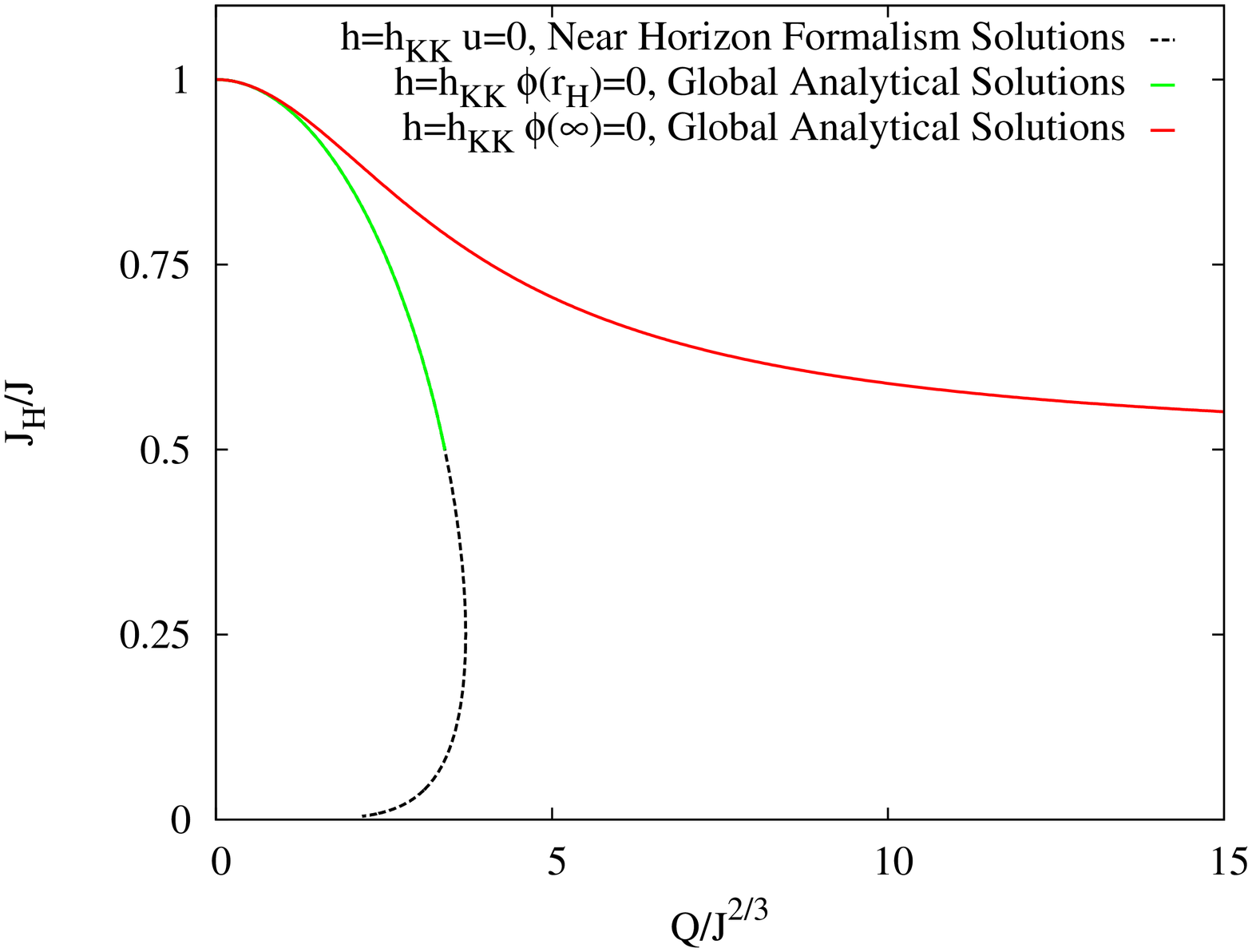}
\label{fig1b}
}
}
\end{center}
\vspace*{-0.5cm}
\caption{\small
(a) Surface of near-horizon EMd solutions in
five dimensions for dilaton coupling constant $h=h_{\rm KK}$, scaled with the angular
momentum: 
The 
horizon angular momenta $J_{\rm H}/J$ 
are shown versus the charge $Q/J^{2/3}$ and the
dilaton parameter $u$.
The solid lines mark the global solutions. (b) Detail of (a) where the global
solutions with $\phi(\infty)=0$ (red line) and with $u=\phi(r_{\rm H})=0$ (green
line) are shown, together with the  near-horizon EMd solutions with $u=0$
(black line).
}
\label{fig1}
\end{figure}

The most intringuing feature of Eq.~(\ref{sol_nh_dilaton}) is the presence of
3 free parameters. However, we know that these extremal black holes possess
only 2 independent parameters (the electric charge $Q$ and the angular
momentum $J$, for instance). One would be tempted to state that the existence
of an extra independent parameter in the near-horizon EMd solutions is related
to the existence of non-globally realized near-horizon EMd solutions. Although
these non-globally realized solutions are really present among the
near-horizon EMd solutions, as we will show below, the question is more subtle. The key point here is the invariance under a scale transformation associated with the dilaton 
\begin{eqnarray}
&&{\tilde x}^\mu = \lambda^{-1} x^\mu \nonumber \ , \\
&&\tilde\phi = \phi + \frac{1}{h}\log\lambda \ , \nonumber \\
&&{\tilde F}_{\mu\nu} = \lambda^2 F_{\mu \nu} \ . \label{scaling_dilaton} 
\end{eqnarray}
Note that under this transformation the value of the dilaton at the horizon
may be set to any given value, so the parameter $u$ may be eliminated by  such
a transformation (i.e., one can set $u=0$ without loss of generality). We illustrate
this fact in Figs.~\ref{fig1}. In Fig.~\ref{fig1a} we plot the scaled horizon
angular momenta $J_{\rm H}/J$ as a function of $u$ and $Q/J^{2/3}$, for the dilaton
coupling constant $h=h_{KK}$ in five dimensions. On the surface we
show black dashed curves corresponding to cuts with constant $u$ and
cuts with constant $J_{\rm H}/J$. 
We also exhibit with solid lines sets of globally realized extremal
solutions.  The curves in light blue, light green, and
pink are solutions with constant $u=\phi(r_{\rm H})$, 
and the curves in dark blue,
dark green, red, and purple are solutions with constant
$\phi(\infty)$. All these globally realized curves range from $J_{\rm H}/J=1$ to
$J_{\rm H}/J=1/2$ (this last limit depending on $h$; its lowest value is 1/4 for EM
$h=0$). The curves with constant $u=\phi(r_{\rm H})$ have finite length while those with
constant $\phi(\infty)$ have infinite length. However, all of them are
equivalent to each other. The curves with constant $u=\phi(r_{\rm H})$ transform into each
other under a transformation of the type Eq.~(\ref{scaling_dilaton}) with a
constant $\lambda$; the same holds among curves with constant
$\phi(\infty)$. A transformation among the curves with constant $u=\phi_H$ and
the curves with constant $\phi(\infty)$ requires a factor $\lambda$ that varies
with $Q$. Since $J_{\rm H}/J$ is invariant under Eq.~(\ref{scaling_dilaton}), the
requirement for two points on the surface to be equivalent is that they have
the same  $J_{\rm H}/J$ value. Note that $Q/J^{2/3}$ is not invariant but transforms
as ${\tilde Q}/{\tilde J}^{2/3} = \lambda Q/J^{2/3}$. That means if one
changes the $Q/J^{2/3}$ axis to $e^{h\phi(\infty)}Q/J^{2/3}$, all the color curves
collapse into the red curve. In fact, the whole upper part of the surface with
$J_{\rm H}/J \ge 1/2$ collapses into the red curve.

Going back to the point of the existence of non-globally realized near-horizon
EMd solutions, we see that solutions with $J_{\rm H}/J \le 1/2$ do not exist
globally for $h=h_{KK}$. We demonstrate this in Fig.~\ref{fig1b}. There only the globally
realized solutions with $\phi(\infty)=0$ and $\phi(r_{\rm H})=0$ are included,
together with the near-horizon solutions with $u=0$. Near-horizon EMd
solutions (in black) with $J_{\rm H}/J < 1/2$ are not realized globally.
(They correspond to the region
where the black and the green curves do not overlap.)

\subsection{Kaluza-Klein case}\label{KK_nh}

Since the black hole for the Kaluza-Klein value of the dilaton coupling is
explicitly known, we can obtain the near-horizon limit of the analytic
solution.

We change the metric Eq.~(\ref{KK_metric_extremal_equal_ai}) to a corrotating frame, shift the
radial coordinate to be centered on the horizon, and make the near-horizon
limit together with an appropriate gauge transformation on the gauge
potential Eq.~(\ref{KK_A_extremal_equal_ai}). The resulting metric can be written in a similar form as in 
Eq.~(\ref{metric_near}). The gauge vector can also be written in the same form as
in Eq.~(\ref{gaugepotential_near}). The dilaton
Eq.~(\ref{KK_dil_extremal_equal_ai}) is given by a parameter $\phi =
u$ in the near-horizon limit. All these parameters satisfy the following relations with the parameters
$a$ and $\alpha$ from the Kaluza-Klein solution:
\begin{eqnarray}
v_1 &=& \frac{1}{2}a^2\frac{(1+N\cosh(\alpha)^2-N)^{\frac{1}{D-2}}}{(N-1)^2} \ , \nonumber\\
v_2 &=& 2N(N-1)v_1 \  ,  \nonumber\\
v_3 &=& \frac{N\cosh(\alpha)^2}{1-N+N\cosh(\alpha)^2} \ , \nonumber\\
k &=& \frac{1}{N\sqrt{N-1}} \ , \nonumber\\
q_1 &=& 0 \ , \nonumber\\
q_2 &=& -\frac{Na\sinh(\alpha)}{1+N\cosh(\alpha)^2-N} \ .
\end{eqnarray}
Note that these relations are compatible with the near-horizon geometry for
generic dilaton coupling Eq.~(\ref{sol1_genD}), in the particular case of
$h=h_{KK}$. 

In this case, since we derive the near-horizon limit of the global solution, we also
obtain a relation for the dilaton parameter,
\begin{equation}
u = -\frac{\sqrt{2(D-1)}}{2\sqrt{D-2}}\ln(1+N\cosh(\alpha)^2-N)\ . \label{dilaton_KK_near}
\end{equation}

We note, that this relation was not found by employing
the near-horizon formalism in the previous section \ref{subsec_EMD_near}. 
Here this relation is only found as a result of 
explictly knowing the global solution and the fact that the dilaton is
required to vanish at infinity. So this solution corresponds to the red curve
in Figs.~\ref{fig1}.

\subsection{Einstein-Maxwell case}

Surprisingly, the EM case is rather different from the EMd case.
In particular, there is not only one set of solutions, as in the EMd case, 
but there are two sets of solutions, which we label currently as the first branch
and the second branch of solutions.
The five-dimensional case was discussed before
\cite{Blazquez-Salcedo:2013yba,Kunduri:2013gce}.
Moreover, the EM solutions depend only on two parameters,
since there is no dilaton parameter.

For a given spacetime dimension $D=2N+1$,
the first set of solutions can be expressed in terms of the
independent parameters $v_1$ and $q_2$.
Thus the first branch is given by
\begin{eqnarray}
\label{sol1_genD_EM}
v_2 &=& 2(N-1)N v_1 \ , \nonumber\\
v_3 &=& N -\frac{2q_2^2}{N(N-1)v_1} \ , \nonumber\\
q_1 &=& 0 \ , \nonumber\\
k &=& \frac{1}{N\sqrt{N-1}} \ , \nonumber\\
J &=& \frac{2^{N+1}\sqrt{2N^{2}(N-1)v_1-4q_2^2}}{
   N(N-1)^{3/2}(N-2)!v1}(\pi N (N-1)v_1)^{N} \ , \nonumber\\
Q &=&  -\frac{2^{N+2}\sqrt{2N^{2}(N-1)v_1-4q_2^2}}{
  N(N-1)^{3/2}N!}(\pi N (N-1)v_1)^{N}\frac{q_2 }{v_1^2} \ .
\end{eqnarray}
As in the EMd case, in this EM solution $q_1$ vanishes and $k$ is constant.
Moreover, by defining $\bar{q_2} = q_2e^{-hu}$ and $\bar{Q} = Qe^{hu}$
in the EMd solution Eq.~(\ref{sol1_genD}),
this first EM branch is obtained.
Because $e^{-2hu}$ is multiplying $q_2$ everywhere, 
we re-obtain the MP solution, when setting $q_2=0$.

Inspecting the horizon charges 
obtained from this solution Eq.~(\ref{sol1_genD_EM}),
we find the horizon mass
\begin{equation}
M_{\rm H} = \frac{(D-1)(D-2)}{2(D-3)} \Omega J_{\rm H} \ ,
\end{equation}
and the horizon angular momenta
\begin{equation}
J_{\rm H} = 2^{N}\sqrt{N-1}(N-2)!\left(N(N-1)\right)^{N-3}\pi^N
\left(2N^2(N-1)v_1-4q_2^2  \right)^{3/2} \ .
\end{equation}

The horizon area is obtained from the entropy function, Eq.~(\ref{at4})
\begin{equation}
A_{\rm H} = \frac{\cal E}{4\pi} = \frac{J}{\sqrt{2(D-3)}} \ .
\end{equation}
Clearly, relation Eq.~(\ref{ah_rel1}) holds along this first branch, 
expressing proportionality of the horizon area and the angular momenta.

The near-horizon equations, however, allow for a second set of solutions,
which can be expressed in terms of the
independent parameters $v_1$ and $v_2$.
These solutions form the second branch and are given by
\begin{eqnarray}
\label{sol2_genD}
v_3 &=& \frac{v_2}{4(N-1)(N-2)v_1}-\frac{1}{N-2} \ , \nonumber\\
k   &=& \frac{2 \sqrt{N-1} v_1}{v_2\sqrt{v_2+4(1-N)v_1}}\sqrt{4(N-1)^2v_1-v_2} \ , \nonumber \\
q_1 &=& \frac{\sqrt{2N-1}}{\sqrt{N-2}}\sqrt{\frac{v_1}{v_2}}\frac{2N(N-1)v_1-v_2}{\sqrt{v_2+4(1-N)v_1}} \ , \nonumber\\
q_2 &=& \frac{\sqrt{2N-1}}{4\sqrt{N-2}(N-1)}\sqrt{\frac{v_2}{v_1}}\sqrt{4(N-1)^{3}v_1+(1-N)v_2} \ , \nonumber\\
J &=& \frac{(\pi
  v_2)^N}{(N-1)!}\frac{v2-4(N-1)v_1}{(N-1)^{3/2}(N-2)^{3/2}v_1^{3/2}\sqrt{v_2}}\sqrt{(N-1)(4(N-1)^2v_1-v_2)} \ , 
\nonumber\\ 
Q &=& \frac{-2(\pi v_2)^N}{(N-1)!}\frac{\sqrt{2N-1}}{\sqrt{N-1}(N-2)}\frac{4(N-1)v_1-v_2}{v_1v_2}  \ .
\end{eqnarray}
The horizon charges of the solutions on the second branch are given by
\begin{equation}
M_{\rm H} = \frac{(D-1)(D-2)}{2(D-3)} \Omega J_{\rm H} \ ,
\end{equation}
\begin{equation}
J_{\rm H} = \frac{J}{D-1} \ ,
\end{equation}
and the horizon area is again obtained from the entropy function, Eq.~(\ref{at4})
\begin{equation}
A_{\rm H} = \frac{\cal E}{4\pi} = \frac{(\pi v_2)^N\sqrt{v_2-4(N-1)v_1}}{\sqrt{N-2}\sqrt{N-1}(N-1)!\sqrt{v_1v_2}} \ .
\end{equation}
For this branch of solutions
the horizon area is not proportional to the angular momenta.
Instead, the horizon angular momenta are proportional
to the total angular momenta 
\cite{Blazquez-Salcedo:2013yba}.

\begin{figure}[t!]
\begin{center}
\mbox{\hspace*{-0.6cm}
\subfigure[][]{
\includegraphics[height=.38\textheight, angle =270]{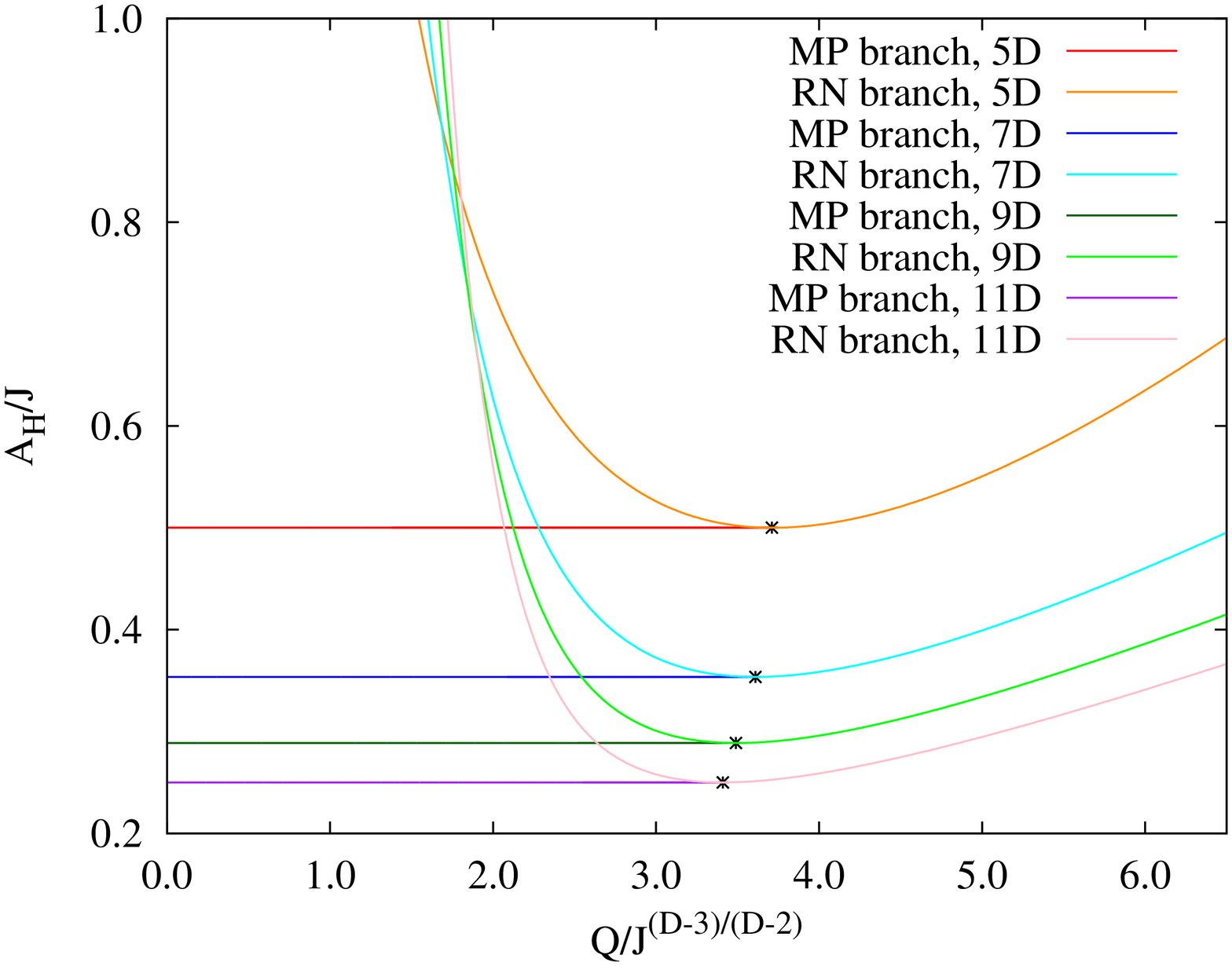}
\label{fig2a}
}
\subfigure[][]{
\hspace*{-0.5cm}
\includegraphics[height=.38\textheight, angle =270]{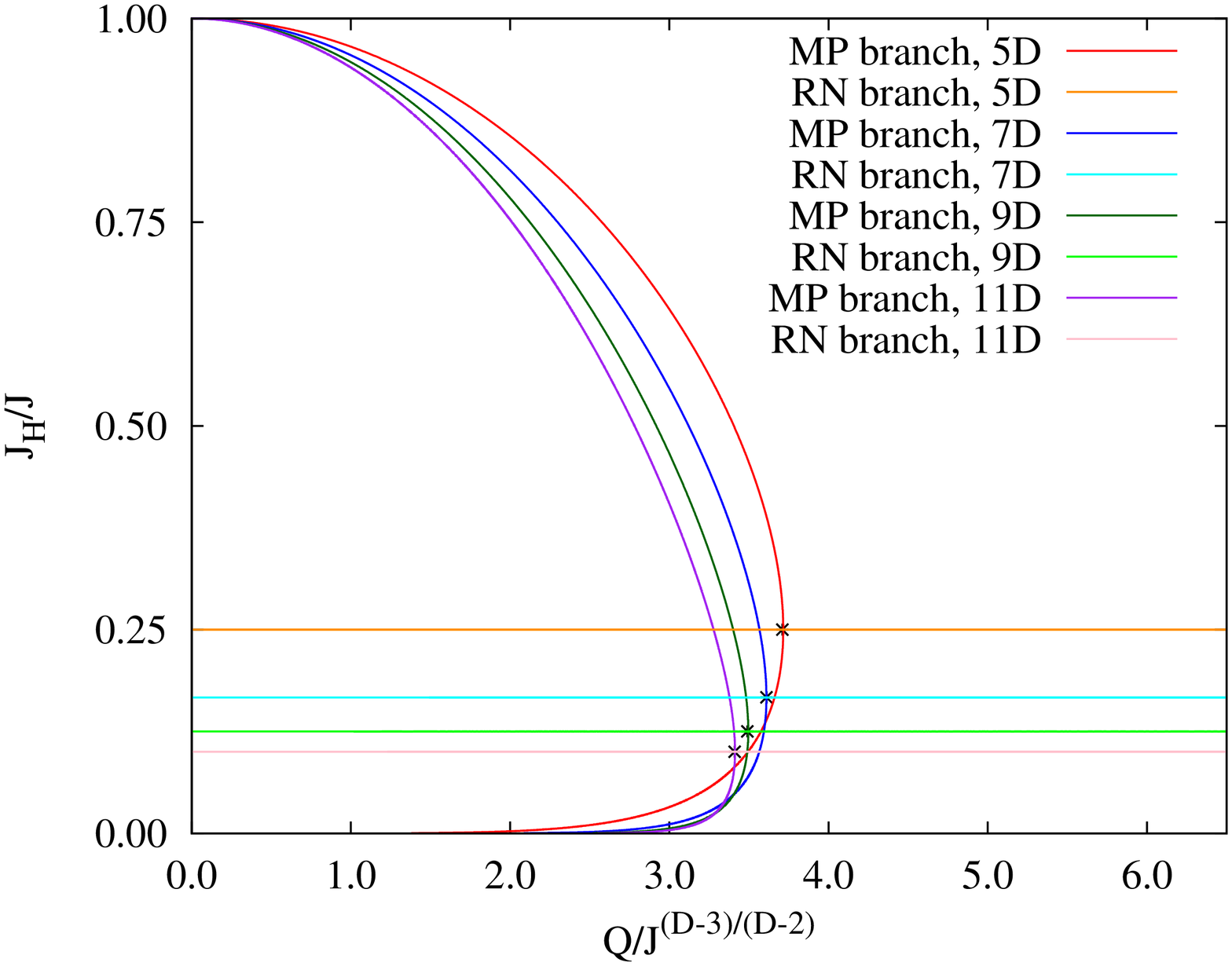}
\label{fig2b}
}
}
\end{center}
\vspace*{-0.5cm}
\caption{\small
Near-horizon EM solutions in several dimensions:
The horizon area $A_{\rm H}$ (a) and the
horizon angular momenta $J_{\rm H}$ (b)
are shown versus the charge $Q$ (quantities are scaled by the angular momenta
$J$.) The asterisks mark the matching points
of the two branches.
The first branch is realized globally 
from the MP solution to the matching point, 
while the second branch is realized globally
from the matching point to the RN solution.
}
\label{fig2}
\end{figure}

Figs.~\ref{fig2} exhibit the two branches of solutions for near-horizon
solutions in odd dimensions, from five to eleven.
In particular, we show
the horizon area $A_{\rm H}$ (a)
and the horizon angular momenta $J_{\rm H}$ (b)
versus the charge $Q$,
scaled by the angular momenta $J$.
The first branch, starting from the MP solution,
extends only up to a maximal value of the charge $Q$,
for given angular momenta $J$.
At this critical value it does not end, however,
but it continues backwards to smaller charges.

The second branch, on the other hand, extends 
over the full axis.
It smoothly reaches the extremal static RN solution,
when $J \to 0$, 
but it also extends all the way to vanishing electric charge $Q$.
Most importantly, however, this second branch 
crosses the first branch precisely at its critical point.
Here all physical quantities match for the two branches.
This matching point is indicated in the figures by an
asterisk.
It is this matching point, which also delimits
the globally realized parts of these branches of solutions.
The first branch is realized globally
from the MP solution to the matching point,
while the second branch is realized globally
from the matching point to the RN solution.

Thus for extremal EM black holes in higher odd dimensions
the branch structure is analogous to the five-dimensional case
\cite{Blazquez-Salcedo:2013yba,Kunduri:2013gce}.
In the following we will denote the branches
as the MP branch and the RN branch, 
since they start at the extremal MP solution and the extremal RN solution,
respectively.

\section{Numerical solutions}

After a short discussion of the numerical procedure
and the boundary conditions,
we present our results for black holes in 5, 7, and 9 dimensions.
We discuss the domains of existence,
the global properties, in particular, the gyromagnetic ratio,
and the horizon properties, where the surface gravity $\kappa$
and the horizon angular velocity $\Omega$ are associated
with a critical behavior.

This critical behavior was observed before for 
static EMd black holes \cite{Gibbons:1987ps}.
Here, at the critical value of the dilaton coupling $h_{\rm cr}$,
\begin{equation}
h_{\rm cr}= \frac{D-3}{\sqrt{2(D-2)}} \ ,
\label{hcr} \end{equation}
the surface gravity $\kappa$ remains finite in the extremal limit.
In contrast, $\kappa$ diverges for $h > h_{\rm cr}$,
while $\kappa \to 0$ for $h < h_{\rm cr}$ in the extremal limit.
Comparing this critical value $h_{\rm cr}$
to the KK value $h_{\rm KK}$, we note,
that 
\begin{eqnarray}
h_{\rm cr} &=& h_{\rm KK} \ {\rm for} \ D=5 \ , \nonumber \\
h_{\rm cr} &>& h_{\rm KK} \ {\rm for} \ D>5 \ . 
\label{hcr_hkk}
\end{eqnarray}

Since the solutions have a scaling symmetry Eq.~(\ref{scaling}),
we typically exhibit scaled physical quantities,
where we scale with respect to appropriate powers of the mass $M$.
In particular, we employ the charge to mass ratio $q$
to demonstrate the dependence on the charge.

The domain of existence of these black holes
increases with increasing dilaton coupling constant $h$,
since the maximal value of the scaled charge $q$ increases with $h$
according to
\begin{equation}
q_{\rm max} = \sqrt{\frac{D-3}{2(D-2)} + h^2} \ ,
\label{qmax} \end{equation}
i.e., for the Kaluza-Klein coupling constant
$h_{\rm KK}$ the maximal value of the scaled charge 
is given by $q_{\rm max}=1$,
independent of the dimension $D$.

The gyromagnetic ratio $g$ of higher dimensional black holes
has drawn much interest, ever since 
perturbations in the charge in the EM case ($h=0$) yielded for MP black holes the
intriguing lowest order
result \cite{Aliev:2004ec}
\begin{equation}
g_{\rm \delta q} = D-2 \ +  o(q^2)\ .
\end{equation}
The same result was obtained when considering first order perturbations in the
angular momentum for the charged static solutions 
in the EM case ($h=0$) \cite{Sheykhi:2008bs}
\begin{equation}
g_{\rm \delta j} = D-2 \ +  o(j^2)\ .
\end{equation}
Nevertheless, this result was shown not to hold in general for higher order
perturbations in the charge
\cite{NavarroLerida:2007ez,Allahverdizadeh:2010xx,Allahverdizadeh:2010fn},  
and non-perturbatively for arbitrary values of $Q$ and $J$
\cite{Kunz:2005nm,Kunz:2006eh}. However, it still represents an important
limiting value for EM and EMd 
black holes, attained for small values of the charge to mass ratio $q$.
In fact, 
in the EMd case ($h\ne 0$) first order perturbations in the
angular momentum for the charged static solutions 
yield, in our conventions in terms of $q = Q/M$ and a generic
dilaton coupling $h$,
the expression \cite{Sheykhi:2008bs}
\begin{equation}
g_{\rm \delta j} = \frac{(D-1)(D-2)(D-3)[D-3 + X]}{4h^2(D-2)^2q^2 + (D-1)(D-3)[D-3 + X]} \ ,
\label{gdeltaj}
\end{equation}
where
\begin{equation}
X^2 = 2(D-2)[2h^2(D-2) + 3-D]q^2 + (D-3)^2 \ .
\end{equation}

\subsection{Numerical procedure and boundary conditions}

In order to solve the coupled system of ODE's, 
we take advantage of the existence of a
first integral of that system, 
\begin{equation}
\frac{r^{D-2} m^{(D-5)/2}}{f^{(D-3)/2}} \sqrt{\frac{m n}{f}} \left(\frac{d
      a_0}{dr} + \frac{\omega}{r} \frac{d a_\vphi}{dr} \right) = - \frac{e^{2h\phi}}{A(S^{D-2})} Q \ ,
\end{equation}
to eliminate $a_0$ from the equations, leaving a 
system of one first order equation (for $n$) and
four second order equations.

For the numerical calculations we take units such that
$16\pi G_D=1$. We introduce
a compactified radial coordinate. For the non-extremal solutions we take the
compactified coordinate to be
$\bar{r}= 1-r_{\rm H}/r$. In the extremal case we employ $\bar{r}= \frac{r}{1+r}$
\cite{Kleihaus:2000kg}. (Note, that we are
using an isotropic coordinate $r$, so $r_{\rm H}=0$ in
the extremal case.)
We employ a collocation method for boundary-value ordinary
differential equations, equipped with an adaptive mesh selection procedure
\cite{COLSYS}.
Typical mesh sizes include $10^3-10^4$ points.
The solutions have a relative accuracy of $10^{-10}$.
The estimates of the relative errors of the global charges
and the magnetic moment are of order $10^{-6}$, 
giving rise to an estimate of the relative error of $g$ of order $10^{-5}$. 

To obtain asymptotically flat solutions
the metric functions should satisfy the boundary conditions
\begin{equation}
f|_{r=\infty}=m|_{r=\infty}=n|_{r=\infty}=1 \ , \ \omega|_{r=\infty}=0 
\ . \label{bc1} \end{equation}
For the gauge potential we choose a gauge, in which it vanishes
at infinity
\begin{equation}
a_0|_{r=\infty}=a_\vphi|_{r=\infty}=0 
\ . \label{bc2} \end{equation}
For the dilaton field we choose
\begin{equation}
\phi|_{r=\infty}=0
\ . \label{bc3} \end{equation}
Note, that any finite value of the dilaton field at infinity
can always be transformed to zero via
$\phi \rightarrow \phi - \phi(\infty)$,
$r \rightarrow r e^{h \phi(\infty)} $, $a_0 \rightarrow a_0  e^{-h \phi(\infty)} $.

Requiring the horizon to be regular, the metric functions must
satisfy the boundary conditions
\begin{equation}
f|_{r=r_{\rm H}}=m|_{r=r_{\rm H}}=n|_{r=r_{\rm H}}=0 \ ,
\ \omega|_{r=r_{\rm H}}=r_{\rm H} \Omega  
\ , \label{bc4} \end{equation}
where $\Omega$ is the horizon angular velocity, 
defined in terms of the Killing vector $\chi$, Eq.~(\ref{chi}).

The gauge potential satisfies at the horizon the conditions
\begin{equation}
\left. \chi^\mu A_\mu \right|_{r=r_{\rm H}} =
\Phi_{\rm H} = \left. (a_0+\Omega a_\vphi)\right|_{r=r_{\rm H}} \ , \ \ \
\left. \frac{d a_\vphi}{d r}\right|_{r=r_{\rm H}}=0
\ , \label{bc5} \end{equation}
with the constant horizon electrostatic potential $\Phi_{\rm H}$.
The boundary condition for the dilaton field reads
\begin{equation}
\left. \frac{d \phi}{d r}\right|_{r=r_{\rm H}}=0
\ . \label{bc6} \end{equation}

Since the KK solutions are known analytically,
we can use them as a test for the accuracy of the numerical calculations,
by choosing the dilaton coupling constant $h=h_{\rm KK}$.
Other general tests are provided by the Smarr formula Eq.~(\ref{smarr}) and
the relation Eq.~(\ref{dilaton_rel}).

\subsection{Expansions}

The asymptotic expansion of the metric, the gauge
potential, and the dilaton field reads
\begin{eqnarray}
f &=& 1-\frac{M}{(D-2)A(S^{D-2})} \frac{1}{r^{D-3}} + \dots \ , \nonumber \\
m &=& 1- \frac{(D-4)M}{(D-2)(D-3) A(S^{D-2})} \frac{1}{r^{D-3}} + \dots \ , \nonumber \\
n &=& 1- \frac{(D-4)M}{(D-2)(D-3) A(S^{D-2})} \frac{1}{r^{D-3}}  + \dots \ , \nonumber \\
\omega &=& \frac{J}{2 A(S^{D-2})} \frac{1}{r^{D-2}}  + \dots \ , \nonumber \\
a_0 &=& \frac{Q}{(D-3)A(S^{D-2})} \frac{1}{r^{D-3}} + \dots \ , \nonumber \\
a_\vphi &=& -\frac{\mu_{\rm mag}}{(D-3)A(S^{D-2})} \frac{1} {r^{D-3}} + \dots \ , \nonumber \\
\phi &=& \frac{\Sigma}{(D-3)A(S^{D-2})} \frac{1}{r^{D-3}} + \dots 
\ . \label{asymp} \end{eqnarray}
The global mass $M$, the global angular momenta
$J$, the electric charge $Q$,
the dilaton charge $\Sigma$, and the magnetic moment $\mu_{\rm mag}$,
can be read off from this expansion. Note that in the non-extremal case, only
three of these parameters are free. In the extremal case there are only two
free parameters.

For the expansion at the horizon we should distinguish explicitly
whether the solutions are non-extremal or extremal.
In the non-extremal case
the expansion of the functions at the horizon reads 
\begin{eqnarray}
f &=& f_2 \delta^2 (1 - \delta) + O(\delta^4) \ , \nonumber \\
m &=& m_2 \delta^2 (1 - 3 \delta) + O(\delta^4) \ , \nonumber \\
n &=& n_2 \delta^2 (1 - 3 \delta)+  O(\delta^4)\ , \nonumber \\
\omega &=& r_{\rm H} \Omega (1 + \delta) + O(\delta^2) \ , \nonumber \\
a_0 &=& a_{0,0} + O(\delta^2) \ , \nonumber \\
a_\varphi &=& a_{\varphi, 0} + O(\delta^2) \ , \nonumber \\
\phi &=& \phi_0 + O(\delta^2)
\ , \label{ex-horizon} \end{eqnarray}
where $\delta = r/r_{\rm H} - 1$ and $f_2$, $m_2$, $n_2$, $a_{0,0}$,
$a_{\varphi,0}$, and $\phi_0$ are constant.

In the extremal case the situation is very different. The expansion near the
horizon $r_{\rm H}=0$ is
\begin{eqnarray}
f &=& f_4 r^4 + f_{\alpha} r^{\alpha} + o(r^6) \ ,
\\
\nonumber
m &=& m_2 r^2 + m_{\beta} r^{\beta} + o(r^4) \ ,
\\
\nonumber
n &=& n_2 r^2 + n_{\gamma} r^{\gamma} + o(r^4) \ ,
\\
\nonumber
\omega &=& \omega_1 r + \omega_2 r^2 + o(r^3) \ ,
\\
\nonumber
a_0 &=& a_{0,0} + a_{0,\lambda} r^{\lambda} + o(r^2) \ ,
\\
\nonumber
a_{\phi} &=& a_{\phi,0} + a_{\phi,\mu} r^{\mu} + o(r^2) \ ,
\\
\nonumber
\phi &=& \phi_0 + \phi_{\nu} r^{\nu} + o(r^2) \ .
\\
\nonumber
\end{eqnarray}
It is interesting to note, that, in general, the exponents $\alpha$, $\beta$, $\gamma$,
$\lambda$, $\mu$, and $\nu$ are non-integer. Note, that $\omega$ is the only function with the usual expansion at the horizon. For the other functions, the next to leading order is given by a term with a non-integer exponent. The ranges are
\begin{eqnarray}
&&4<\alpha<6 \ , \quad 2<\beta<4 \ ,\quad  2<\gamma<4 \ , \nonumber \\
&&0<\lambda<2 \ , \quad  0<\mu<2 \ ,\quad  0<\nu<2 \ .
\end{eqnarray}
In the pure Einstein-Maxwell case ($h=0$) the expansion contains similarly non-integer exponents (but there is no
dilaton function). This feature is found in both EM branches 
(i.e., in the MP branch and in the RN branch).

For the numerical integration the following reparametrization of the functions was 
made,
\begin{eqnarray}
&&f = \hat f x^2 \ ,\quad  m = \hat m \ , \quad n = \hat n \ , \nonumber
\\
&&\omega = \hat \omega (1-x)^2 \ , \quad a_k = \hat a_{k}/x^2 \ , 
\quad \phi = \hat \phi/x^2 \ .
\\
\nonumber
\end{eqnarray}
Note, that all the redefined functions except for $\hat \omega$ now start with an
$x^2$-term in the compactified coordinate $x = r/(r+1)$. (The
reparametrization of $\omega$ is not related to the expansion at the horizon. It
is done in order to be able to fix the angular momentum by a boundary
condition). Numerically this reparametrization is used to deal with the possible
divergence of the first- and second-order derivatives of any
field functions. 

\boldmath
\subsection{$D=5$}
\unboldmath

Here we present our numerical results for black holes with equal
magnitude angular momenta in 5 dimensions.
We begin with a discussion of the solutions for a
generic value of the dilaton coupling constant, $h=2$.
Then we discuss the dependence of the solutions on the
dilaton coupling constant, $0 \le h < \infty$,
including the Einstein-Maxwell case ($h=0$) and the
Kaluza-Klein case ($h=h_{\rm KK}=\sqrt{2/3}$).
The formulae for the latter are collected in Section \ref{sec_KK}.
We note, that
uniqueness of the stationary black holes
in $5D$ Einstein-Maxwell and Einstein-Maxwell-dilaton gravity
were discussed by Yazadjiev \cite{Yazadjiev:2011fg}.

\boldmath
\subsubsection{Dilaton coupling constant $h=2$}
\unboldmath

Let us first address the domain of existence
of rotating EMd black holes with equal-magnitude angular momenta.
For such black holes the domain of existence is always bounded.
The boundary is provided by the set of extremal solutions.
However, when we consider quantities, that do not depend
on the direction of the rotation,
the set of static solutions provides a part of the boundary.
The non-extremal rotating solutions then reside within this boundary,
while solutions outside this boundary exhibit naked singularities.

\begin{figure}[p]
\begin{center}
\vspace{-0.7cm}
\mbox{\hspace{-1.5cm}
\subfigure[][]{
\includegraphics[height=.35\textheight, angle =270]{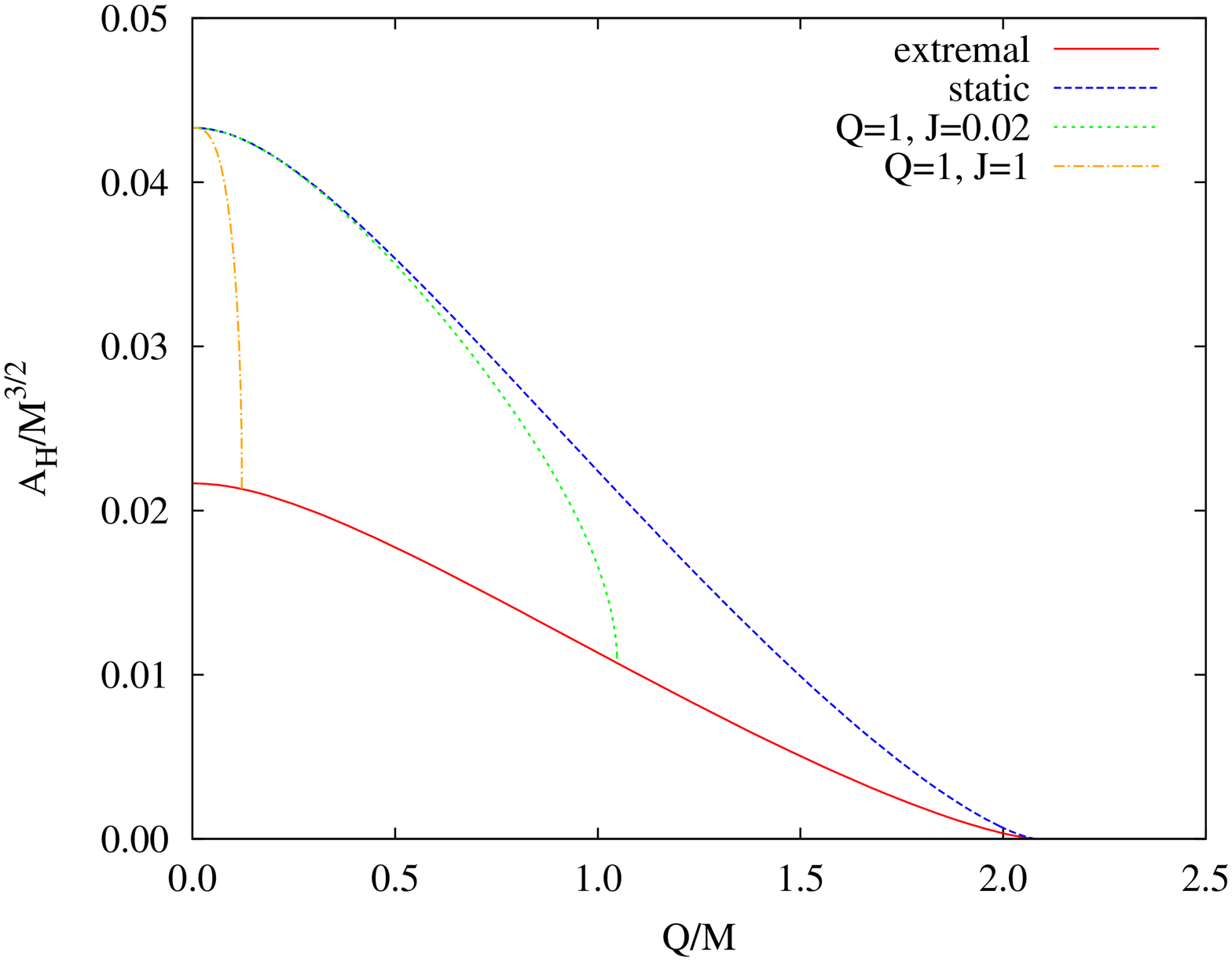}
\label{fig3a}
}
\subfigure[][]{\hspace{-0.5cm}
\includegraphics[height=.35\textheight, angle =270]{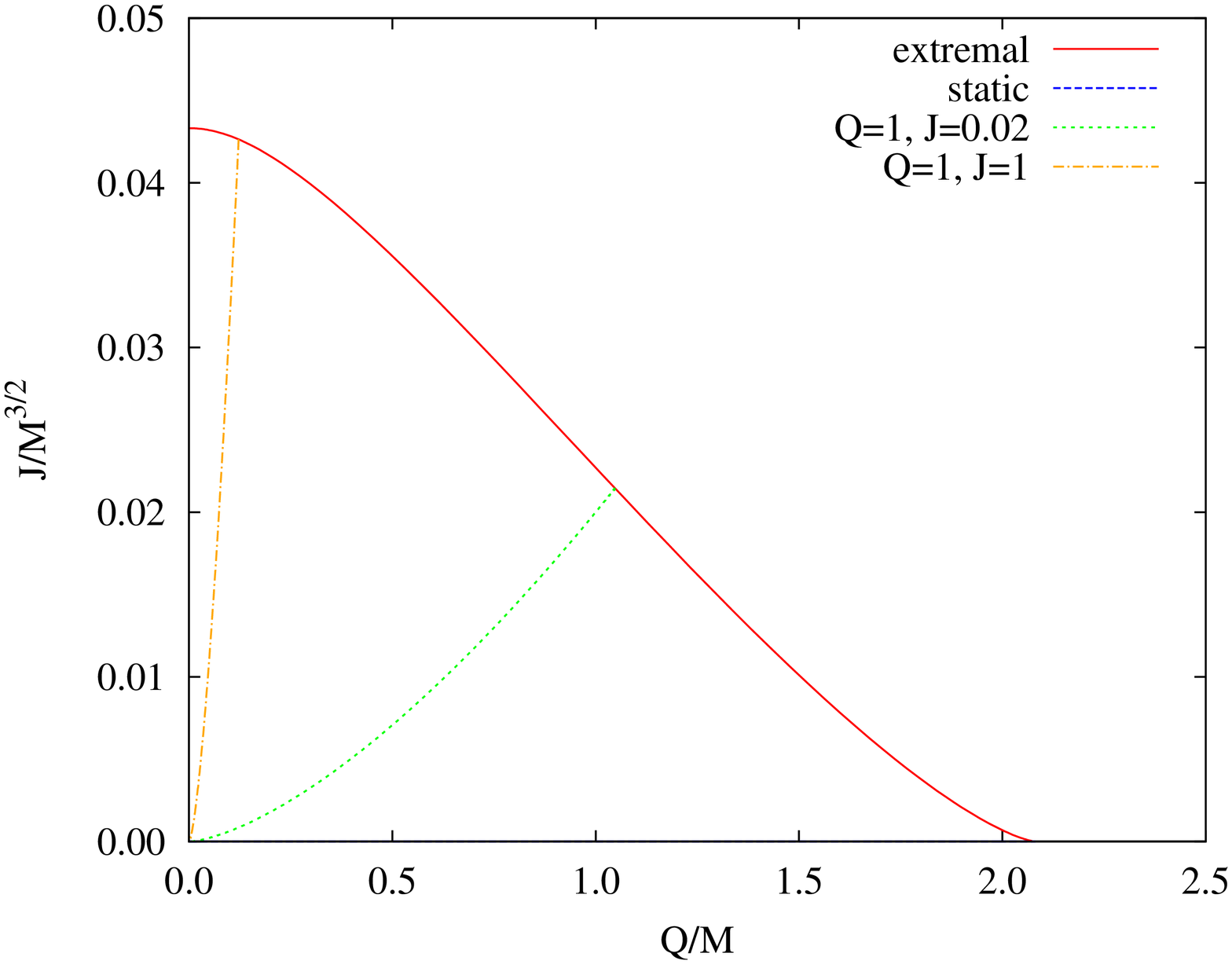}
\label{fig3b}
}
}
\mbox{\hspace{-1.5cm}
\subfigure[][]{
\includegraphics[height=.35\textheight, angle =270]{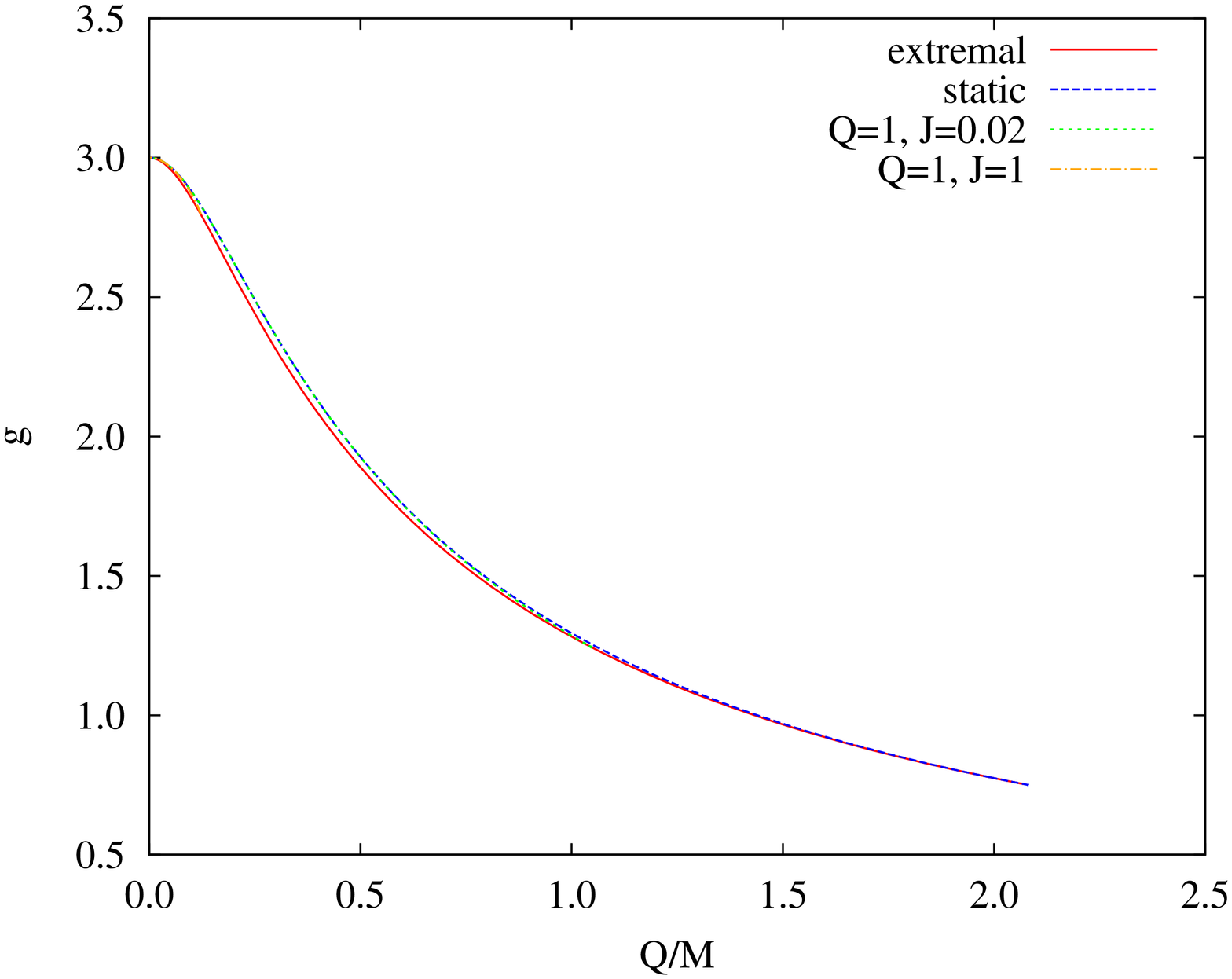}
\label{fig3c}
}
\subfigure[][]{\hspace{-0.5cm}
\includegraphics[height=.35\textheight, angle =270]{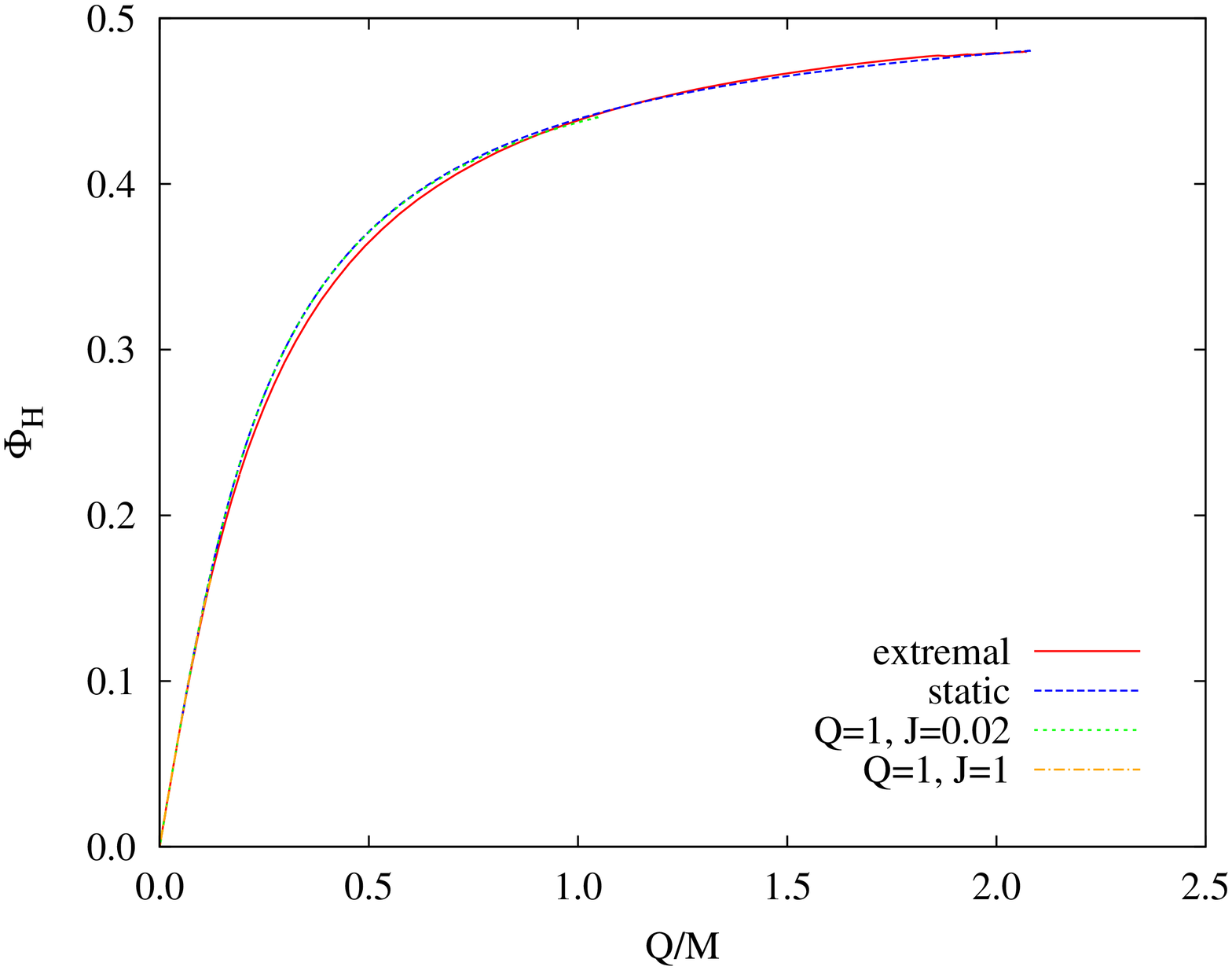}
\label{fig3d}
}
}
\mbox{\hspace{-1.5cm}
\subfigure[][]{
\includegraphics[height=.35\textheight, angle =270]{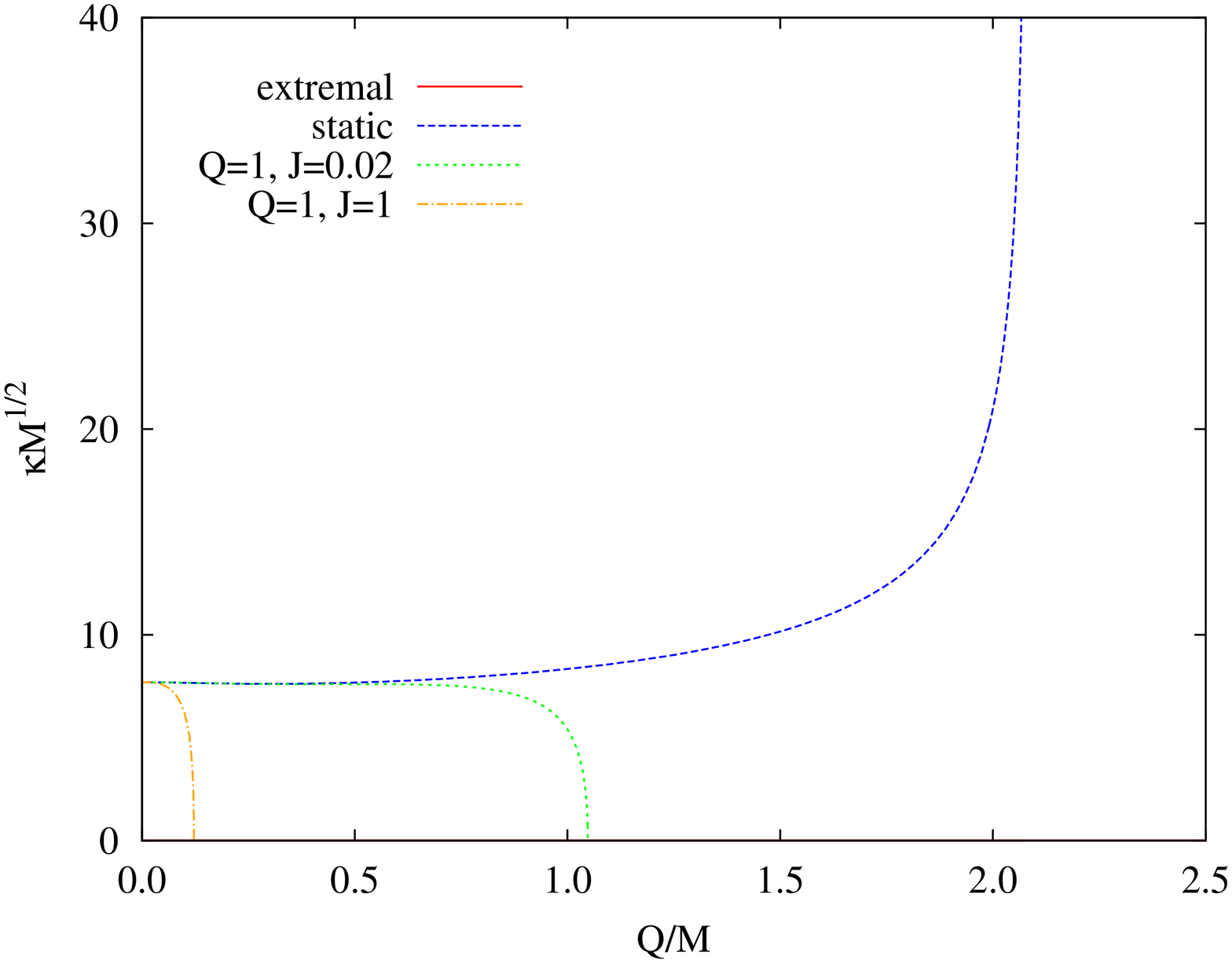}
\label{fig3e}
}
\subfigure[][]{\hspace{-0.5cm}
\includegraphics[height=.35\textheight, angle =270]{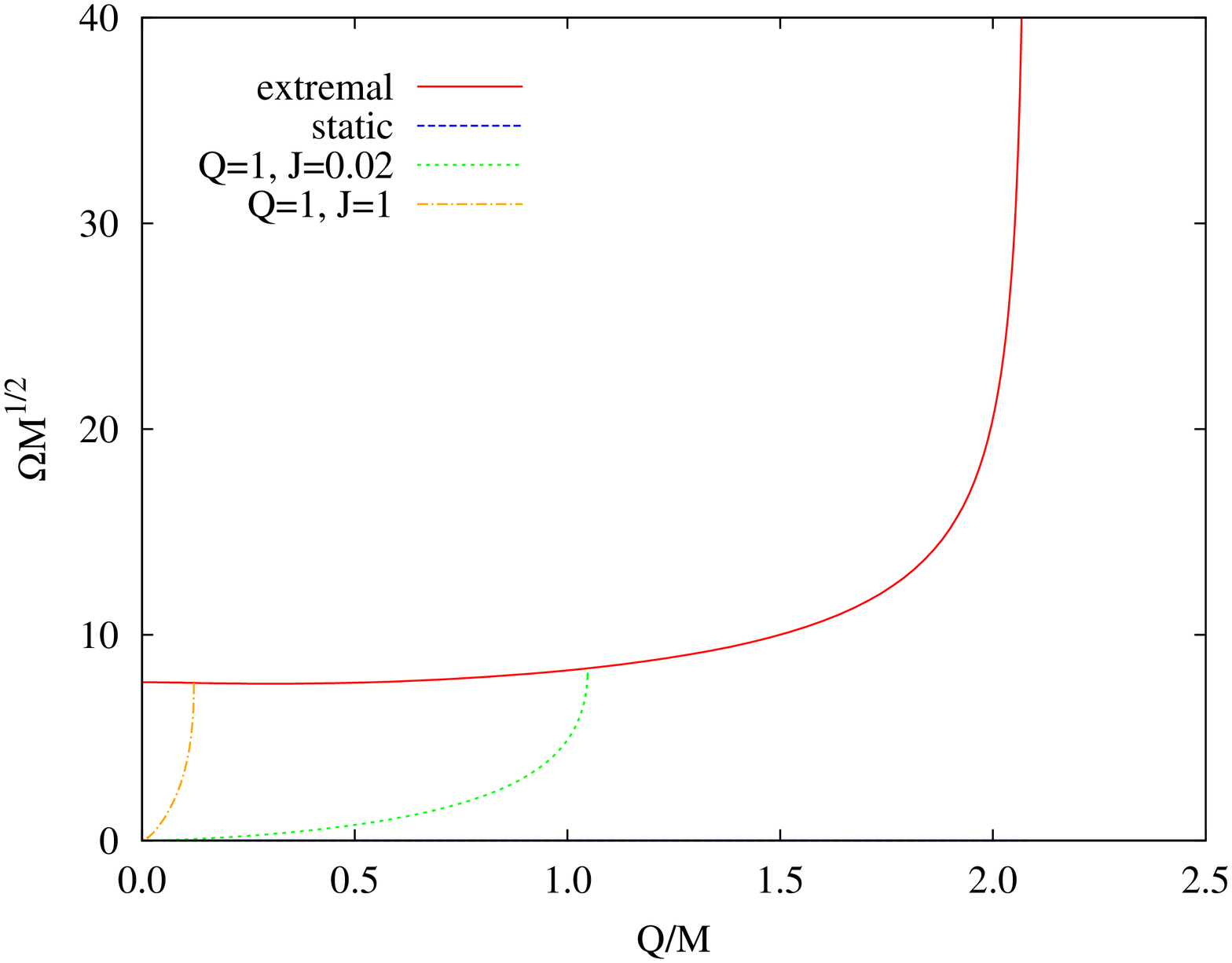}
\label{fig3f}
}
}
\end{center}
\caption{Properties of EMd black hole solutions  in five dimensions
for dilaton coupling constant $h=2$.
(a) The scaled area $a_{\rm H} = A_{\rm H}/M^{3/2}$
versus the scaled charge $q=|Q|/M$
for extremal and for static solutions,
providing the boundary of the domain of existence.
Also shown is $a_{\rm H}$ versus $q$ for two further sets of non-extremal solutions 
($Q=1$, $J=0.02$ and $Q=1$, $J=1$).
For the same sets of (extremal and non-extremal) solutions 
we exhibit in
(b) the scaled angular momenta $j=J/M^{3/2}$ versus $q$,
(c) the gyromagnetic ratio $g$ versus $q$,
(d) the horizon electrostatic potential $\Phi_{\rm H}$ versus $q$
(e) the scaled surface gravity $\bar \kappa = \kappa M^{1/2}$ versus $q$,
(f) the scaled horizon angular velocity $\bar \Omega = \Omega M^{1/2}$
    versus $q$.
\label{fig3}
}
\end{figure}

We illustrate the domain of existence of EMd black holes with
dilaton coupling constant $h=2$ 
for the scaled horizon area $a_{\rm H} = A_{\rm H}/M^{3/2}$
in Fig.~\ref{fig3a}. Considering $a_{\rm H}$
versus the scaled charge $q$,
the extremal and static solutions
form the lower and upper boundaries of the domain of existence,
respectively.
In the extremal limit, the horizon area 
of the static black holes vanishes,
while the rotating extremal black holes have finite horizon area.
Thus only static extremal black holes have vanishing area, while
all other black holes have finite area.

We also exhibit in Fig.~\ref{fig3a} two further sets of solutions,
which are non-extremal, except at their respective endpoints.
Here the charge $Q$ and the angular momenta $J$ are kept fixed,
in particular, the choices are $Q=1$, $J=0.02$ and $Q=1$, $J=1$, 
while the (isotropic) horizon radius $r_{\rm H}$ is varied.
These two sets start at a small value of $q$ and end at the respective extremal solutions.

Next to the area
we illustrate the scaled angular momenta $j=J/M^{3/2}$ versus $q$
in Fig.~\ref{fig3b}.
We note the proportionality of the  scaled angular momenta $j$
and the area $a_{\rm H}$,
Eq.~(\ref{ah_rel1}), for the extremal solutions.
Since we have chosen $J \ge 0$, the non-extremal rotating solutions
reside within the boundary formed by the extremal and the static solutions,
which have $j=0$.

Fig.~\ref{fig3c} exhibits the gyromagnetic ratio $g$ of these black holes.
For small charge, the perturbative value $g_{\rm \delta q}=3$ is found
\cite{Aliev:2004ec},
while for larger values of the charge considerable deviations from this value arise,
as shown in higher order perturbation theory
\cite{NavarroLerida:2007ez,Allahverdizadeh:2010xx,Allahverdizadeh:2010fn}.
The curves formed by the gyromagnetic ratio $g$ of the extremal black holes 
and by the gyromagnetic ratio $g_{\rm \delta j}$, Eq.~(\ref{gdeltaj}),
obtained for black holes in the static limit $J\to 0$ \cite{Sheykhi:2008bs},
enclose the domain, where the gyromagnetic ratio can take its values.
Since the extremal and the `static' curve are very close to each other,
the `static' values represent a good approximation for a given value of $q$.
Consequently, the gyromagnetic ratio is not well resolved for
the non-extremal sets of solutions in the figure.
We recall, that in the KK case, $g$ is given by a single curve, i.e.,
the KK domain of $g$ is only one-dimensional.

The situation is similar for the horizon electrostatic potential $\Phi_{\rm H}$,
which we exhibit in Fig.~\ref{fig3d}.
Again the static and the extremal solutions
forming  the boundaries for the admissible values of this quantity
are very close to each other, thus the values of the
non-extremal black holes are not well discernable here, either.
In the KK case $\Phi_{\rm H}$ is again given by a single curve, i.e.,
its KK domain is only one-dimensional.

The surface gravity $\kappa$
of these black holes is addressed in Fig.~\ref{fig3e},
where the scaled surface gravity $\bar \kappa = \kappa M^{1/2}$
is exhibited versus the scaled charge $q$.
The static set of solutions here forms the upper boundary,
while the set of extremal solutions, having $\kappa=0$ (for finite $J$),
forms the lower boundary. 
At the static extremal black hole,
the static curve diverges, $\kappa = \infty$
\cite{Gibbons:1987ps,Horowitz:1991cd}, 
since $h > h_{\rm cr}=\sqrt{2/3}$, Eq.~(\ref{hcr}).
Thus for the set of extremal black holes the surface gravity
jumps from zero, its value for finite angular momentum,
to infinity in the static limit.

Finally, we exhibit the scaled horizon angular velocity 
$\bar \Omega = \Omega M^{1/2}$ in Fig.~\ref{fig3f}.
Here the static black holes, possessing $\Omega=0$, form the lower boundary,
while the extremal black holes form the upper boundary,
with all other black holes assuming values in-between.
Thus the scaled horizon angular velocity
and the surface gravity show an analogous behavior,
where the static and extremal solutions, however,
have switched roles.

\boldmath
\subsubsection{$h$-dependence}
\unboldmath

\begin{figure}[p]
\begin{center}
\vspace{-0.7cm}
\mbox{\hspace{-1.5cm}
\subfigure[][]{
\includegraphics[height=.32\textheight, angle =270]{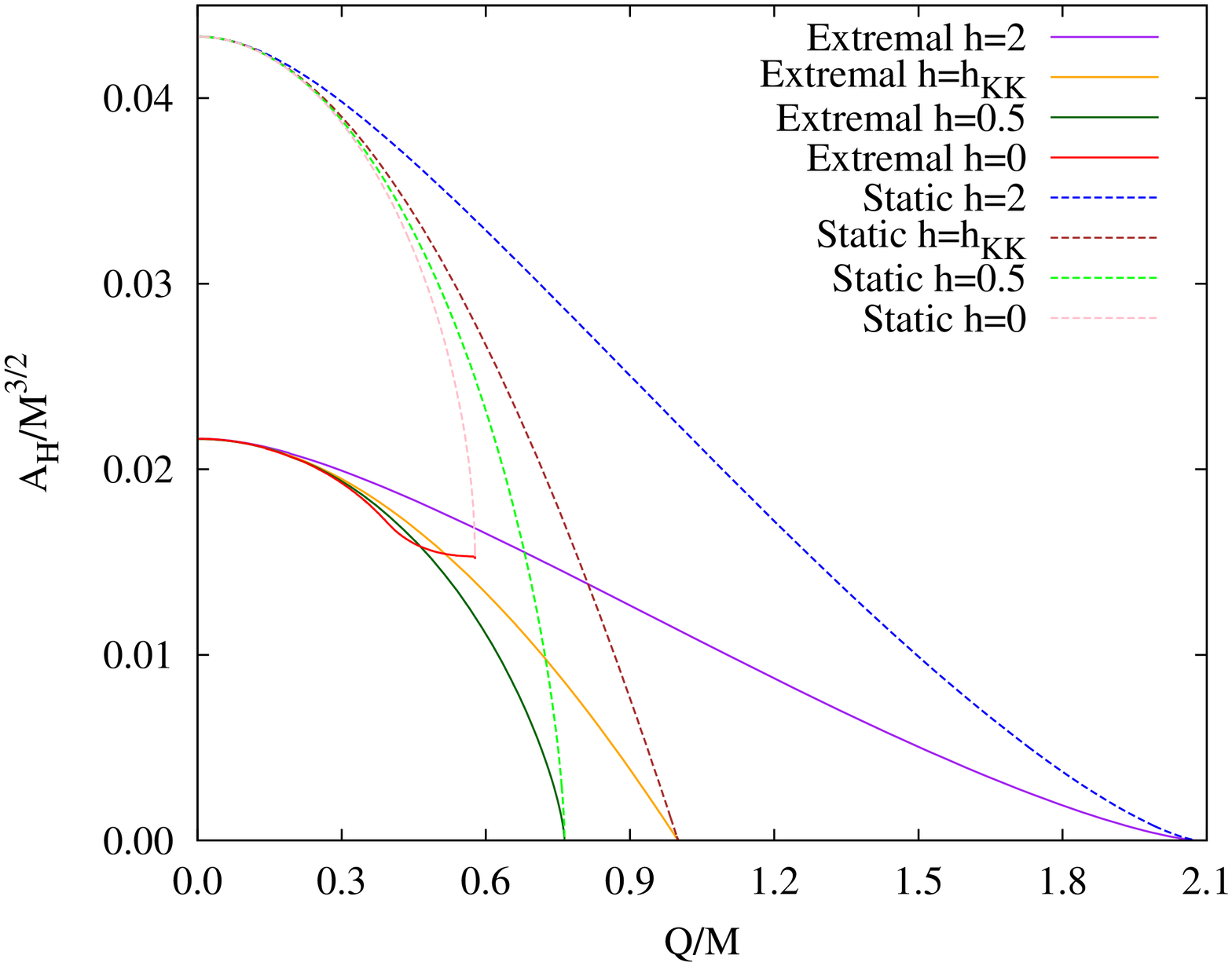}
\label{fig4a}
}
\subfigure[][]{\hspace{-0.5cm}
\includegraphics[height=.32\textheight, angle =270]{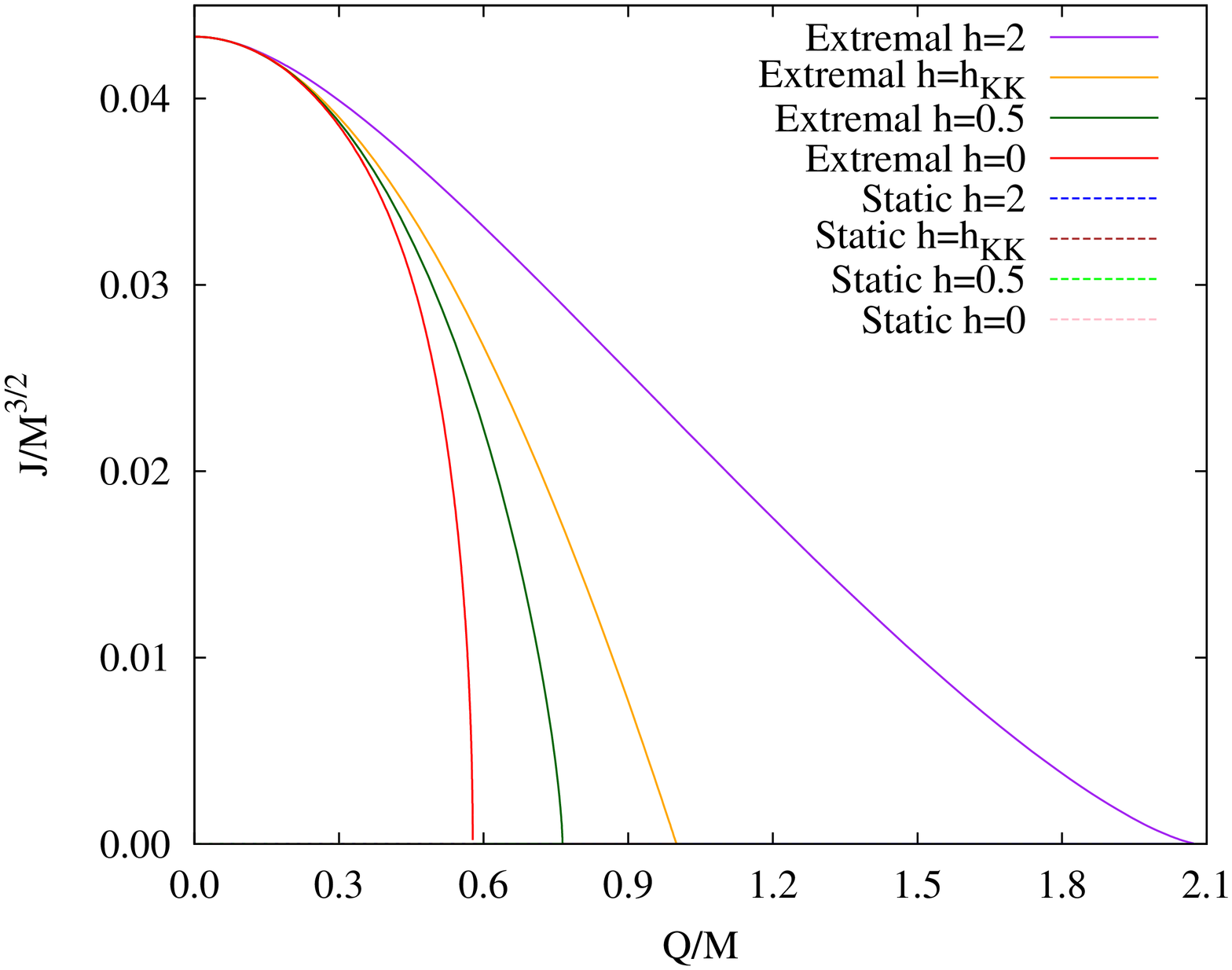}
\label{fig4b}
}
}
\mbox{\hspace{-1.5cm}
\subfigure[][]{
\includegraphics[height=.32\textheight, angle =270]{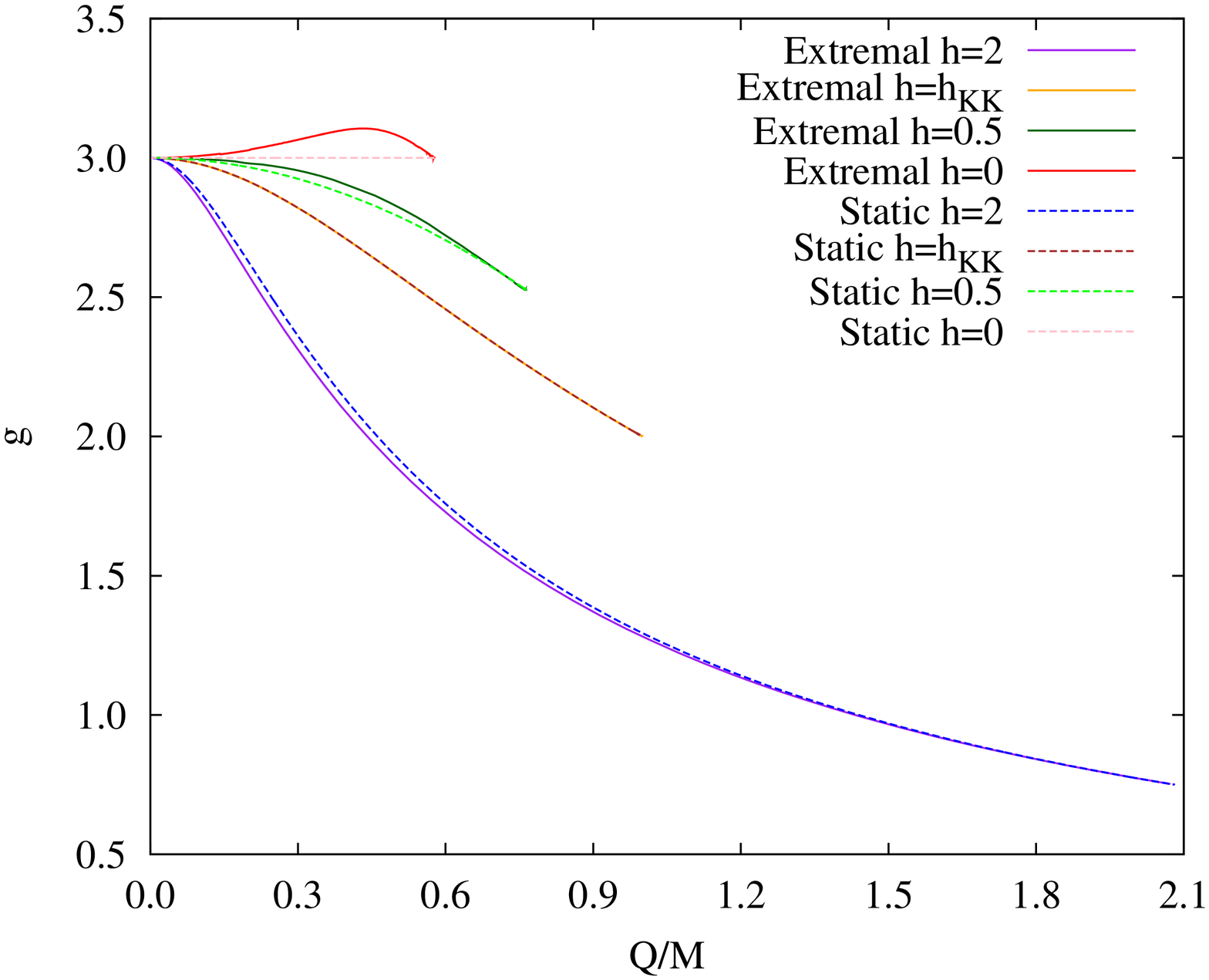}
\label{fig4c}
}
\subfigure[][]{\hspace{-0.5cm}
\includegraphics[height=.32\textheight, angle =270]{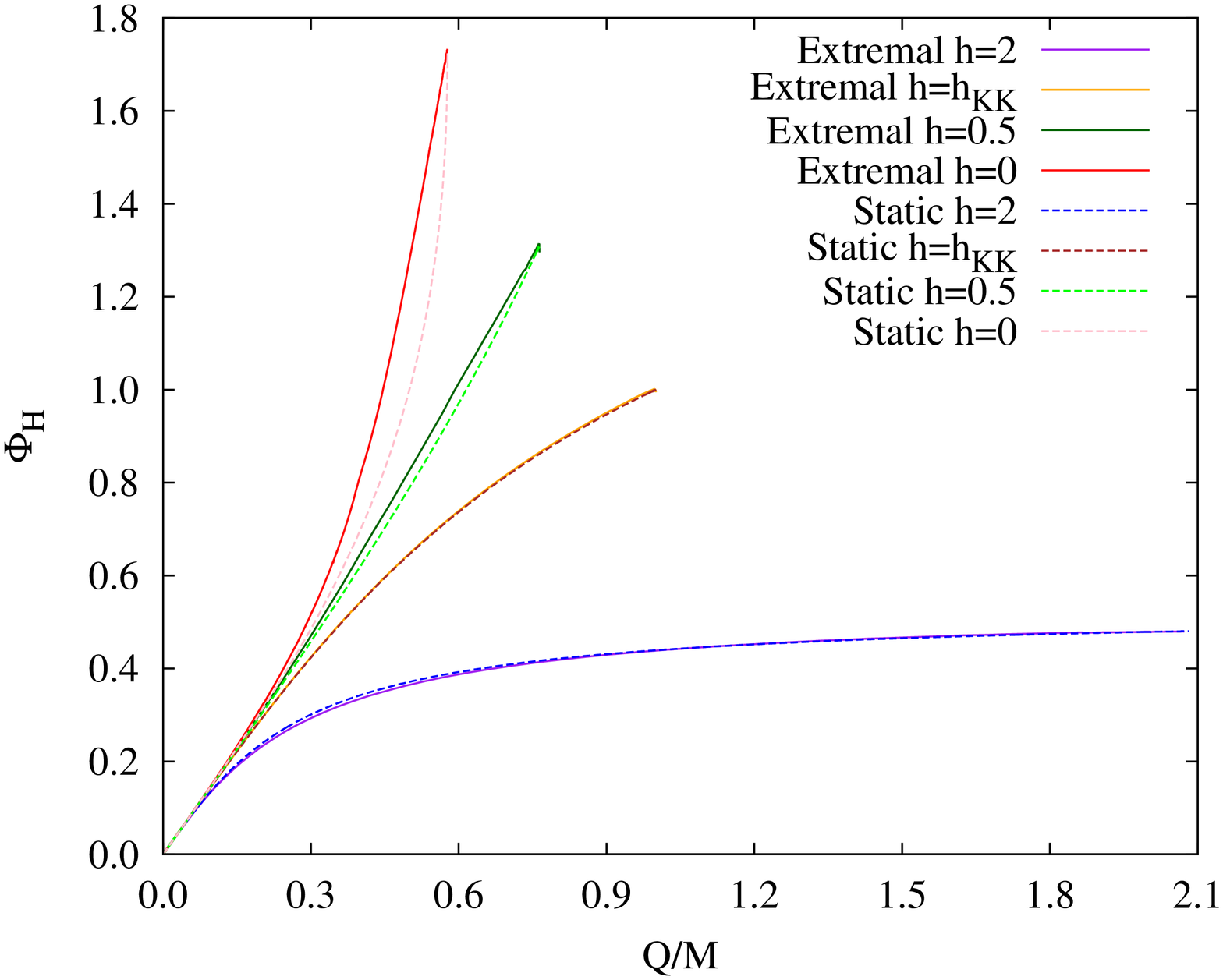}
\label{fig4d}
}
}
\mbox{\hspace{-1.5cm}
\subfigure[][]{
\includegraphics[height=.32\textheight, angle =270]{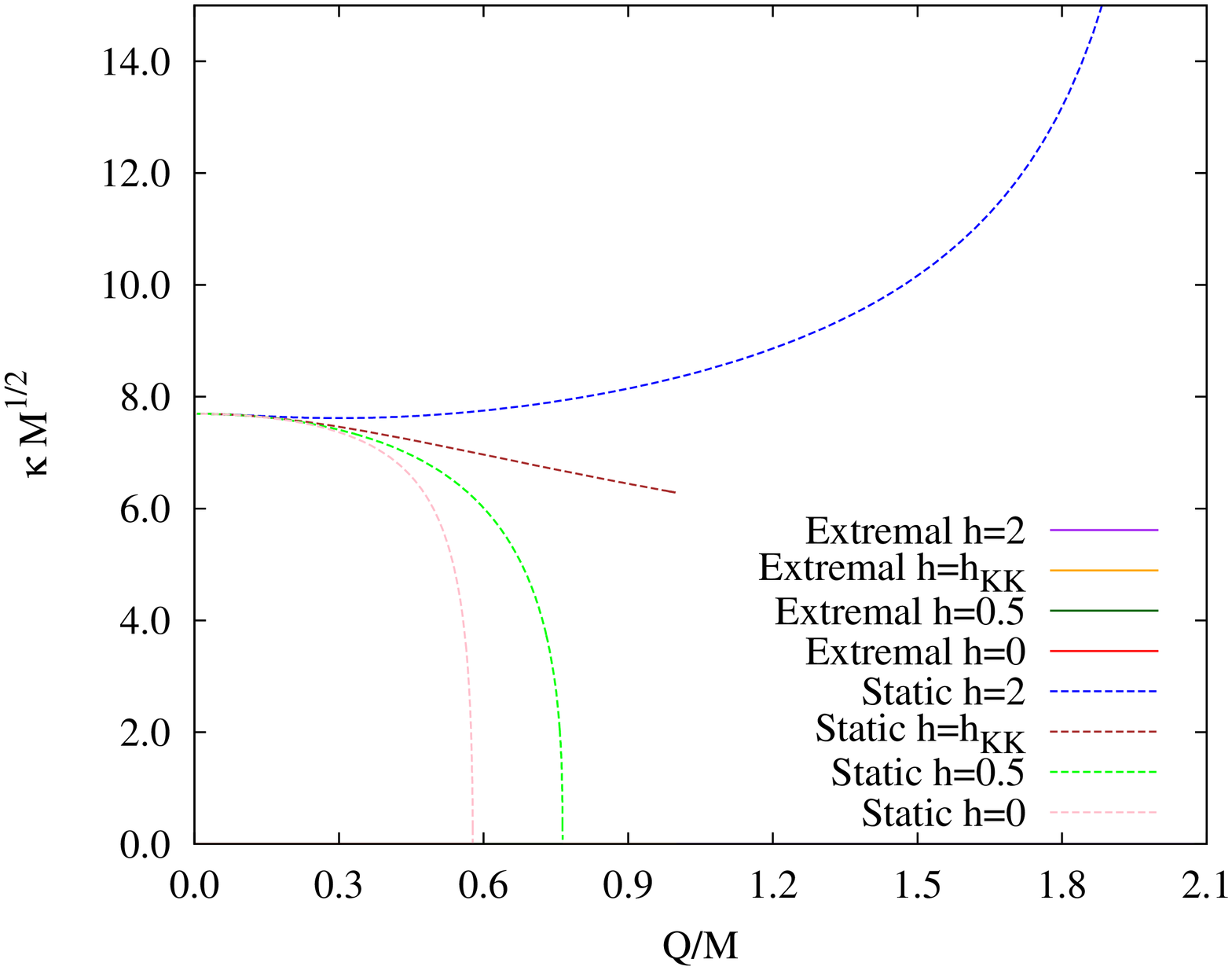}
\label{fig4e}
}
\subfigure[][]{\hspace{-0.5cm}
\includegraphics[height=.32\textheight, angle =270]{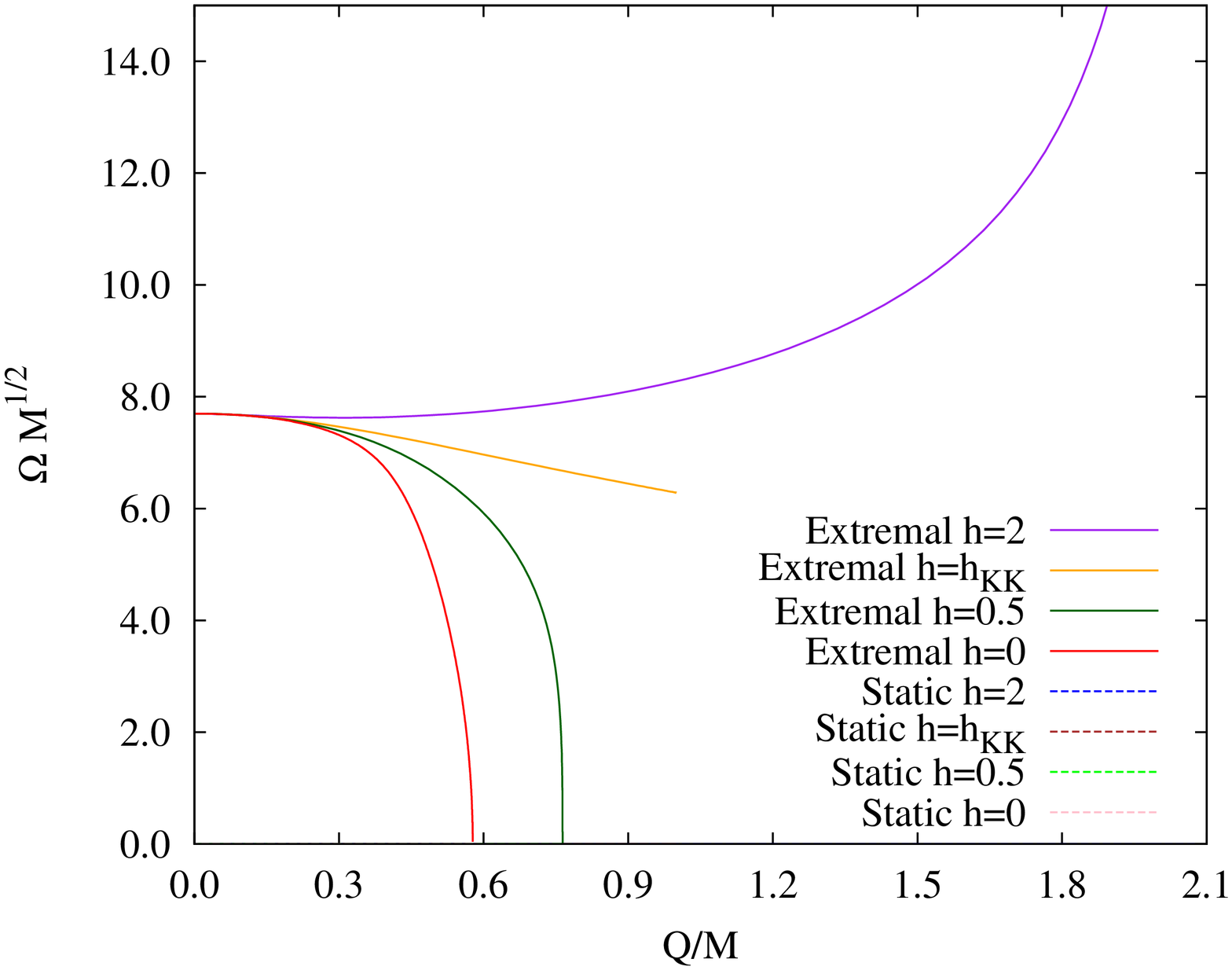}
\label{fig4f}
}
}
\end{center}
\caption{Properties of EMd black hole solutions 
in five dimensions
are shown for several values of the dilaton coupling constant $h$:
$h=0$ (EM), $0.5$, $\sqrt{2/3}$ (KK), $2$.
(a) The scaled area $a_{\rm H} = A_{\rm H}/M^{3/2}$
versus the scaled charge $q=|Q|/M$
for extremal and for static solutions,
providing the boundary of the domain of existence.
For the same sets of (extremal and non-extremal) solutions 
we exhibit in
(b) the scaled angular momenta $j=J/M^{3/2}$ versus $q$,
(c) the gyromagnetic ratio $g$ versus $q$,
(d) the horizon electrostatic potential $\Phi_{\rm H}$ versus $q$
(e) the scaled surface gravity $\bar \kappa = \kappa M^{1/2}$ versus $q$,
(f) the scaled horizon angular velocity $\bar \Omega = \Omega M^{1/2}$
    versus $q$.
\label{fig4}
}
\end{figure}

To study the dependence of these black holes on
the dilaton coupling constant $h$, we now consider
several fixed values of $h$:
$h=0$, corresponding to the EM case, $h=0.5$, 
$h=\sqrt{2/3}$, corresponding to the KK case,
and $h=2$, the case studied above in more detail.
We exhibit properties of these solutions in Figs.~\ref{fig4}.
In particular, we show the extremal and the static solutions,
which form the boundary of the domain of existence
for these black holes.
All non-extremal rotating black holes are located inside this boundary.
The quantities shown are
the scaled horizon area $a_{\rm H}$ (Fig.~\ref{fig4a}),
the scaled angular momenta $j$ (Fig.~\ref{fig4b}),
the gyromagnetic ratio $g$ (Fig.~\ref{fig4c}),
the horizon electrostatic potential $\Phi_{\rm H}$ (Fig.~\ref{fig4d}),
the scaled surface gravity $\bar \kappa$ (Fig.~\ref{fig4e}),
and the scaled horizon angular velocity $\bar \Omega$ (Fig.~\ref{fig4f}).

\begin{figure}[h!]
\begin{center}
\mbox{\hspace{-1.5cm}
\subfigure[][]{
\includegraphics[height=.32\textheight, angle =270]{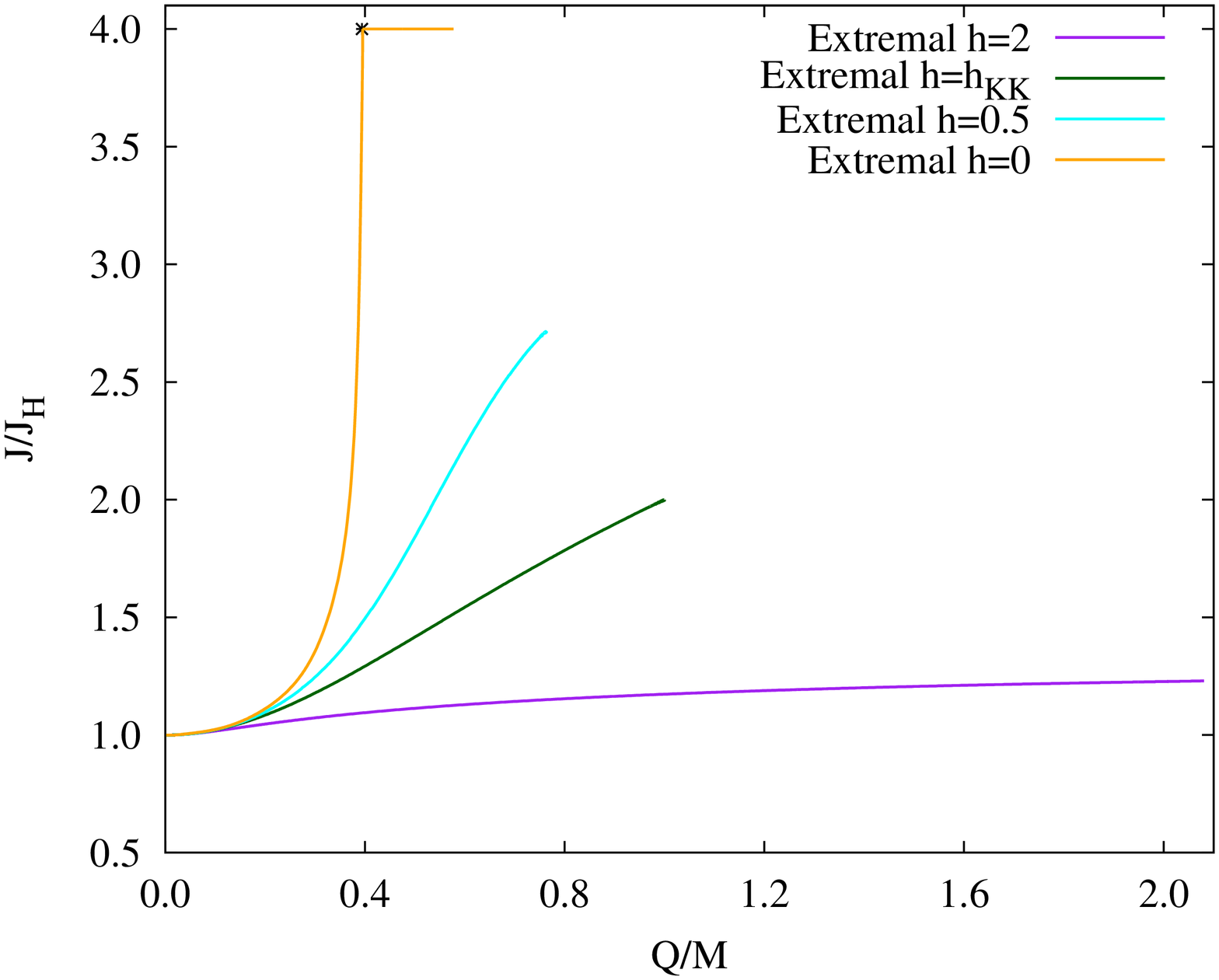}
\label{fig5a}
}
\subfigure[][]{\hspace{-0.5cm}
\includegraphics[height=.32\textheight, angle =270]{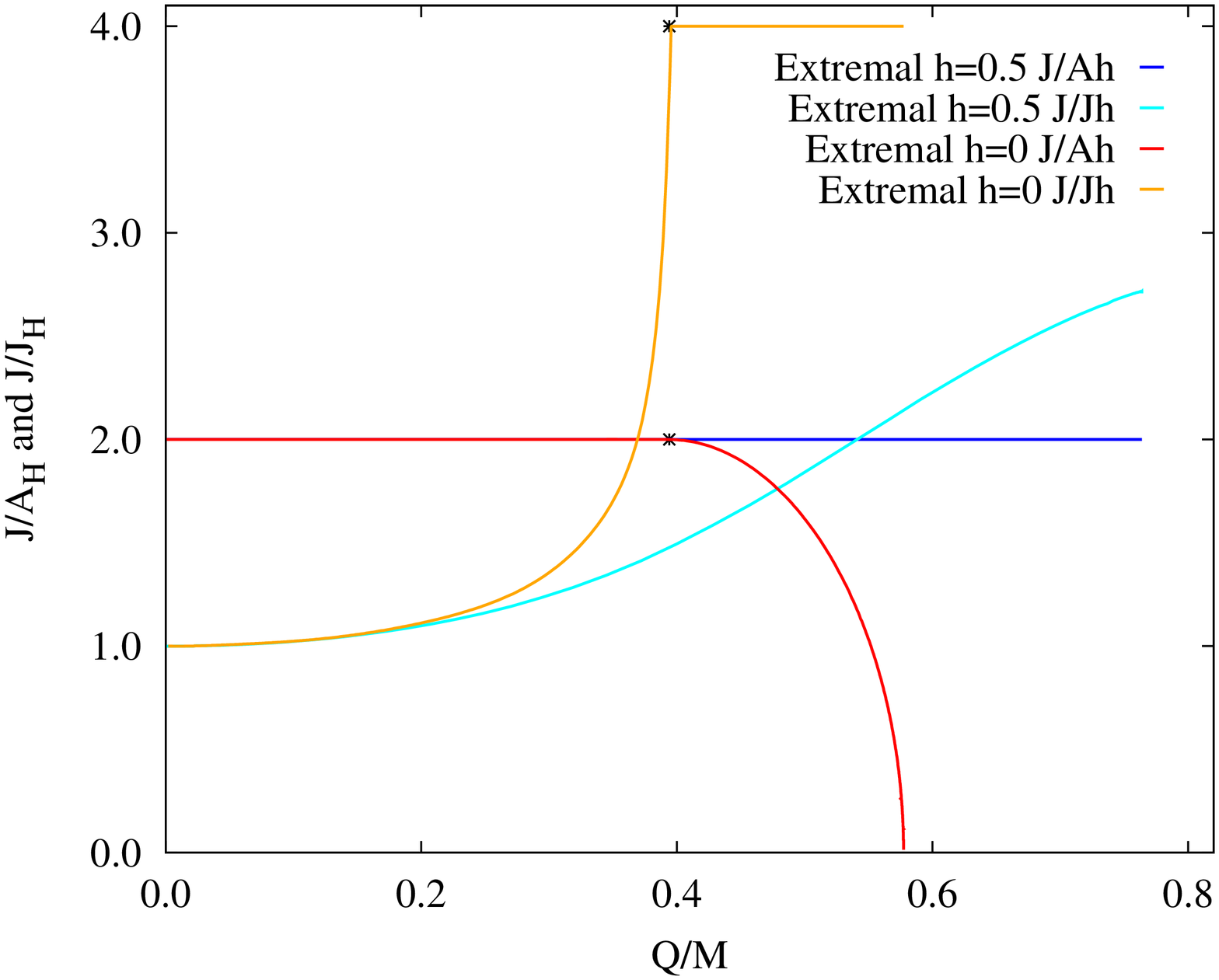}
\label{fig5b}
}
}
\end{center}
\caption{\small
(a) $J/J_{\rm H}$ is shown versus the scaled charge $q$ for extremal EMd black
hole solutions in five dimensions for $h=0$ (EM), $0.5$,
$\sqrt{2/3}$ (KK), and $2$.
(b) The ratios $J/A_{\rm H}$ and
$J/J_{\rm H}$ versus $q$ 
for $h=0$ and $h=0.5$. The asterisks mark the matching point
of the two EM branches.  
}
\label{fig5}
\end{figure}

For the area and the angular momenta, we note again the 
proportionality for finite dilaton coupling.
For the EM case, $h=0$, this proportionality holds only along
the MP branch, while it is violated on the RN branch of the solutions.
We demonstrate this further in Figs.~\ref{fig5},
where we exhibit the branch structure for these solutions. 
In particular, we show the ratio $J/J_{\rm H}$ versus the scaled
charge $q$ in Fig.~\ref{fig5a} for the
above set of coupling constants and the ratios $J/A_{\rm H}$ and $J/J_{\rm H}$
versus $q$ for $h=0$ and $h=0.5$ in Fig.~\ref{fig5b}.

The figures clearly reveal the two-branch structure of the
extremal EM solutions, together with their matching point,
and the single-branch structure of the EMd solutions. 
Comparison with Figs.~\ref{fig2} shows, that for the EM solutions
only the first part of the near-horizon MP branch 
and the second part of the near-horizon RN branch are realized globally.
For the EMd solutions, on the other hand, the surfaces of
near-horizon solutions are reduced to single curves,
that are - in part - realized globally, for arbitrary finite value of the
dilaton coupling $h$ (compare Figs.~\ref{fig1}).

In the extremal limit, the horizon area
of the static black holes vanishes as long as the
dilaton coupling constant $h$ is non-vanishing, no matter how small it is.
In contrast, for the static EM black holes the area remains finite.
The EM limit, $h \rightarrow 0$,
of the extremal static black hole
is therefore not smoothly approached.

Turning to 
the gyromagnetic ratio $g$, exhibited in Fig.~\ref{fig4c},
we note, that it
attains the perturbative value $g_{\rm \delta q} = 3$ \cite{Aliev:2004ec}
for small $q$,
independent of the dilaton coupling constant $h$.
For a fixed value of $h$,
the curves formed by the gyromagnetic ratio $g$ of the extremal black holes 
and by the gyromagnetic ratio $g_{\rm \delta j}$, Eq.~(\ref{gdeltaj}),
obtained for black holes in the static limit $J\to 0$ \cite{Sheykhi:2008bs},
enclose the domain, where the gyromagnetic ratio
can take its values.

In the EM case, the perturbative value obtained for small $q$, $g_{\rm \delta q}=3$, 
coincides with the perturbative value for small $j$, $g_{\rm \delta j}=3$, 
which forms a horizontal line.
At the same time, this `static' limit constitutes a lower bound for the
gyromagnetic ratio of all EM black holes with equal
magnitude angular momenta
\cite{NavarroLerida:2007ez,Kunz:2005nm,Kunz:2006eh}.

For small but finite values of the dilaton coupling constant $h$, 
the gyromagnetic ratio is no longer constant in the static limit.
Instead it decreases monotonically with
increasing $q$, Eq.(\ref{gdeltaj}).
As the dilaton coupling constant $h$ is increased,
the boundary curves of the domain of $g$,
approach each other, 
until at the Kaluza-Klein value $h_{\rm KK}$
both curves coincide, 
as seen in Eq.~(\ref{gyromagnetic_ratios}).
Considering the gyromagnetic ratio $g$ versus $q$,
all KK black holes fall on a single curve with $3 \ge g \ge 2$.

For $h > h_{\rm KK}$, the
boundary curves formed by the
extremal and the `static' black holes
separate again, retaining common end points.
In all cases, however, the static value of $g$ for a given $q$
is a rather good approximation for the true value,
which becomes exact in the KK case.

The horizon electrostatic potential $\Phi_{\rm H}$ is
shown in Fig.~\ref{fig4d}.
As always, the static and the extremal solutions
form the boundaries for the admissible values of this quantity.
Analogous to the case of the gyromagnetic ratio,
the two boundary curves are always rather close to each other.
They approach each other with increasing $h$,
form a single curve for the Kaluza-Klein value $h_{\rm KK}$ 
and then separate again. 
Thus the static value of $\Phi_{\rm H}$ for a given $q$
approximates the true value well,
and becomes exact in the KK case.

\begin{figure}[t!]
\begin{center}
\mbox{\hspace{-1.5cm}
\subfigure[][]{
\includegraphics[height=.32\textheight, angle =270]{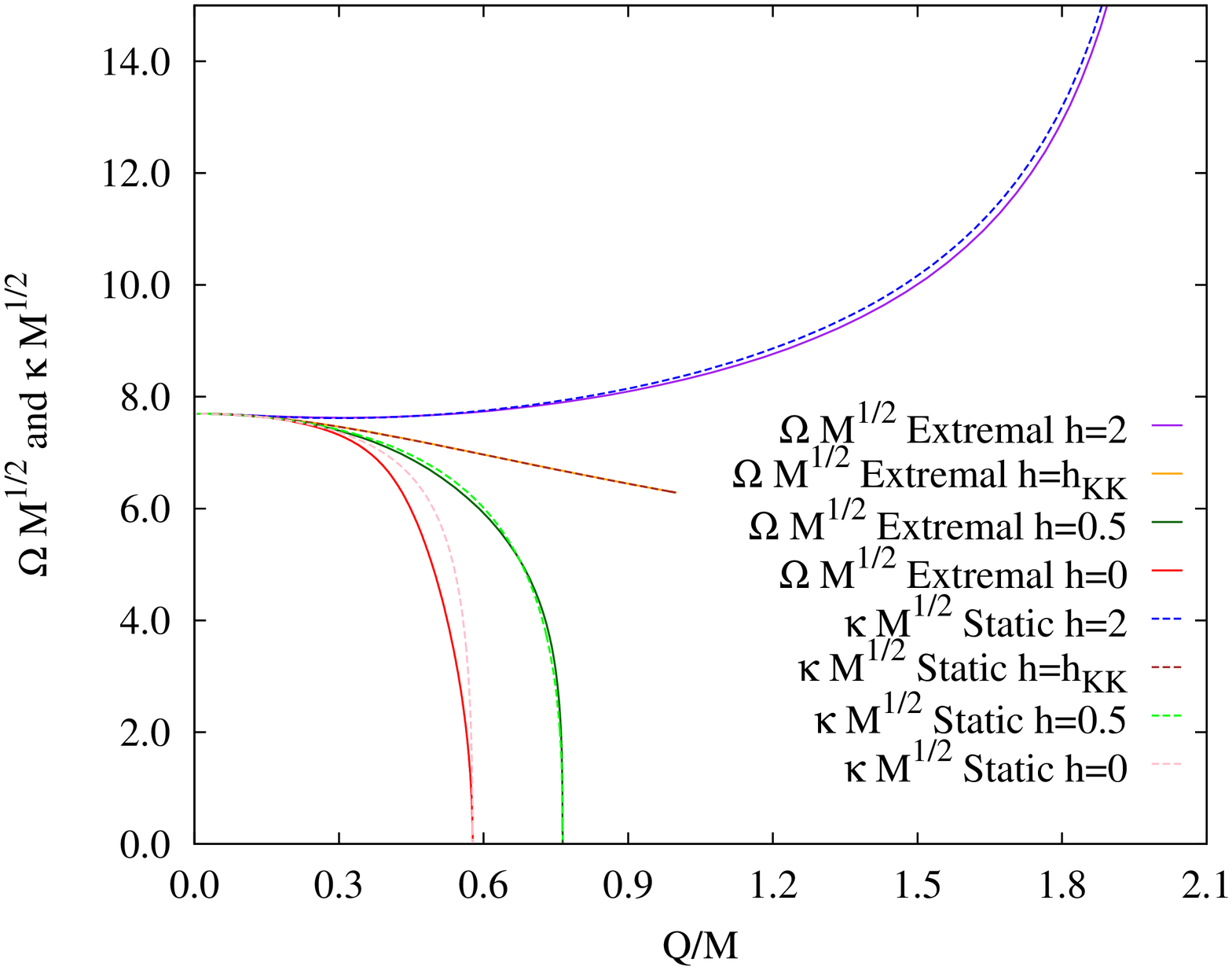}
\label{fig6a}
}
\subfigure[][]{\hspace{-0.5cm}
\includegraphics[height=.32\textheight, angle =270]{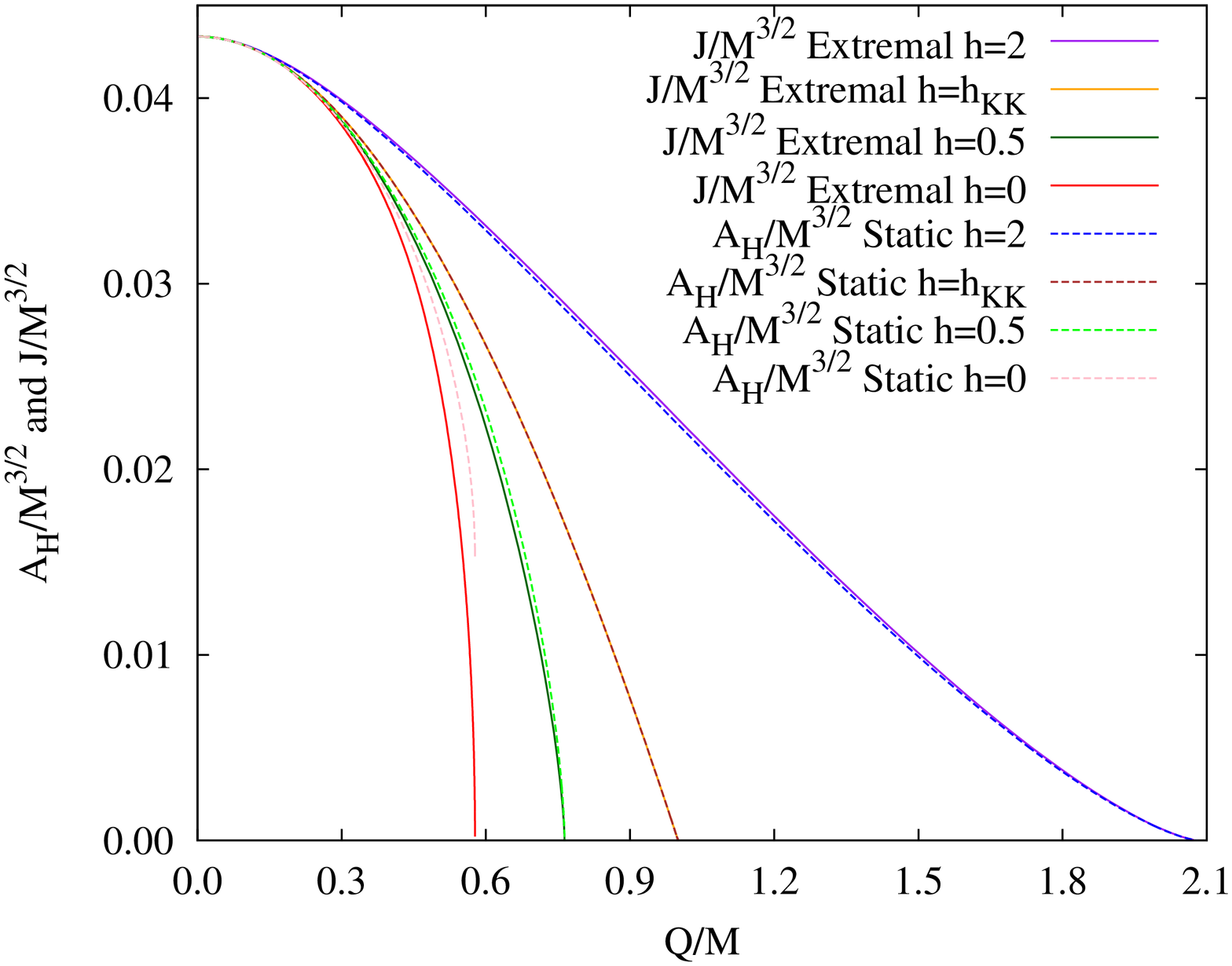}
\label{fig6b}
}
}
\end{center}
\caption{\small
Properties of EMd black hole solutions 
in five dimensions
are shown for several values of the dilaton coupling constant $h$:
$h=0$ (EM), $0.5$, $\sqrt{2/3}$ (KK), $2$.
(a) The scaled surface gravity $\bar \kappa = \kappa M^{1/2}$
of the static solutions and
the scaled horizon angular velocity $\bar \Omega = \Omega M^{1/2}$
of the extremal solutions
versus the scaled charge $q=|Q|/M$,
(b) the scaled area $a_{\rm H} = A_{\rm H}/M^{3/2}$
of the static solutions and
the scaled angular momenta $j=J/M^{3/2}$
of the extremal solutions.
}
\label{fig6}
\end{figure}

The scaled surface gravity $\bar \kappa$
vanishes for the extremal solutions.
The curves seen in Fig.~\ref{fig4e}
thus represent the upper boundary of the domain
of existence, formed by the static solutions
for the various values of the dilaton coupling $h$.
Of particular interest is the extremal endpoint of these static curves.
For $h < h_{\rm cr}=\sqrt{2/3}$ the surface gravity
is zero at the endpoint. 
At the critical value $h = h_{\rm cr}$ the surface gravity
assumes a finite value, $\kappa_{\rm cr}$,
whereas for  $h > h_{\rm cr}$ the surface gravity
diverges at the endpoint \cite{Gibbons:1987ps,Horowitz:1991cd}.
Thus for the set of extremal rotating black holes the surface gravity
jumps from zero, its value for finite angular momentum,
to the finite value $\kappa_{\rm cr}$ for $h = h_{\rm cr}$
and to infinity for $h > h_{\rm cr}$ in the static limit,
as indicated in the figure.

The figure for the scaled horizon angular velocity 
$\bar \Omega$, shown in Fig.~\ref{fig4f},
looks completely analogous to the one for the scaled surface gravity.
However, here the static solutions form the lower boundary
of the domain of existence, since they have $\Omega=0$,
while the upper boundary is formed by the extremal rotating solutions.
Inspecting again the endpoint of the set of extremal solutions,
we note the same dependence on  $h$.
For $h < h_{\rm cr}=\sqrt{2/3}$ the horizon angular
velocity vanishes at the endpoint.
At the critical value $h = h_{\rm cr}$ the horizon angular velocity
assumes a finite value, $\Omega_{\rm cr}$,
whereas for  $h > h_{\rm cr}$ the horizon angular velocity
diverges at the endpoint.

To better understand this analogy, let us inspect Fig.~\ref{fig6a},
where we compare
the scaled surface gravity $\bar \kappa$
of the static solutions to
the scaled horizon angular velocity $\bar \Omega$
of the extremal solutions for several values of the dilaton coupling $h$.
We note, that they agree for $h=h_{\rm KK}$,
while they are close for other values of $h$.
The situation is analogous, when we compare
the scaled area $a_{\rm H}$
of the static solutions to
the scaled angular momenta $j$
of the extremal solutions, as seen in Fig.~\ref{fig6b}.

We now recall our discussion  in section \ref{sec_KK_rel}.
There we showed, that for KK black holes in five dimensions 
the scaled surface gravity of static black holes
indeed agrees with the scaled horizon angular velocity of extremal black holes.
Considering the scaled Smarr formula,
we furthermore showed relation Eq.~(\ref{rel_kaOm}) for KK black holes,
which derived from the fact, that for KK black holes
the horizon electrostatic potential only depends on the scaled charge.
Since we have seen, that the horizon electrostatic potential
depends only little on the angular momenta also for other values of 
the dilaton coupling $h$,
we conclude that relation Eq.~(\ref{rel_kaOm}) holds approximately
also for other values of $h$.
This then leads to the similarity of the static $\bar \kappa$
and the extremal $\bar \Omega$,
as well as the similarity of the static $a_{\rm H}$ 
and the extremal $j$, observed in Figs.~\ref{fig6}.
We conclude, that we can learn much about extremal rotating solutions
by only inspecting static solutions.

\begin{figure}[t!]
\begin{center}
\mbox{\hspace{-1.5cm}
\subfigure[][]{
\includegraphics[height=.32\textheight, angle =270]{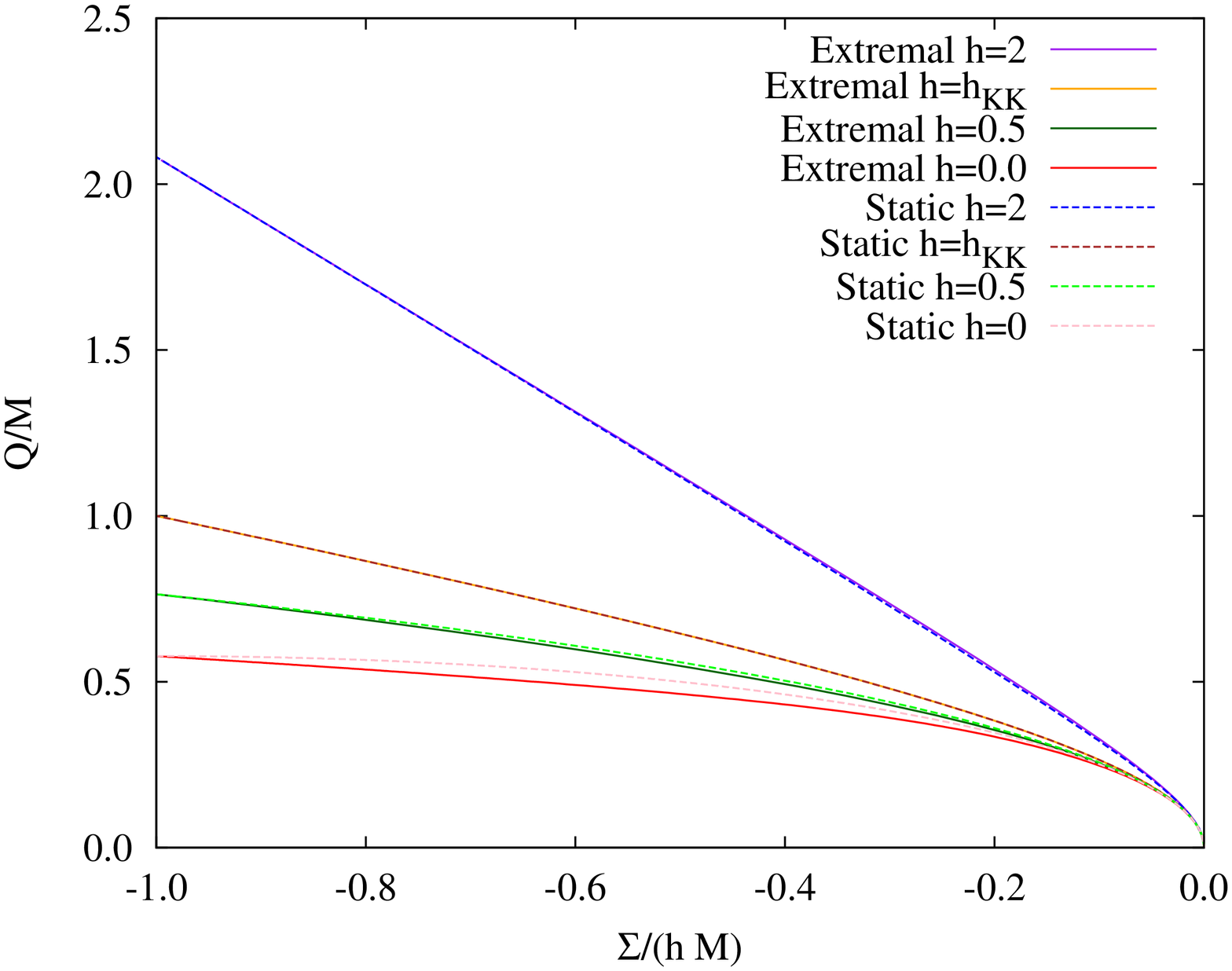}
\label{fig7a}
}
\subfigure[][]{\hspace{-0.5cm}
\includegraphics[height=.32\textheight, angle =270]{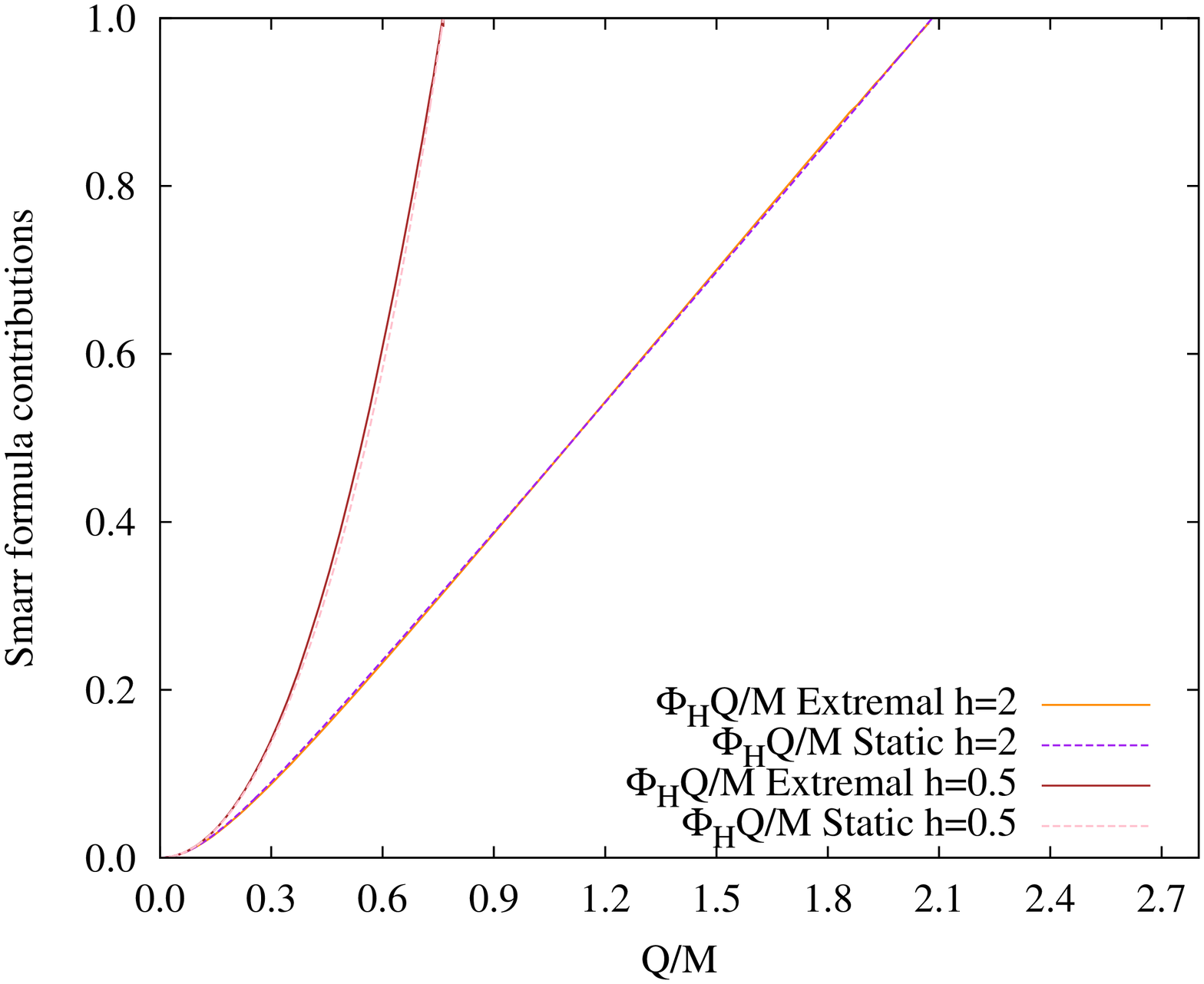}
\label{fig7b}
}
}
\end{center}
\caption{\small
(a) The scaled charge $q=Q/M$ versus the scaled relative dilaton charge $\Sigma/(hM)$ for
static and extremal solutions with $h=0$, $0.5$, $h_{KK}$ and $2$. The
quadratic relation Eq.~(\ref{quadratic_relation_static})
between both scaled charges is fulfilled  for every value of $h$ in the static case,
and for $h_{KK}$ Eq.~(\ref{quadratic_relation}) in the extremal case. (b) Contributions to
the Smarr formula Eq.~(\ref{mass3}) for static and extremal solutions, for both $h=0.5$
and $h=2$. 
}
\label{fig7}
\end{figure}

In Fig.~\ref{fig7a} we show the scaled charge $Q/M$ versus the scaled relative dilaton
charge $\Sigma/(hM)$ for static and extremal solutions and several values of
the dilaton coupling. Note that the quadratic relation
Eq.~(\ref{quadratic_relation}) is only fulfilled in the KK case, while the
static solutions satisfy a similar quadratic relation,
\begin{equation}
\DS \frac{Q^2}{M + \left[\DS 1-\frac{2(D-2)h^2}{D-3}\right] \DS \frac{\Sigma}{2h}}=
-\frac{D-3}{D-2}\frac{\Sigma}{h}  
\ , \label{quadratic_relation_static} 
\end{equation}
valid for arbitrary $h$. It is interesting to note that this relation coincides with the quadratic
relation Eq.~(\ref{quadratic_relation}) of the KK solution, when $h=h_{KK}$. In
Fig.~\ref{fig7a} we see that for extremal solutions with general dilaton
coupling, this quadratic relation is almost 
fulfilled: the extremal curves of Fig.~\ref{fig7a} are very close to the static
curves, where relation Eq.~(\ref{quadratic_relation_static}) holds exactly. The approximation of the 
(in general unknown)
extremal relation by the static quadratic relation is very good for large dilaton
couplings (and exact in the KK case).

In Fig.~\ref{fig7b} we present the Smarr formula contributions for the
extremal and static solutions. Here the situation is similar. Again we note that
the extremal and static 
curves are very close, and we can use the information from the static
solutions to gain insight on the extremal solutions.

\boldmath
\subsection{$D>5$}
\unboldmath

Let us now turn to black holes in more than five dimensions.
Here we show, that the basic features observed for EMd
black holes with equal-magnitude angular momenta
are retained in higher odd dimensions. To exhibit the
dependence on the number of dimensions $D$,
we compare sets of solutions in $D=5$, 7, and 9 dimensions.

\boldmath
\subsubsection{$D=7$}
\unboldmath

\begin{figure}[t!]
\begin{center}
\vspace{-0.7cm}
\mbox{\hspace{-1.5cm}
\subfigure[][]{
\includegraphics[height=.35\textheight, angle =270]{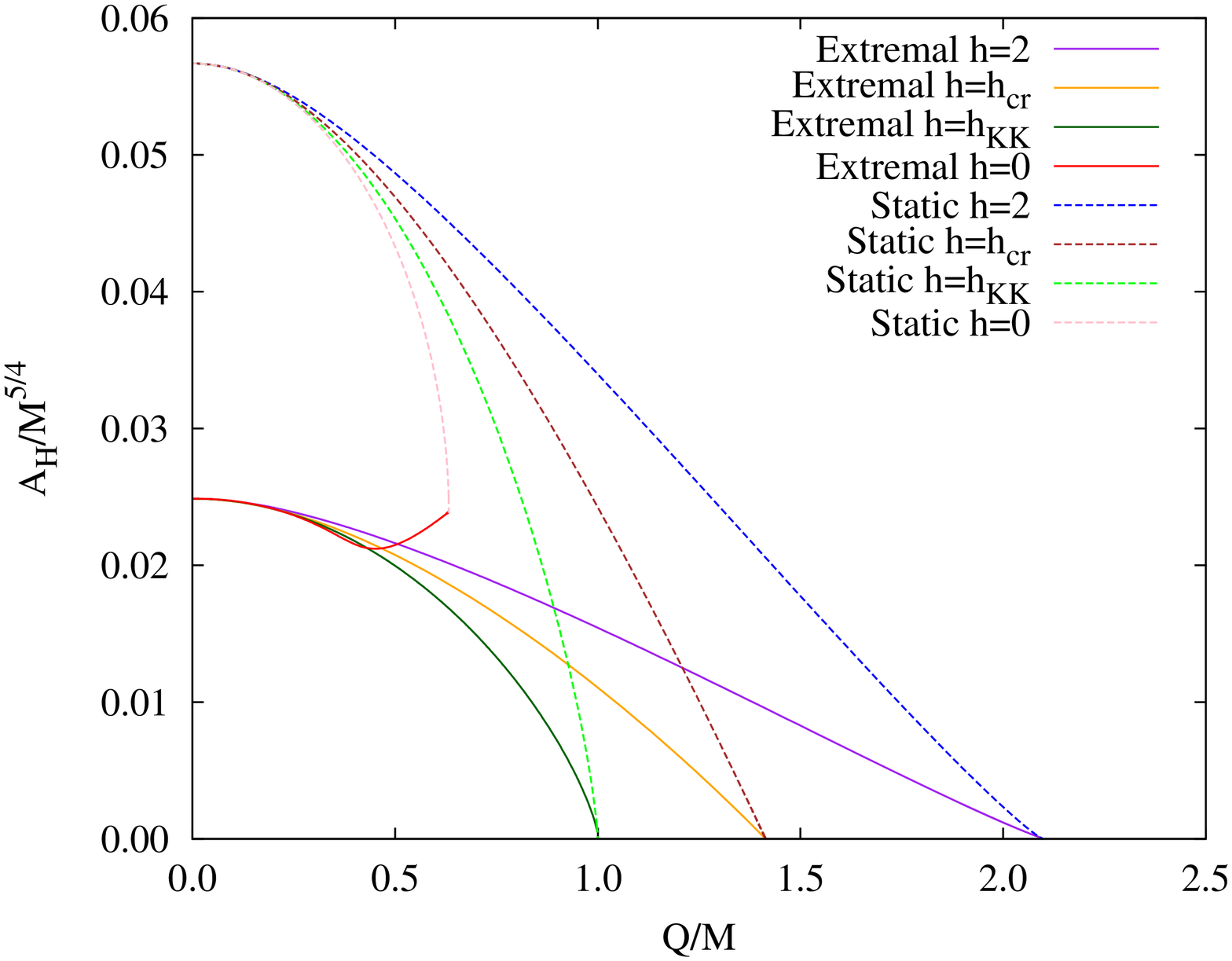}
\label{fig8a}
}
\subfigure[][]{\hspace{-0.5cm}
\includegraphics[height=.35\textheight, angle =270]{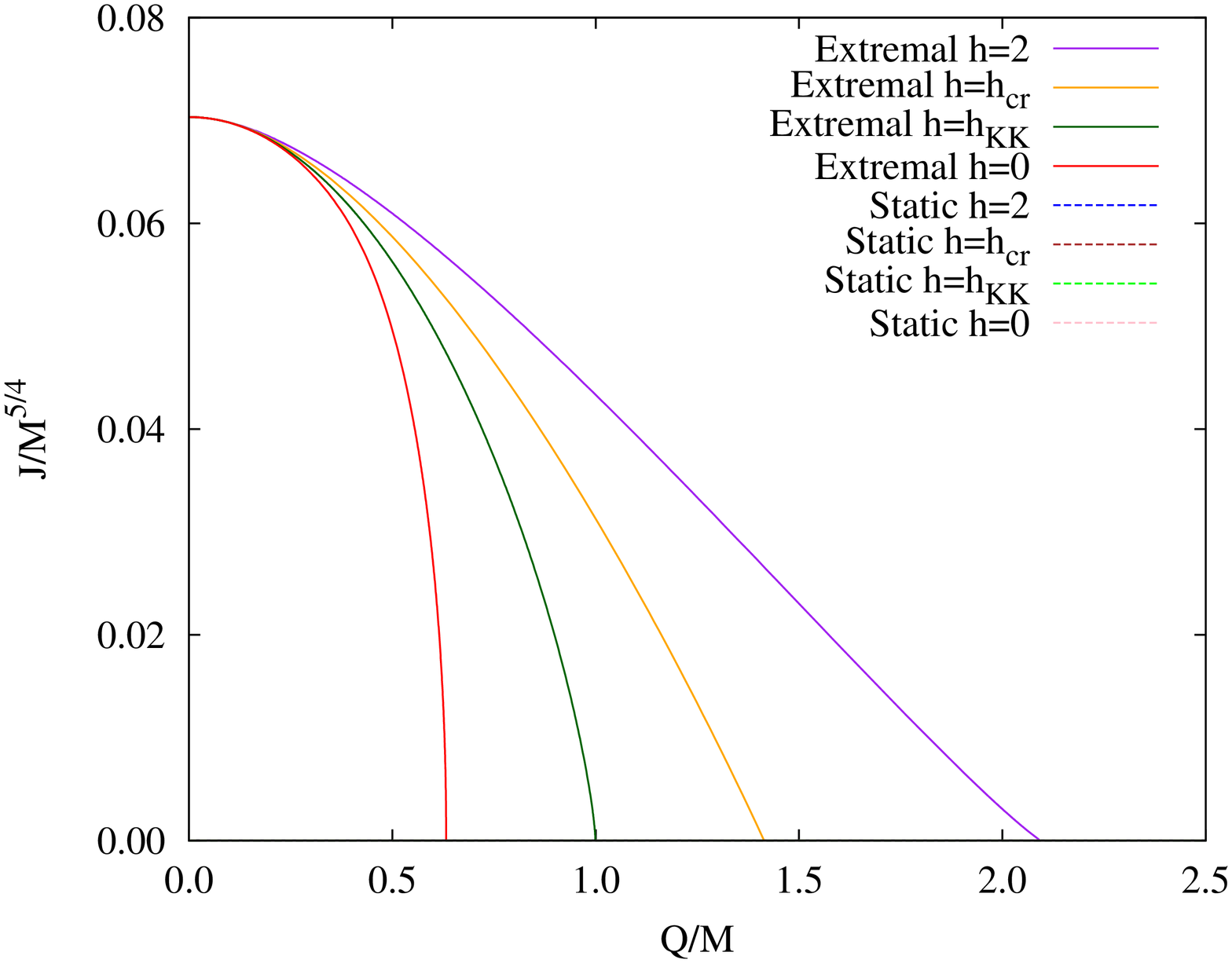}
\label{fig8b}
}
}
\vspace{-0.5cm}
\mbox{\hspace{-1.5cm}
\subfigure[][]{
\includegraphics[height=.35\textheight, angle =270]{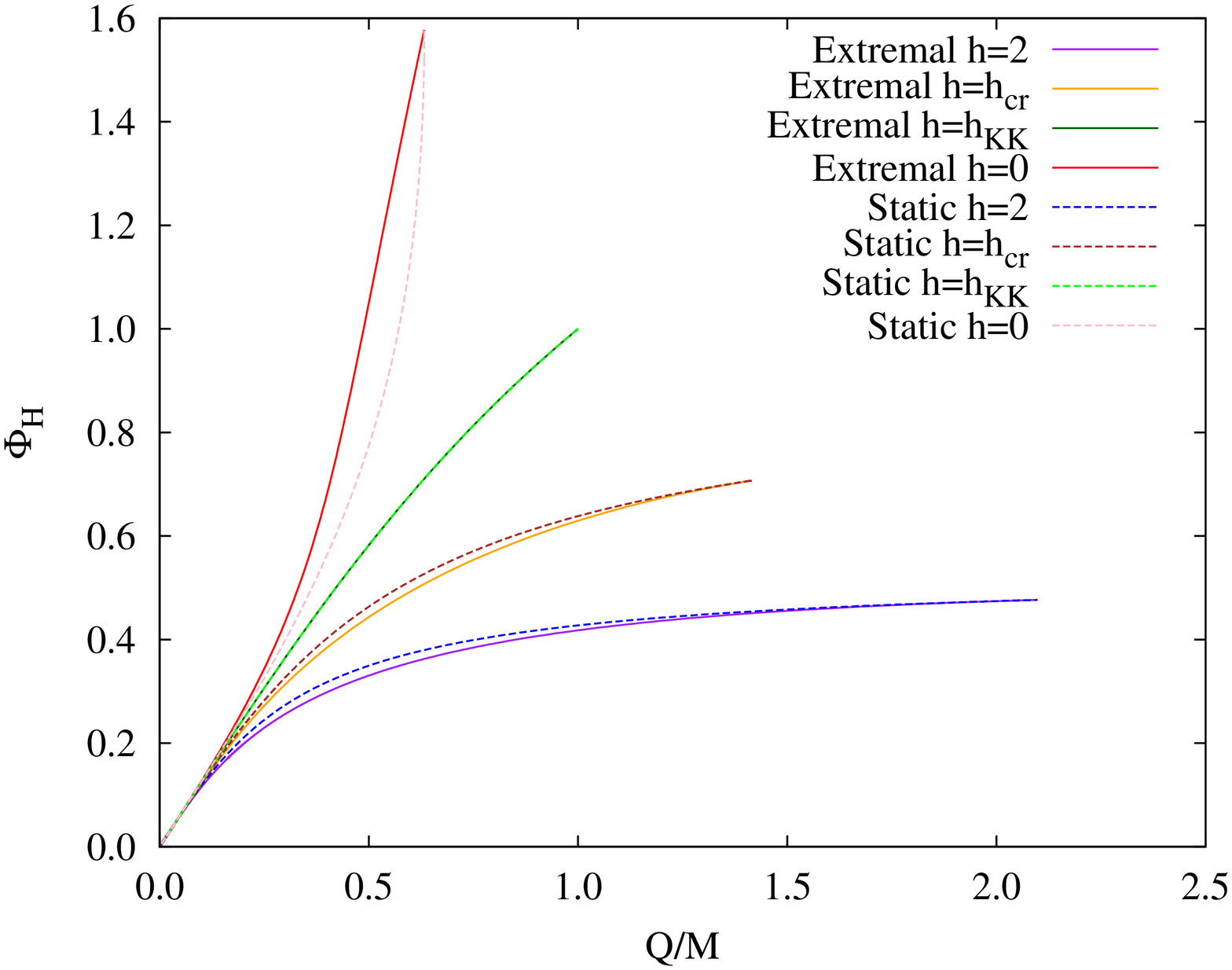}
\label{fig8c}
}
\subfigure[][]{\hspace{-0.5cm}
\includegraphics[height=.35\textheight, angle =270]{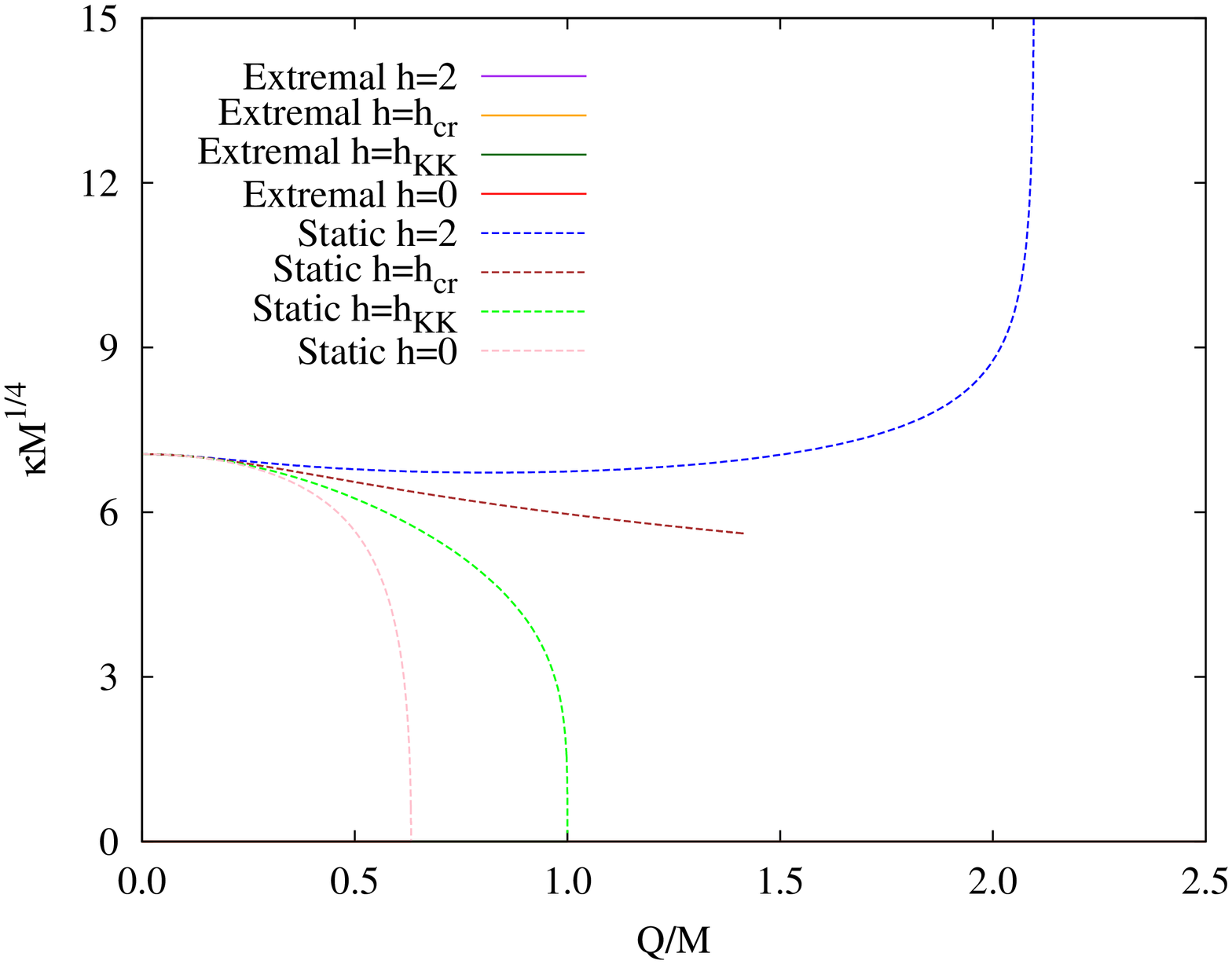}
\label{fig8d}
}
}
\end{center}
\caption{
Properties of EMd black hole solutions in seven dimensions
are shown for several values of the dilaton coupling constant $h$:
$h=0$ (EM), $h_{\rm KK}=\sqrt{3/5}$, $h_{\rm cr}=\sqrt{8/5}$, and $h=2$.
(a) The scaled area $a_{\rm H} = A_{\rm H}/M^{5/4}$
versus the scaled charge $q=|Q|/M$
for extremal and for static solutions,
providing the boundary of the domain of existence.
For the same sets of (extremal and static) solutions 
we exhibit in
(b) the scaled angular momenta $j=J/M^{5/4}$ versus $q$,
(c) the horizon electrostatic potential $\Phi_{\rm H}$ versus $q$,
(d) the scaled surface gravity $\bar \kappa = \kappa M^{1/4}$ versus $q$.
\label{fig8}
}
\end{figure}

Let us first address the domain of existence again
and recall, that unlike the case of a single non-vanishing angular momentum,
where no extremal solutions exist in $D>5$ dimensions \cite{Myers:1986un},
extremal solutions always exist for odd $D$ black holes with
equal-magnitude angular momenta.

To study the dependence of these black holes on
the dilaton coupling constant $h$, we again consider
several fixed values of $h$:
$h=0$, corresponding to the EM case, 
$h_{\rm KK}=\sqrt{3/5}$,
corresponding to the KK case,
$h_{\rm cr}=\sqrt{8/5}$, 
corresponding to the critical case of the surface gravity
and the horizon angular momentum,
and finally $h=2$.

We exhibit a number of interesting properties of these 
black hole solutions in Figs.~\ref{fig8}.
In particular, we show the extremal and the static solutions,
which form the boundary of the domain of existence
for these seven-dimensional black holes,
while all non-extremal rotating black holes are located inside this boundary.
The quantities shown are
the scaled horizon area $a_{\rm H}$ (Fig.~\ref{fig8a}),
the scaled angular momenta $j$ (Fig.~\ref{fig8b}),
the horizon electrostatic potential $\Phi_{\rm H}$ (Fig.~\ref{fig8c}),
and the scaled surface gravity $\bar \kappa$ (Fig.~\ref{fig8d}).

We note, that the scaled horizon area $a_{\rm H}=A_{\rm H}/M^{5/4}$
is proportional to the scaled angular momenta,
as long as $h$ is non-vanishing.
For $h=0$, we find again the two-branch structure of the
extremal EM solutions, where
only the first part of the near-horizon MP branch 
and the second part of the near-horizon RN branch are realized globally.
For the EMd solutions again the surfaces of
near-horizon solutions are reduced to single curves,
that are - in part - realized globally, for arbitrary non-vanishing value of the
dilaton coupling $h$.

The gyromagnetic ratio $g$ of these black holes
attains the perturbative value $g_{\rm \delta q} = 5$ \cite{Aliev:2004ec}
for small values of $q$,
independent of the dilaton coupling constant $h$.
In all cases the `static' value $g_{\rm \delta j}$, Eq.~(\ref{gdeltaj}), 
is a rather good approximation for the true value,
which becomes exact in the KK case.
Likewise, for the horizon electrostatic potential $\Phi_{\rm H}$
the static value for a given $q$
approximates the true value well,
and becomes exact in the KK case.

The scaled surface gravity $\bar \kappa=\kappa M^{1/4}$
of the static solutions forms the upper boundary
of the domain of existence, while the
scaled surface gravity of the extremal solutions
vanishes.
At the extremal endpoints of the static curves
the surface gravity is zero
for $h < h_{\rm cr}=\sqrt{8/5}$. 
This includes the KK case in seven dimensions,
where $h_{\rm KK}=\sqrt{3/5}$.
At the critical value $h = h_{\rm cr}$ the surface gravity
assumes a finite value, $\kappa_{\rm cr}$,
and for  $h > h_{\rm cr}$ the surface gravity
diverges at the endpoint \cite{Gibbons:1987ps,Horowitz:1991cd}.

As discussed in five dimensions,
the situation is analogous for 
the scaled horizon angular velocity 
$\bar \Omega$ (not shown in the figure).
Only here the static solutions form the lower boundary
of the domain of existence, since they have $\Omega=0$,
while the upper boundary is formed by the extremal rotating solutions.

Again, we can understand this analogy
by recalling our formulae in section \ref{sec_KK_rel}.
There we showed, that for KK black holes in $D$ dimensions 
the scaled surface gravity of static black holes
is proportional to the scaled horizon angular velocity of extremal black holes.
We furthermore showed relation Eq.~(\ref{rel_kaOm}) for KK black holes,
which derived from the fact, that for KK black holes
the horizon electrostatic potential only depends on the scaled charge.

Since the horizon electrostatic potential
depends only little on the angular momenta also for other values of 
the dilaton coupling $h$ in seven dimensions,
relation Eq.~(\ref{rel_kaOm}) holds approximately
also for other values of $h$.
The similarity of the static $\bar \kappa$
and the extremal $\bar \Omega$,
as well as the similarity of the static $a_{\rm H}$ 
and the extremal $j$ thus also holds in seven dimensions.
Again
we conclude, that we can learn much about extremal rotating solutions
from the static solutions.

\boldmath
\subsubsection{$D$-dependence}
\unboldmath

\begin{figure}[p!]
\begin{center}
\vspace{-0.7cm}
\mbox{\hspace{-1.5cm}
\subfigure[][]{
\includegraphics[height=.35\textheight, angle =270]{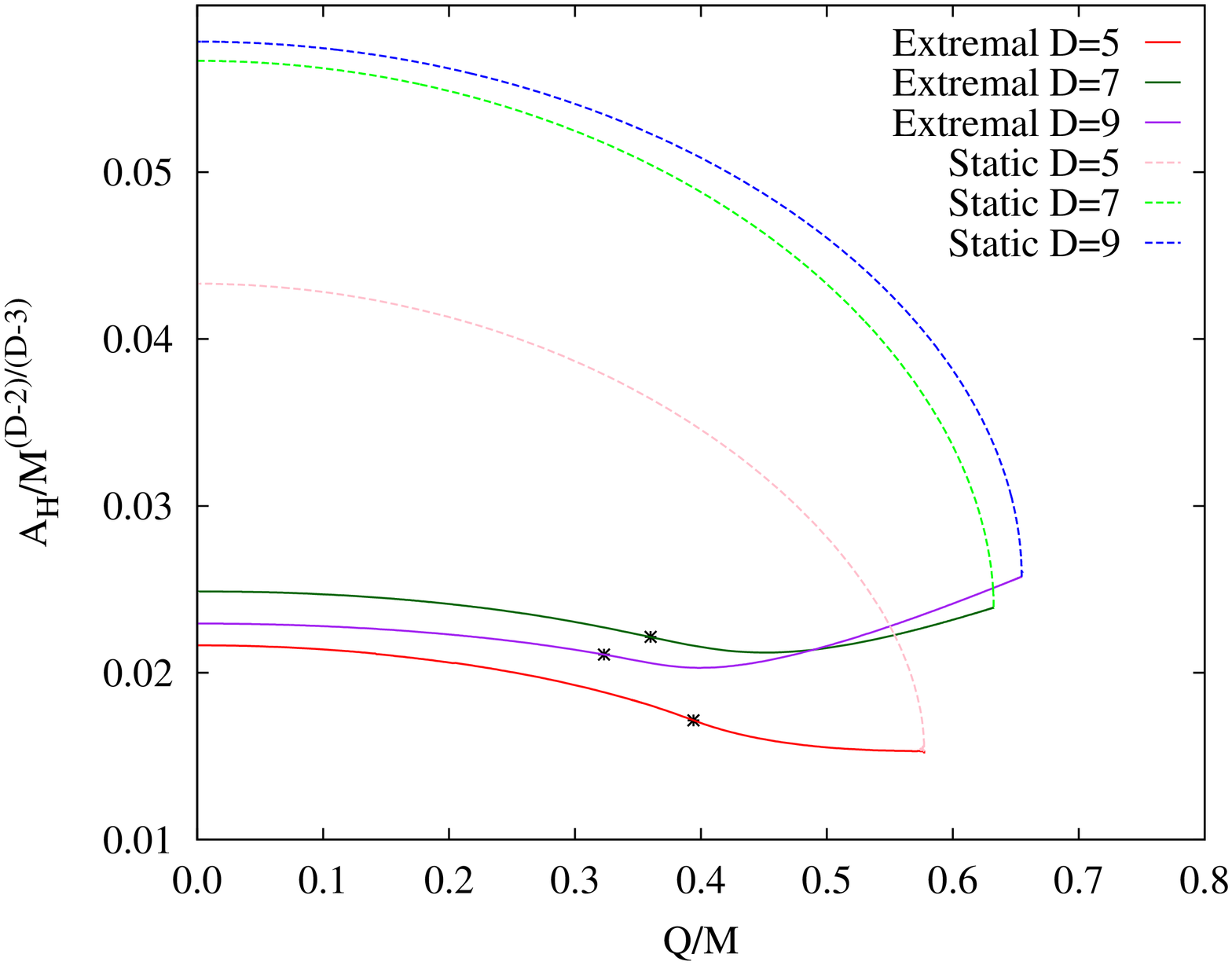}
\label{fig9a}
}
\subfigure[][]{\hspace{-0.5cm}
\includegraphics[height=.35\textheight, angle =270]{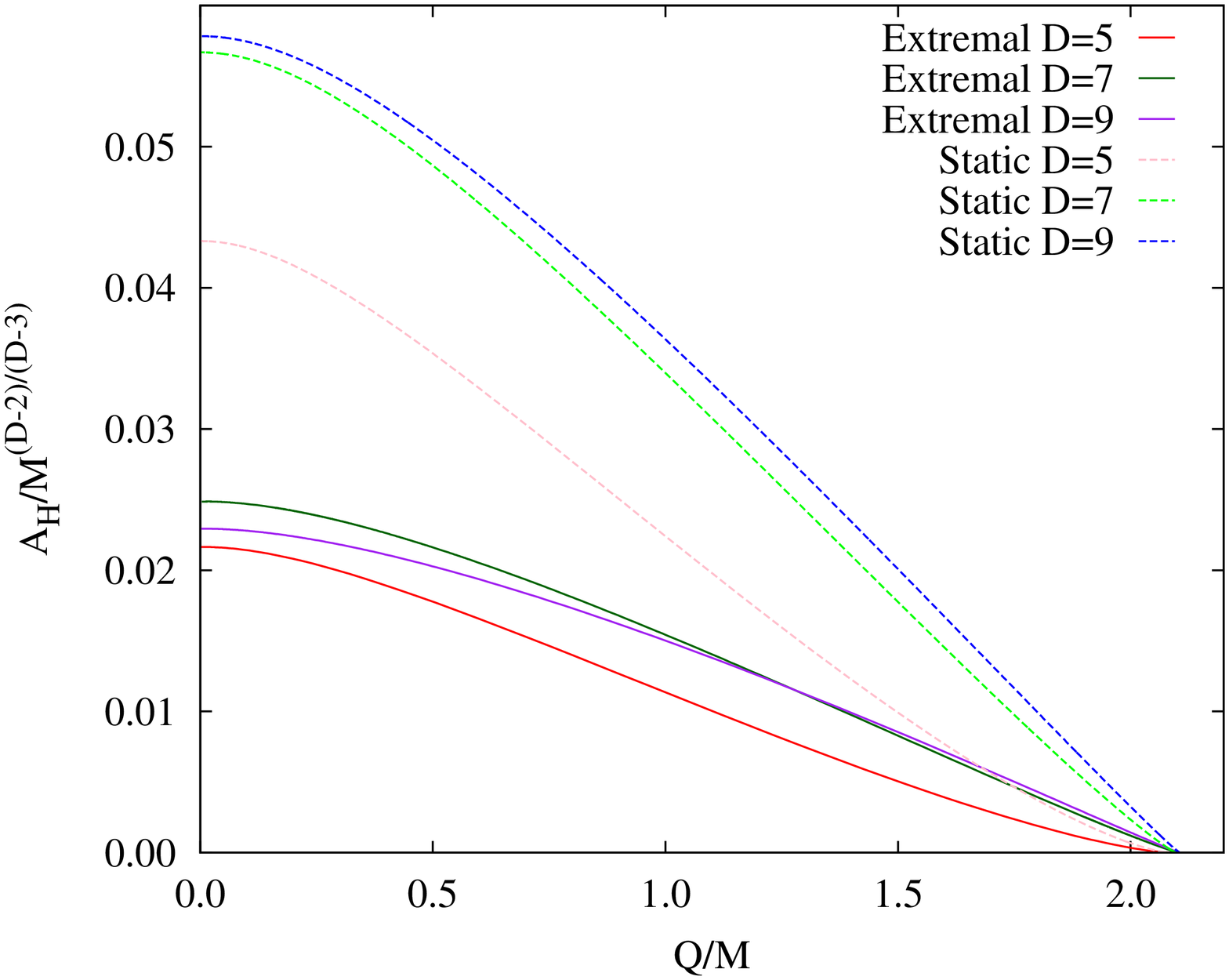}
\label{fig9b}
}
}
\mbox{\hspace{-1.5cm}
\subfigure[][]{
\includegraphics[height=.35\textheight, angle =270]{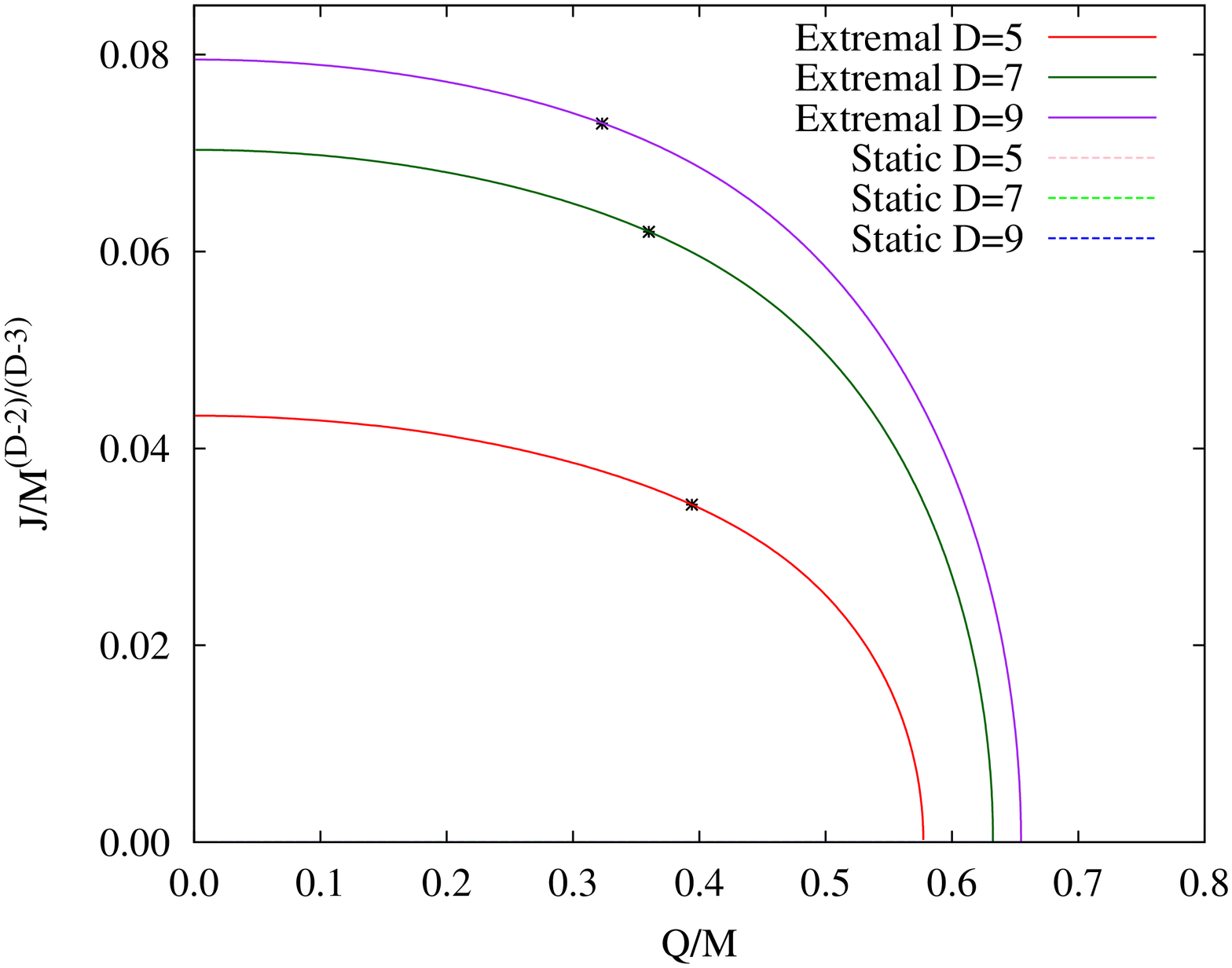}
\label{fig9c}
}
\subfigure[][]{\hspace{-0.5cm}
\includegraphics[height=.35\textheight, angle =270]{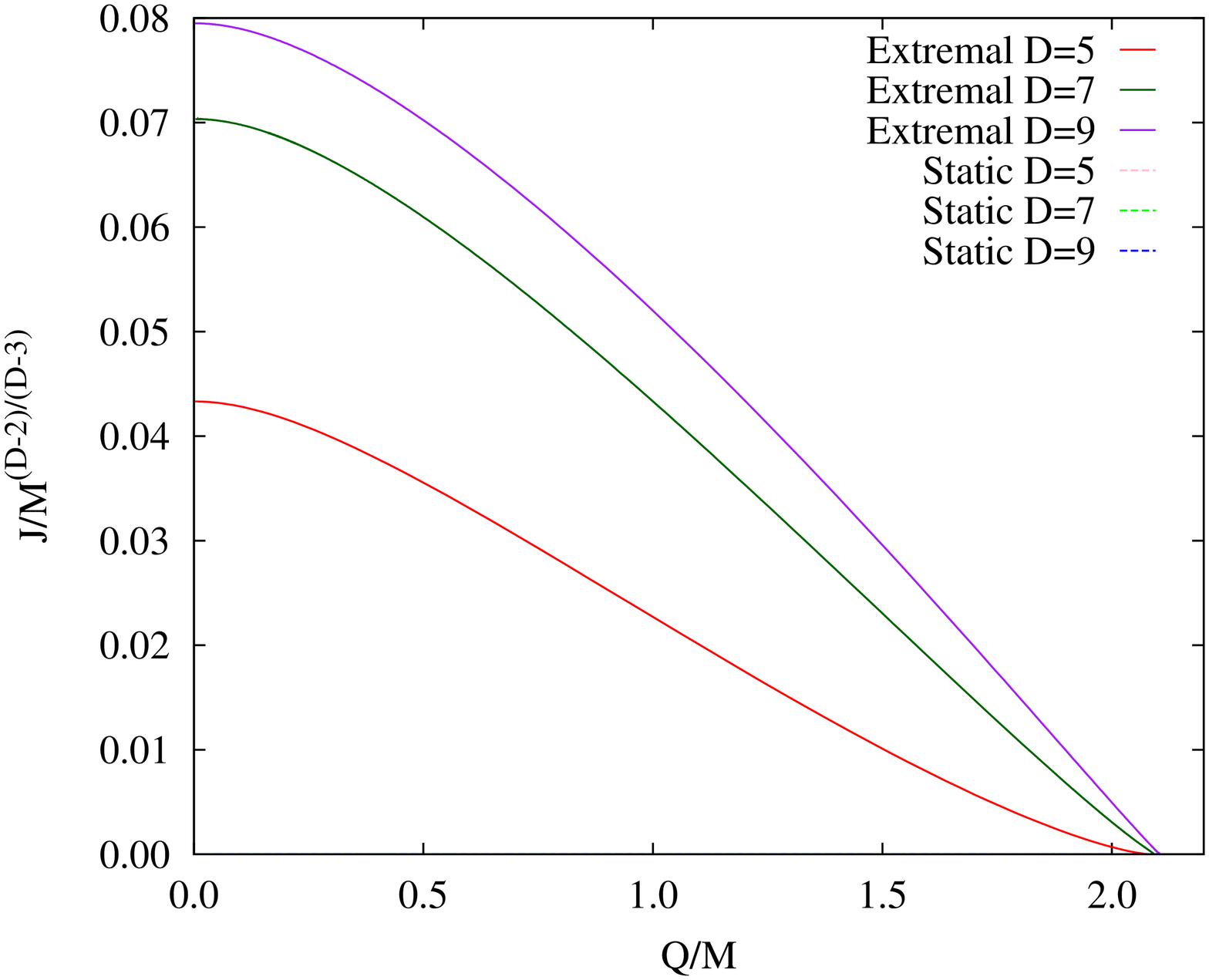}
\label{fig9d}
}
}
\mbox{\hspace{-1.5cm}
\subfigure[][]{
\includegraphics[height=.35\textheight, angle =270]{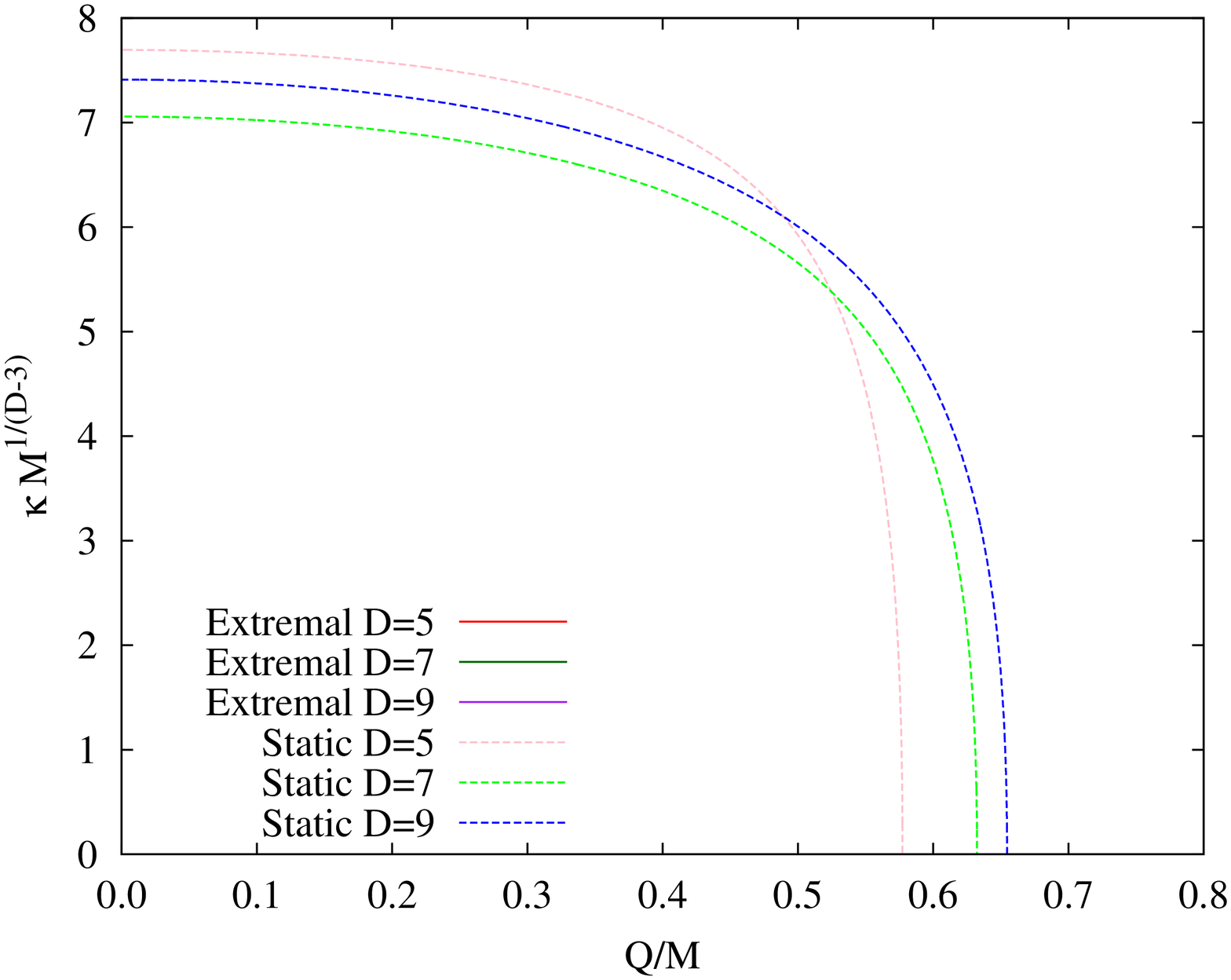}
\label{fig9e}
}
\subfigure[][]{\hspace{-0.5cm}
\includegraphics[height=.35\textheight, angle =270]{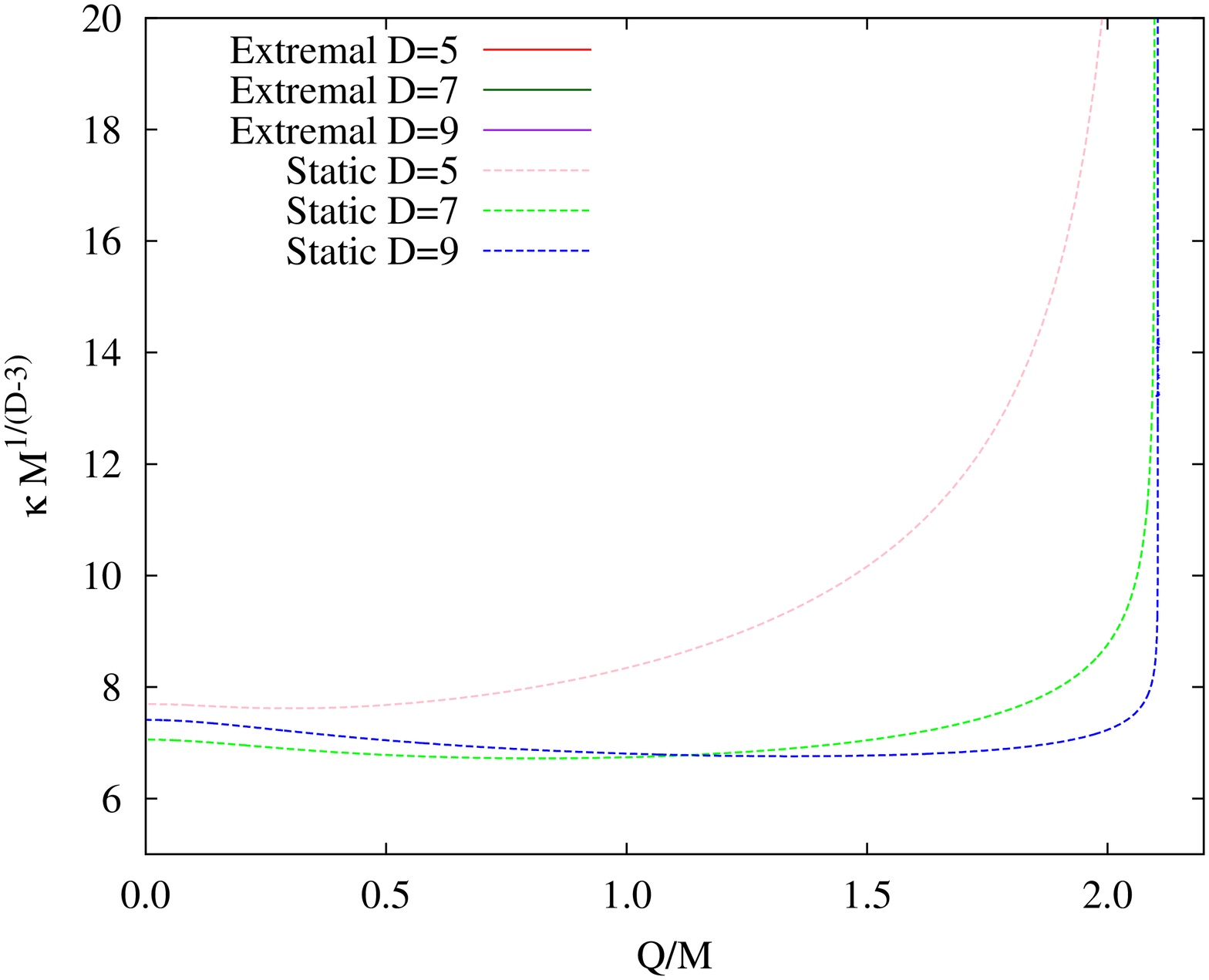}
\label{fig9f}
}
}

\end{center}
\caption{
Properties of EMd black hole solutions in five, seven, and nine dimensions
are shown for the EM case ($h=0$) (left column)
and one example of the EMd case ($h=2$)(right column).
The scaled area $a_{\rm H} = A_{\rm H}/M^{(D-2)/(D-3)}$
versus the scaled charge $q=|Q|/M$
(a) for $h=0$ and (b) for $h=2$
for extremal and for static solutions,
providing the boundary of the domain of existence.
For the same sets of 
solutions 
we exhibit
the scaled angular momenta $j=J/M^{(D-2)/(D-3)}$ versus $q$,
(c) for $h=0$ and (d) for $h=2$, 
and
the scaled surface gravity $\bar \kappa = \kappa M^{1/{(D-3)}}$ versus $q$
(e) for $h=0$ and (f) for $h=2$. 
The asterisks mark the matching points
of the two EM branches.
\label{fig9}
}
\end{figure}

The solutions in nine dimensions do not reveal any unexpected
behavior, but repeat the pattern observed in five
and seven dimensions.
Here the special values of the
dilaton coupling constant beside the EM case
$h=0$, are the KK case
$h_{\rm KK}=\sqrt{4/7}$,
known analytically, and the critical case $h_{\rm cr}=\sqrt{18/7}$.
As in lower dimensions,
for $h=h_{\rm cr}$
the surface gravity of the static solutions
and the horizon angular velocity of the extremal rotating solutions
remain finite at $q_{\rm max}$,
while they diverge at $q_{\rm max}$ for $h>h_{\rm cr}$,
and tend to zero at $q_{\rm max}$ for $h<h_{\rm cr}$.

We compare the solutions in five, seven, and nine dimensions
in Figs.~\ref{fig9}.
In particular, we exhibit the EM case ($h=0$) in the left column
and a generic value of the EMd case ($h=2$) in the right column.

The scaled area $a_{\rm H} = A_{\rm H}/M^{(D-2)/(D-3)}$
of the extremal and static EM solutions is shown in Fig.~\ref{fig9a},
enclosing the domain of existence of the respective sets of 
black holes.
For the extremal solutions it
again reveals the two-branch structure,
as indicated by the asterisks.
As before, the near-horizon solutions are only partly realized globally.
The horizon area of the EM solutions is always finite.
For the EMd solutions the angular momenta and the
horizon area are proportional for all extremal solutions.
A single branch of near-horizon solutions is - in part - realized globally.
The horizon area of EMd solutions vanishes in the extremal static case. 

The gyromagnetic ratio $g$ of these black holes
attains the perturbative value $g_{\rm \delta q} = D-2$ \cite{Aliev:2004ec}
for small $q$,
independent of the dilaton coupling constant $h$.
This indicates already, that
$g$ increases with increasing dimension $D$. 
For a given $q$ the `static' value $g_{\rm \delta j}$, Eq.~(\ref{gdeltaj}),
is a rather good approximation for the true value, 
that becomes exact for $h_{\rm KK}$.
For the EM case $g_{\rm \delta q} =D-3$ 
is a rather good approximation in general.

The horizon electrostatic potential $\Phi_{\rm H}$ (not shown in Figs.~\ref{fig9}),
is remarkably independent of the dimension $D$,
as may be expected from its KK expression, Eq.~(\ref{electros_pot}).
It is significantly influenced only by the dilaton coupling $h$.
Since its dependence on the angular momenta is small,
the horizon electrostatic potential
found in the static limit for a given value of $q$
represents a rather good approximation for
the horizon electrostatic potential,
which becomes exact in the KK case.

The scaled surface gravity $\bar \kappa = \kappa M^{1/{(D-3)}}$
is exhibited in Figs.~\ref{fig9e} and \ref{fig9f}.
Here the static black holes form the upper boundary of the domain of 
existence, while the extremal rotating black holes have vanishing 
$\bar \kappa$.
At the extremal endpoints of the static curves
the surface gravity is zero
for $h < h_{\rm cr}$, corresponding to the EM case.
At the critical value $h = h_{\rm cr}$ the surface gravity
assumes a finite value, $\kappa_{\rm cr}$.
For  $h > h_{\rm cr}$ the surface gravity
diverges at the endpoint, as seen for the EMd case, $h=2$.

The scaled horizon angular velocity $\bar \Omega = \Omega M^{1/{(D-3)}}$ 
(not shown in Figs.~\ref{fig9}) shows an analogous critical behavior,
where the static and the extremal rotating solutions
have switched roles.
Here the extremal rotating black holes 
form the upper boundary of the domain of      
existence, while the static black holes have vanishing 
$\bar \Omega$.
For any dimension
the scaled horizon angular velocity 
of the extremal rotating black holes can be inferred
to good approximation from the
scaled surface gravity of the static black holes.

\section{Conclusions}

We have considered rotating black holes
in Einstein-Maxwell-dilaton theory,
which are asymptotically flat,
and possess a spherical horizon topology.
Restricting to odd dimensions,
$D=2N+1$, and angular momenta with equal-magnitude,
$J=|J_i|$, $i=1,...,N$,
the symmetry of the solutions is enhanced,
and the resulting cohomogeneity-1 problem is
more amenable to approximate analytical treatment
and to numerical analysis.

Treating the dilaton coupling constant $h$
as a parameter, we have studied the dependence of these
solutions on $h$.
Global analytical solutions for these rotating charged black holes
are only available in the Kaluza-Klein case.
For extremal solutions, however, the near-horizon formalism
can be employed to obtain exact solutions,
describing a rotating squashed $AdS_2\times S^{D-2}$ spacetime.
These near-horizon solutions are then interpreted 
as the neighborhood of the event horizon of
extremal black holes.

In the EM case, we found two sets of near-horizon solutions
for all odd dimensions.
Denoting them as the MP branch and the RN branch, 
since they start at the extremal MP solution and the extremal RN solution,
respectively,
we noted that the solutions on the MP branch
possess proportionality of the angular momenta and the horizon area,
whereas the solutions on the RN branch
do not. Instead these exhibit proportionality 
of the angular momenta and the horizon angular momenta.

Interestingly, the branches cross at a critical point,
which we denoted as the matching point.
Numerical construction of the extremal solutions then revealed,
that only parts of these near-horizon solutions are realized globally.
The sets of global solutions consists of the
first part of the MP branch, reaching up to the matching point,
and the second part of the RN branch, starting from the matching point.

For the EMd solutions, on the other hand,
we found a single set of near-horizon solutions.
But because of the presence of the dilaton field 
this set depends on one more parameter. However, this parameter can be 
eliminated by rescaling, so the physical dependence reduces to two independent
parameters. 
Nevertheless, as in the pure EM case,
analytical treatment of the extremal KK solutions
and numerical construction of the extremal solutions
for other values of $h$ have revealed that there are near-horizon solutions
that are not realized globally. All extremal EMd solutions 
possess proportionality of the angular momenta and the horizon area.

We have studied the physical properties of these black holes
numerically,
in particular, their global charges and horizon properties.
The scaling symmetry Eqs.~(\ref{scale1})-(\ref{scaling})
of the solutions has allowed us to give a
comprehensive account of all physical properties,
by scaling these quantities with the appropriate powers of the mass $M$ or
angular momentum $J$. 

For a given dimension $D$ and dilaton coupling constant $h$,
any considered physical property of the corresponding
family of black holes then has a domain of existence,
which is determined by the set of static black holes
on the one hand, and the set of extremal rotating black holes
on the other hand.
A generic non-extremal rotating black hole will be found
within this domain of existence, whereas outside this domain
singular solutions or no solutions at all should be found.

Addressing some properties, in particular, we note that
the gyromagnetic ratio $g$
increases with increasing dimension $D$. 
For the EM case $g=D-3$ 
is a rather good approximation in general,
while in the EMd case, the gyromagnetic ratio 
can depend strongly on the scaled charge $q$.
But the `static' value, obtained perturbatively in the limit $J \to 0$,
is a rather good approximation,
that becomes exact for $h_{\rm KK}$.

The horizon electrostatic potential $\Phi_{\rm H}$,
on the other hand,
is remarkably independent of the dimension $D$.
It is significantly influenced only by the dilaton coupling $h$.
The static limit for a given $q$ also
represents a rather good approximation for
the horizon electrostatic potential.

For the scaled surface gravity $\bar \kappa = \kappa M^{1/{(D-3)}}$
the static black holes form the upper boundary of the domain of 
existence, while the extremal rotating black holes have vanishing 
$\bar \kappa$.
At the extremal endpoints of the static curves
the surface gravity is zero
for $h < h_{\rm cr}$
At the critical value $h = h_{\rm cr}$ the surface gravity
assumes a finite value, $\kappa_{\rm cr}$,
and for  $h > h_{\rm cr}$ the surface gravity
diverges at the endpoint

For the scaled horizon angular velocity $\bar \Omega = \Omega M^{1/{(D-3)}}$,
on the other hand, 
the extremal rotating black holes form the upper boundary of the domain of      
existence, while the static black holes have vanishing 
$\bar \Omega$.
Interestingly,
the scaled horizon angular velocity 
of the extremal rotating black holes can be inferred
to good approximation from the
scaled surface gravity of the static black holes.

While this relation between the scaled horizon angular velocity
and the scaled surface gravity is exact 
in the KK case, for general $h$ it can be seen to arise
from the Smarr law, the low dependence of
the horizon electrostatic potential on the angular momenta,
and the closeness of the horizon area of the static solutions
to the angular momenta of the extremal rotating solutions.
The surprising finding here is thus, that one can learn much
about the extremal rotating solutions from the much simpler
static solutions.

We note, that in four dimensions analogous relations hold,
when the extremal and the static black hole solutions
of EM theory are considered, i.e., the extremal Kerr-Newman (KN)
and the static RN solutions,
or the KK black holes of $4D$ EMd theory.
We conjecture, that these observations hold also for
general dilaton coupling $h$. 
Since these EMd black holes represent a cohomogeneity-2 problem,
we have not yet performed the corresponding numerical calculations
to obtain the extremal solutions for general $h$.

\begin{figure}[t!]
\begin{center}
\mbox{\hspace{-1.5cm}
\subfigure[][]{
\includegraphics[height=.26\textheight, angle =270]{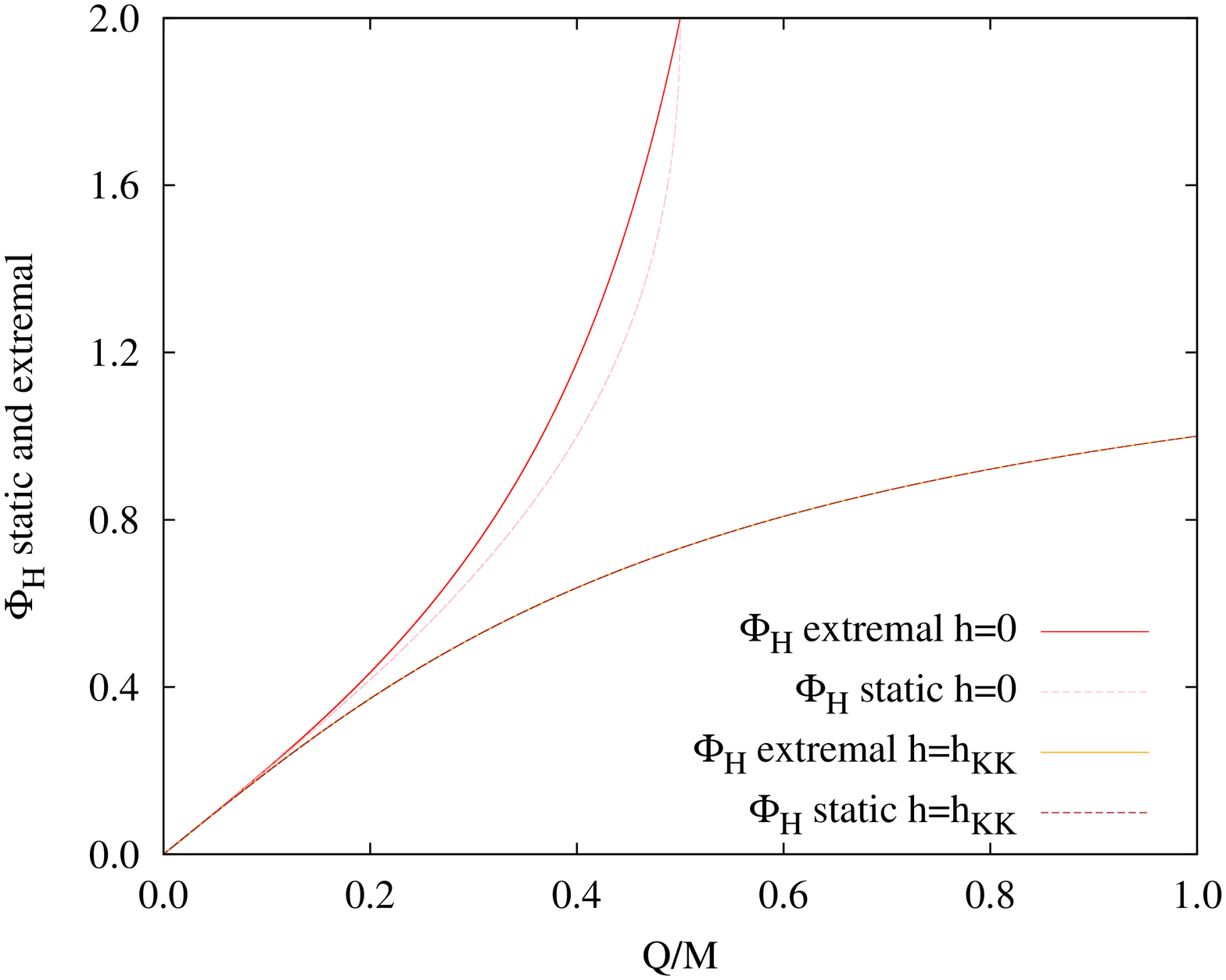}
\label{fig10a}
}
\subfigure[][]{\hspace{-0.5cm}
\includegraphics[height=.26\textheight, angle =270]{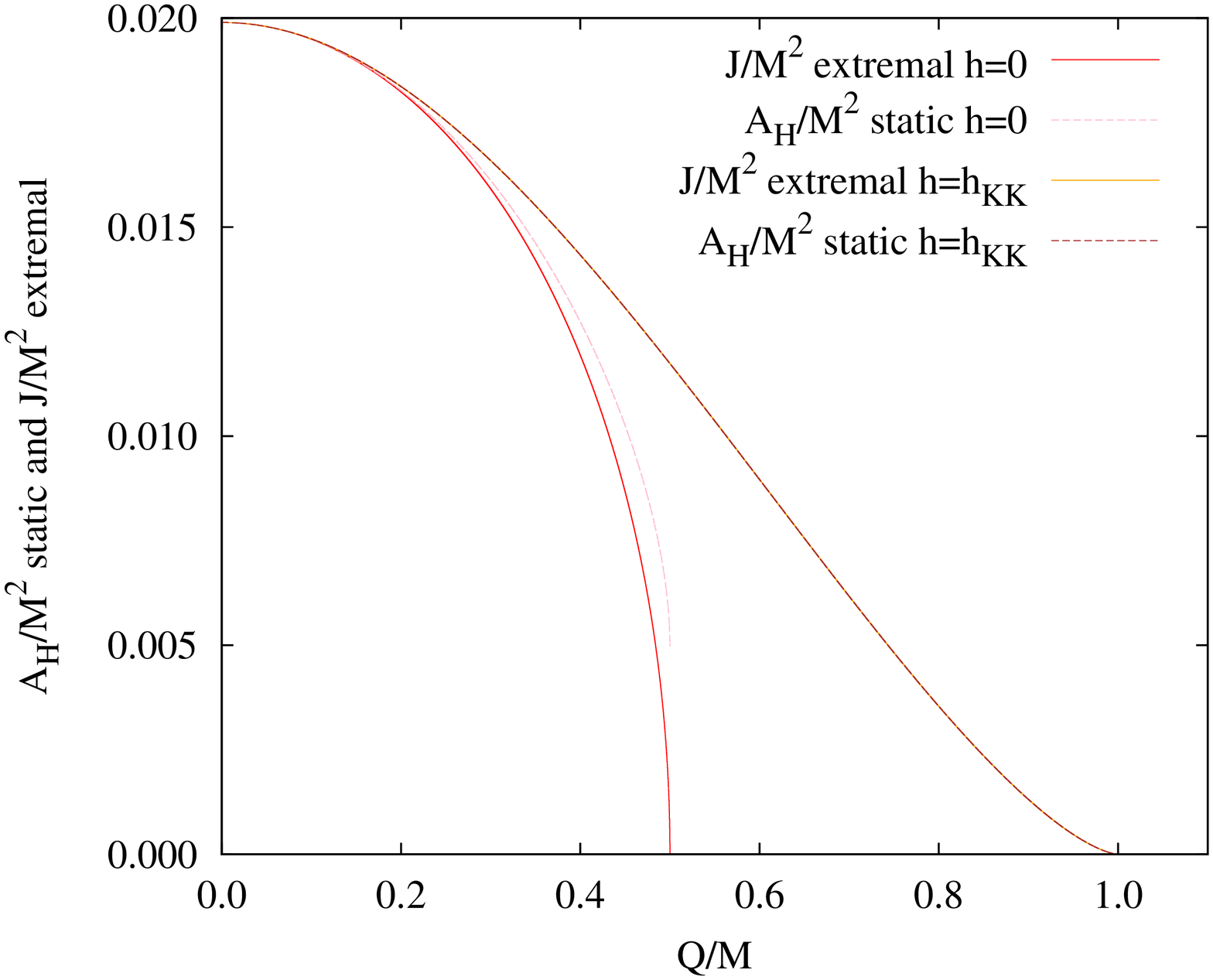}
\label{fig10b}
}
\subfigure[][]{\hspace{-0.5cm}
\includegraphics[height=.26\textheight, angle =270]{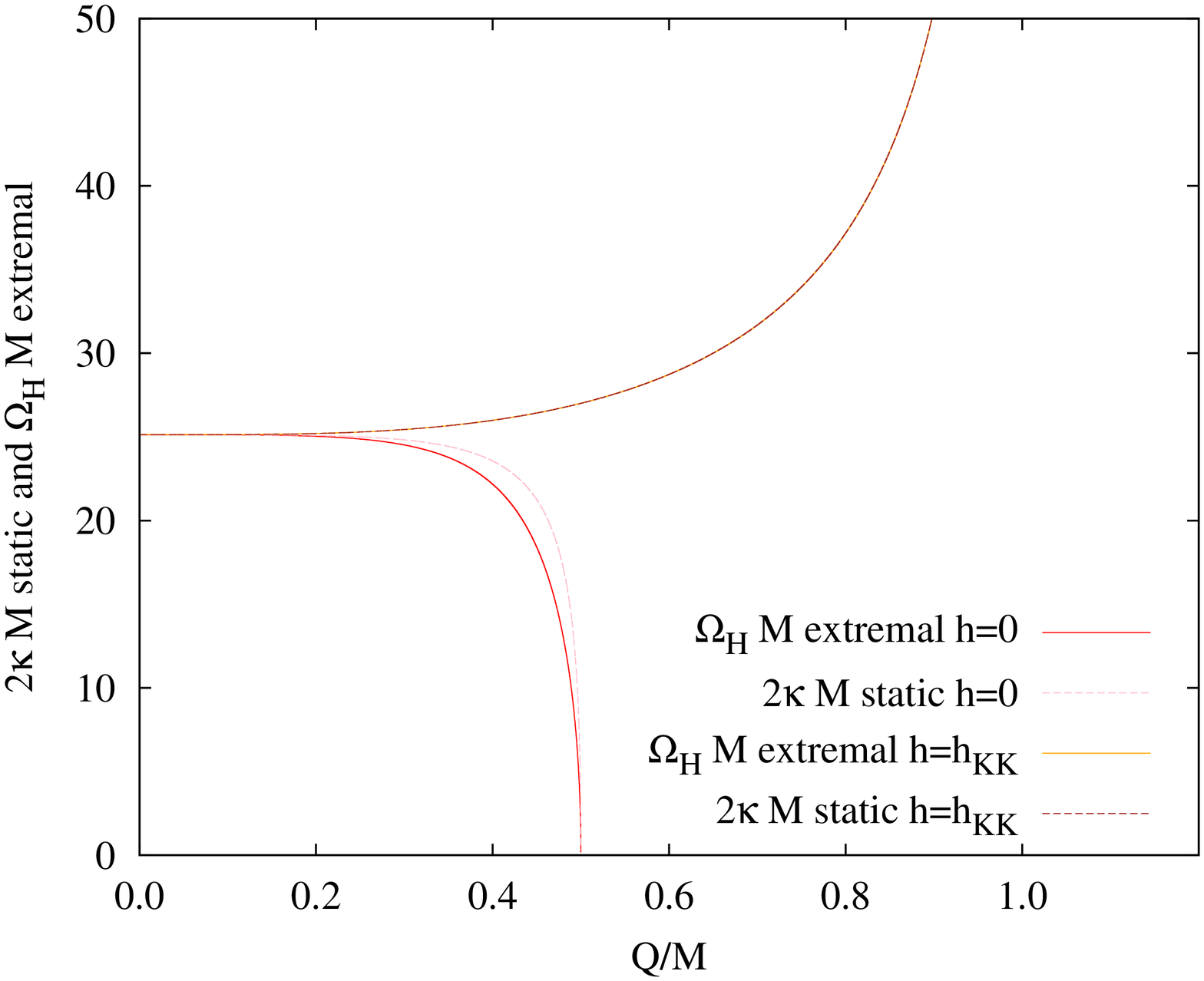}
\label{fig10c}
}
}
\end{center}
\caption{\small
Properties of EMd black hole solutions 
in four dimensions
are shown for the values of the dilaton coupling constant $h$:
$h=0$ (EM), $3/\sqrt{2}$ (KK).
The horizon electrostatic potential $\Phi_{\rm H}$ 
of the extremal and static solutions (a),
the scaled horizon area $a_{\rm H} = A_{\rm H}/M^{2}$
of the static solutions
and the scaled angular momentum $j=J/M^{2}$ of the extremal solutions (b),
and the scaled surface gravity $\bar \kappa = \kappa M$ 
of the static solutions 
and the scaled horizon angular velocity $\bar \Omega = \Omega M$ 
(c)
versus the scaled charge $q=|Q|/M$.
}
\label{fig10}
\end{figure}

Fig.~\ref{fig10a} shows, that the electrostatic potential
is rather independent of the angular momentum, also for
the four-dimensional EM case, while the KK solutions 
again have no dependence on the angular momentum.
Moreover, for a given scaled charge,
the scaled horizon area of the static RN solutions 
is very close to the scaled angular momentum
of the extremal KN solutions,
as long as the scaled charge is not too close to its maximal value,
as seen in Fig.~\ref{fig10b}.
For the KK solutions, 
on the other hand, the scaled horizon area of the static solutions
and the extremal solutions is identical.

Consequently, it follows from the Smarr relation in four dimensions
that the scaled surface gravity of the static RN solutions
is very close to one-half of the scaled horizon angular velocity
of the extremal KN solutions.
This is illustrated in Fig.~\ref{fig10c}.
For the KK solutions
the relation
$2 \bar \kappa_{\rm st} = \bar \Omega_{\rm ex}$ holds exactly,
and we expect that this relation should hold
approximately for all values of the dilaton coupling.

Concerning the near-horizon geometry, 
in \cite{Astefanesei:2006dd} four-dimensional 
Kaluza-Klein black holes are studied. 
However, these solutions possess both electric and magnetic charge
\cite{Rasheed:1995zv,Kleihaus:2003df}.
Interestingly, here (for fixed magnetic charge)
also two branches of black holes are present: the
ergo-free branch, which connects to the static RN solution, and the
ergo branch, which connects to the extremal Kerr solution. 

The near-horizon geometry of the ergo-free branch is independent of the
particular value of the dilaton at infinity, therefore this branch
represents an attactor.
In contrast, the near-horizon geometry of
the ergo branch does depend on the value of the dilaton at infinity
and so does the value of the dilaton at the horizon. 
Of course, for a given angular momentum and charge
all these solutions are equivalent under the scaling symmetry. But the
scaling relation depends on the asymptotic value of the dilaton.

This property of the ergo branch of four-dimensional EMd black holes
is similar to what we have found for the branch of extremal EMd black holes
in odd $D$-dimensions, where the geometry at the horizon also
depends on the asymptotic value of the dilaton at infinity. But since the area
is proportional to the total angular momentum, if the total angular momentum
is fixed, the attractor mechanism works
and the entropy does not depend on the value of the dilaton at infinity.
We plan to investigate these four-dimensional EMd solutions further,
allowing for general values of the dilaton coupling constant 
\cite{Kleihaus:2003df}.


Another case of interest in this connection is represented by the black holes
of Einstein-Maxwell-Chern-Simons (EMCS) theory
\cite{Breckenridge:1996is,Cvetic:2004hs,Chong:2005hr,Kunz:2005ei,Blazquez-Salcedo:2013muz}.
In the supergravity case, these black holes are known analytically.
The extremal black holes then again exhibit two branches, an ergo-free branch
and an ergo branch.
When the Chern-Simons (CS) coupling constant 
is increased beyond the supergravity value,
one of the branches of extremal black holes becomes counterrotating.
When the CS coupling constant is increased even further,
beyond a critical value further branches of extremal black holes arise.

Comparing these global EMCS solutions to near-horizon solutions,
one realizes that the relation between global solutions and
near-horizon solutions becomes even more diverse than observed 
in EMd theory.
In particular, such an EMCS near-horizon solution can correspond
to i) more than one global solution, 
ii) precisely one global solution, 
or iii) no global solution at all.
Clearly, only a study of near-horizon solutions is insufficient to clarify
the domain of existence of extremal solutions.
Thus the construction of the global solutions is indeed essential,
as was first observed for the extremal dyonic black holes of
$D=4$ Gau\ss -Bonnet gravity
\cite{Chen:2008hk}.

\subsection*{Acknowledgment}

We would like to thank J.~Viebahn for initial collaboration on this
project, and B.~Kleihaus and E.~Radu for helpful discussions.
We gratefully acknowledge support by the Spanish Ministerio de Ciencia e
Innovacion, research project FIS2011-28013, and by the DFG, in particular, 
the DFG Research Training Group 1620 ``Models of Gravity''. 
J.~L.~B.-S. was supported by the Spanish Universidad Complutense de Madrid.

\end{document}